\documentclass[sn-mathphys,Numbered]{sn-jnl}

\usepackage{graphicx}
\usepackage{paralist}
\usepackage{multirow,multicol}
\usepackage{amsmath,amssymb,amsfonts,mathdots}
\usepackage{mathrsfs}
\usepackage{xcolor}
\usepackage{textcomp}
\usepackage{manyfoot}
\usepackage{footnote}
\usepackage{algorithm}
\usepackage{algorithmicx}
\usepackage{algpseudocode}
\usepackage{listings}
\usepackage{makecell}
\usepackage{xspace}
\usepackage{float}
\usepackage{longtable}
\usepackage{subcaption}
\floatstyle{plaintop}
\restylefloat{table}
\usepackage{multirow}
\usepackage{url}
\usepackage{mathtools}
\usepackage[utf8]{inputenc}
\usepackage{booktabs}
\usepackage[]{siunitx}
\DeclareSIUnit{\pp}{\textup{pp}}
\usepackage{longtable}
\usepackage{tablefootnote}
\usepackage{nicematrix, tikz}
\usepackage{adjustbox}
\usepackage{anyfontsize}
\usepackage{bookmark}
\usepackage{natbib}

\newcommand{\new}[1]{\textcolor{black}{#1}}
\newcommand{\minor}[1]{\textcolor{black}{#1}}

\newcommand{\hdfs}{HDFS\xspace}
\newcommand{\hadoop}{Hadoop\xspace}
\newcommand{\fdataset}{F-dataset\xspace}
\newcommand{\openstack}{OpenStack\xspace}
\newcommand{\bgl}{BGL\xspace}
\newcommand{\tbird}{Thunderbird\xspace}
\newcommand{\spirit}{Spirit\xspace}
\newcommand{\hades}{Hades\xspace}
\newcommand{\hdfsMessages}{11175629}
\newcommand{\hdfsBlocks}{575061}
\newcommand{\hdfsBadBlocks}{16838}
\newcommand{\hdfsGoodBlocks}{558223}

\newcommand{\ep}{execution path\xspace}

\newcommand{\hadesMinIR}{0.13\%\xspace}
\newcommand{\hadesMaxIR}{1.60\%\xspace}
\newcommand{\hadesMinWS}{10}
\newcommand{\hadesMaxWS}{300}

\newcommand{\spiritMinIR}{29.27\%\xspace}
\newcommand{\spiritMaxIR}{49.23\%\xspace}
\newcommand{\spiritMinWS}{10}
\newcommand{\spiritMaxWS}{250}
\newcommand\footnotewithoutSymbol[1]{\begingroup
  \renewcommand\thefootnote{}\footnote{#1}\addtocounter{footnote}{-1}\endgroup
}

\begin{document}

\title{A Comprehensive Study of Machine Learning Techniques for Log-Based Anomaly Detection}

\author*[1]{\fnm{Shan} \sur{Ali}}\email{shan.ali@uottawa.ca}

\author[2]{\fnm{Chaima} \sur{Boufaied}$\dagger$\footnotewithoutSymbol{$\dagger$This work was done while the author was with the University of Ottawa.}}\email{cboufaied@psu.edu.sa}

\author[3]{\fnm{Domenico} \sur{Bianculli}}\email{domenico.bianculli@uni.lu}
\author[1]{\fnm{Paula} \sur{Branco}}\email{pbranco@uottawa.ca}
\author[1,4]{\fnm{Lionel} \sur{Briand}}\email{lbriand@uottawa.ca}

\affil*[1]{\orgname{University of Ottawa}, \country{Canada}}

\affil[2]{\orgname{Prince Sultan University}, \country{Saudi Arabia}}

\affil[3]{\orgname{University of Luxembourg}, \country{Luxembourg}}

\affil[4]{\orgname{Research Ireland Lero Centre, University of Limerick}, \country{Ireland}}

\abstract{Growth in system complexity increases the need for automated techniques dedicated to different log analysis tasks such as Log-based Anomaly Detection (LAD). The latter has been widely addressed in the literature, mostly by means of a variety of deep learning techniques.
However, despite their many advantages, that focus on deep learning techniques is somewhat arbitrary as traditional Machine Learning (ML) techniques may perform well in many cases, depending on the context and datasets. In the same vein, semi-supervised techniques deserve the same attention as supervised techniques since the former have clear practical advantages.
Further, current evaluations mostly rely on the assessment of detection accuracy. However, this is not enough to decide whether or not a specific ML technique is suitable to address the LAD problem in a given context. Other aspects to consider include training and prediction times as well as the sensitivity to hyperparameter tuning, which in practice matters to engineers.

In this paper, we present a comprehensive empirical study, in which we evaluate a wide array of supervised and semi-supervised, traditional and deep ML techniques w.r.t. four evaluation criteria: detection accuracy, time performance, sensitivity of detection accuracy and time performance to hyperparameter tuning. Our goal is to provide much stronger and comprehensive evidence regarding the relative advantages and drawbacks of alternative techniques for LAD. 

The experimental results show that supervised traditional and deep ML techniques fare similarly in terms of their detection accuracy and prediction time \new{on most of the benchmark datasets considered in our study}. Moreover, overall, sensitivity analysis to hyperparameter tuning with respect to detection accuracy shows that supervised traditional ML techniques are less sensitive than deep learning techniques. 
Further, semi-supervised techniques yield significantly worse detection accuracy than supervised techniques.}

 
\maketitle
\section{Introduction}\label{intro}

Systems typically produce execution logs recording execution information about the state of the system, inputs and outputs, and operations performed. These logs are typically used during testing campaigns to detect failures, or after deployment and at runtime, to identify abnormal system behaviors; these are referred to as \emph{anomalies}.   

The Log-based Anomaly Detection (LAD) problem consists of detecting anomalies from execution logs recording normal and abnormal system behaviors. 
It has been widely addressed in the literature by means of deep learning techniques~\cite{du2017deeplog,zhu2020approach,xie2020attention, huang2020hitanomaly, liu2021lognads, meng2019loganomaly, yang2021semiPlelog, zhang2019robust, lu2018detectingCNN, wang2022lightlog, le2021logNeural, guo2021logbert, qi2023logencoder, chen2022tcn,qi2022adanomaly, catillo2022autolog,zhang2023layerlog,almodovar2023logfit,xia2021loggan,hashemi2021onelog,du2021log,li2022swisslog,xie2022loggd,huang2023improving, han2021interpretablesad,lee2023heterogeneous,chen2021experience,le2022logHowFar,wu2023effectiveness,yu2024deep,li2024graph,xiao2024contexlog,guo2024logformer,zang2024mlad,yin2024semi,lin2024fastlogad,gong2024logeta,wang2024loggt}.
Since logs are typically unstructured, many of the supervised and semi-supervised LAD techniques (except NeuralLog~\cite{le2021logNeural}, LayerLog~\cite{zhang2023layerlog}, Logfit~\cite{almodovar2023logfit}\new{,} LogGD~\cite{xie2022loggd}\new{, ContexLog~\cite{xiao2024contexlog} and  SaRLog~\cite{adeba2024sarlog}}) rely on log parsing (e.g., using Drain~\cite{he2017drain}) to identify and extract log templates (also called log events\new{~\cite{landauer2024critical, yang2024try, guo2024logformer, yu2024deep, yang2021semiPlelog, qi2023logencoder, xie2022loggd, gong2024logeta, lee2023heterogeneous, yin2024semi, li2024graph, huang2023improving, chen2022tcn} or log keys~\cite{chen2021experience, du2017deeplog, lu2018detectingCNN, guo2021logbert, zhang2023layerlog, han2021interpretablesad}}).
The extracted templates can be grouped into different windows (i.e., fixed, sliding, or session windows) forming different template sequences.

Features first need to be extracted from different template sequences to enable the use of machine learning (ML) techniques.
DeepLog~\cite{du2017deeplog}, for example, extracts features using sequential vectors where each component is an index-based encoding of a single template within a template sequence. The remaining deep learning techniques rely on semantic vectors to capture the semantic information from the different templates within a sequence. Semantic vectors are obtained by means of different semantic vectorization techniques such as Template2Vec~\cite{meng2019loganomaly}, word2vec~\cite{han2021interpretablesad} (augmented by a Post-Processing Algorithm (PPA)~\cite{wang2022lightlog}), FastText~\cite{zhang2019robust} complemented by Term Frequency - Inverse Document Frequency (TF-IDF~\cite{salton1988term}), Recurrent Neural Network (RNN)-based encoders (e.g., the attention Bi-directional Long Short-Term Memory Bi-LSTM encoder LogVec~\cite{zhang2023layerlog}) and Transformer-based encoders~\cite{huang2020hitanomaly,guo2021logbert,le2021logNeural}.
Based on the above features, existing deep learning techniques detect log anomalies using different types of neural networks such as Recurrent Neural Network (RNN)~\cite{du2017deeplog,zhu2020approach,xie2020attention,liu2021lognads,meng2019loganomaly,yang2021semiPlelog,zhang2019robust,qi2023logencoder, li2022swisslog,zhang2023layerlog,han2021interpretablesad,gong2024logeta, nguyen2024efficient}, Convolutional Neural Network (CNN)~\cite{lu2018detectingCNN,wang2022lightlog, chen2022tcn,hashemi2021onelog,yin2024semi}, Transformer-based deep learning models~\cite{huang2020hitanomaly,le2021logNeural,guo2021logbert,almodovar2023logfit,du2021log,huang2023improving,lee2023heterogeneous,guo2024logformer,zang2024mlad,xiao2024contexlog, adeba2024sarlog}, Auto Encoders (AE)~\cite{catillo2022autolog}, Graph Neural Network (GNN)~\cite{xie2022loggd,li2024graph,wang2024loggt}, and Generative Adversarial Network (GAN)~\cite{xia2021loggan,qi2022adanomaly,lin2024fastlogad}.

Some empirical studies~\cite{le2021logNeural, le2022logHowFar,yin2024semi} investigate the impact of log parsing methods on the detection accuracy of deep learning anomaly detection techniques.
Others~\cite{zhang2022logst} study the impact of different semantic vectorization techniques on the detection accuracy of deep learning techniques.

Detection accuracy has also been further evaluated to assess the impact of several factors~\cite{le2022logHowFar}, such as training data selection strategies, data grouping methods, data imbalance and data noise (e.g., log mislabelling). \new{High detection accuracy often comes with longer training and prediction times, which can be a challenge at run-time for large-scale applications. In such cases, a model with slightly lower detection accuracy but faster time performance may be more practical. The trade-off between detection accuracy and time performance depends on the specific needs of the application, such as the need for real-time detection or available computational resources.\label{timeperformance-updated}}
Thus, some empirical studies (e.g.,~\cite{huang2020hitanomaly, yang2021semiPlelog, wang2022lightlog, le2021logNeural,guo2024logformer,li2024graph,yin2024semi,lin2024fastlogad,xiao2024contexlog}) assess the time performance of alternative LAD techniques.
Further, a technique with an overall high detection accuracy and practical time performance, may be very sensitive to hyperparameter settings and exhibit widely different results across datasets.

Based on the above discussion, we contend that four evaluation criteria should be systemically considered to assess the overall performance of any ML technique for LAD, regardless of the type of learning they involve.
These criteria are i) detection accuracy, ii) time performance,  sensitivity of iii) detection accuracy and iv) time performance to different hyperparameter settings. 

Most of the existing empirical studies focus on supervised deep learning techniques~\cite{huang2020hitanomaly, liu2021lognads, zhang2019robust, le2021logNeural, du2021log, li2022swisslog, xie2022loggd,huang2023improving,zhang2023layerlog,hashemi2021onelog, lee2023heterogeneous,chen2022tcn,han2021interpretablesad}.
Although many studies~\cite{huang2020hitanomaly, liu2021lognads,zhang2019robust, le2021logNeural, guo2021logbert,catillo2022autolog,zhang2023layerlog,du2021log,li2022swisslog,xie2022loggd,huang2023improving,yu2024deep,li2024graph,xiao2024contexlog,guo2024logformer,wang2024loggt} compare some deep learning techniques to some traditional ones, none of these studies systematically evaluates these techniques w.r.t. the four aforementioned evaluation criteria. Indeed, the strong focus on deep learning is rather arbitrary as traditional ML may indeed fare well in many situations and offer practical advantages.  Further, including semi-supervised learning in such studies is also important given the usual scarcity of anomalies in many logs.   

In this paper, we report on the first comprehensive, systematic empirical study that includes not only deep learning techniques but also traditional ones, both supervised and semi-supervised, considering the four aforementioned evaluation criteria.
More precisely, we systematically evaluate and compare, on \new{seven} benchmark datasets, a) supervised traditional (Support Vector Machine SVM~\cite{cortes1995support} and Random Forest RF~\cite{breiman2001randomRF}) and deep learning techniques (Long Short-Term Memory LSTM~\cite{hochreiter1997LSTM}\new{,} LogRobust~\cite{zhang2019robust} \new{and NeuralLog~\cite{le2021logNeural}}), as well as b) semi-supervised traditional (One Class SVM OC-SVM~\cite{scholkopf2001estimating}) and deep learning techniques (DeepLog~\cite{du2017deeplog} \new{and Logs2Graphs~\cite{li2024graph}}). We compare them in terms of i) detection accuracy, ii) time performance, sensitivity of iii) detection
accuracy and iv) time performance to hyperparameter tuning.

Our experimental results show that supervised traditional and deep ML techniques perform very closely in terms of detection accuracy and prediction time. Further, supervised traditional ML techniques show less sensitivity to hyperparameter tuning than deep learning techniques. Last, semi-supervised techniques, both traditional and deep learning, do not fare well in terms of detection accuracy, when compared to supervised ones.

The results suggest that, despite the strong research focus on deep learning solutions for LAD, traditional ML techniques such as Random Forest can fare much better with respect to our four criteria and therefore be a solution of choice in practice. Semi-supervised techniques, however, do not seem to be a good option at this point, resulting in practical challenges to collect sufficient anomalous log data. 

The rest of the paper is organized as follows. 
Section~\ref{background} explains and formalizes the background concepts used in the rest of the paper and provides a brief overview of the ML techniques considered in the study. 
Section~\ref{state} reports on state-of-the-art empirical studies that are related to the study presented in the paper.
Section~\ref{encoding} explains the semantic vector embedding techniques we used to extract features from input log data. 
Section~\ref{evaluationMehodology} describes the design of our empirical study.
Section~\ref{results} reports and discusses the results of the different supervised and semi-supervised ML techniques. 
Section~\ref{conclusion} concludes the paper, providing directions for future work.
 \section{Background }
\label{background} 
In this section, we first introduce the different concepts used in the remainder of the paper (\S~\ref{executionLog}). 
\new{We then briefly describe the common workflow of LAD using DL models(\S~\ref{ladWorkflow}).}
\new{Finally, }\new{w}e describe three traditional ML techniques (\S~\ref{traditional}) and \new{five} deep learning techniques (\S~\ref{deep}) that have been previously used to address the LAD problem and are considered in our study.

\subsection{Execution Logs}\label{executionLog} 
Information about system executions is usually stored in log files, called execution logs. These logs help with troubleshooting and hence help system engineers understand the behavior of the system under analysis across its different executions. We distinguish between normal and abnormal system executions. The former represents an expected behavior of the system, while the latter represents an anomalous system behavior, possibly leading to a failure. These system executions are therefore stored in labeled execution logs, where the label refers to whether the execution is normal or not. 

An execution log can be defined as a sequence of consecutive log entries that capture the behavior of the system over a given time period. 
A log entry contains: i) an ID; ii) the timestamp at which the logged event was recorded; iii) the log message denoting the occurrence of an event, called \textit{log event occurrence}~\cite{le2021logNeural, zhang2019robust} \new{(also called occurrence of log template~\cite{wu2023effectiveness})};
and iv) the parameter value(s) recorded for that specific log event occurrence. 
\figurename~\ref{log} shows an example of an execution log containing ten log entries (seven of which are displayed in the figure). For instance,
the log entry with ID=4 in the figure contains the timestamp ``16:05:14'', an occurrence of log event \emph{gyroscope\_sensor\_reading}, and the corresponding parameter values ($0.0012$, $-0.0086$ and $0.0020$).
The different log entries collected in a log  are chronologically ordered w.r.t. their recorded timestamps.

\begin{figure}[tb]
\centering
\begin{NiceTabular}
{m{1cm} m{2.5cm} m{5.5cm} m{1.75cm}}[hvlines-except-borders] 
\CodeBefore
\rowcolor{gray!15}{1-1}
\Body
\hline
\Block{1-1}{\emph{ID}} & \Block{1-1}{\emph{Timestamp}} & \Block{1-1}{\emph{Log Event Occurrence}} & \Block{1-1}{\emph{Parameter Value}} \\
\hline
$1$ & $16:05:12$ & $battery\_filtered\_voltage\_reading$ & $16.182$ \\ \hline
$2$ & $16:05:13$ & $gyroscope\_sensor\_reading$ & $-0.0013$, $-1.135$, $-0.002$ \\ \hline
$3$ & $16:05:13$ & $ekf2\_attitude\_pitch\_reading$ & $-0.026$ \\ \hline
$4$ & $16:05:14$ & $gyroscope\_sensor\_reading$ & $0.0012$, $ -0.0086$, $0.0020$ \\ \hline
$5$ & $16:05:14$ & $battery\_filtered\_voltage\_reading$ & $15.687$ \\ \hline
$6$ & $16:05:19$ & $battery\_filtered\_voltage\_reading$ & $14.921$ \\ \hline
$\dots$ & $\dots$ & $\dots$ & \dots\\ \hline
$10$ & $16:05:48$ & $ekf2\_attitude\_pitch\_reading$ & $-0.025$ \\ \hline
\end{NiceTabular}
\caption{Example of an Execution Log}
\label{log}
\end{figure}
 \begin{figure}[tb]
\includegraphics[width=13.5cm]{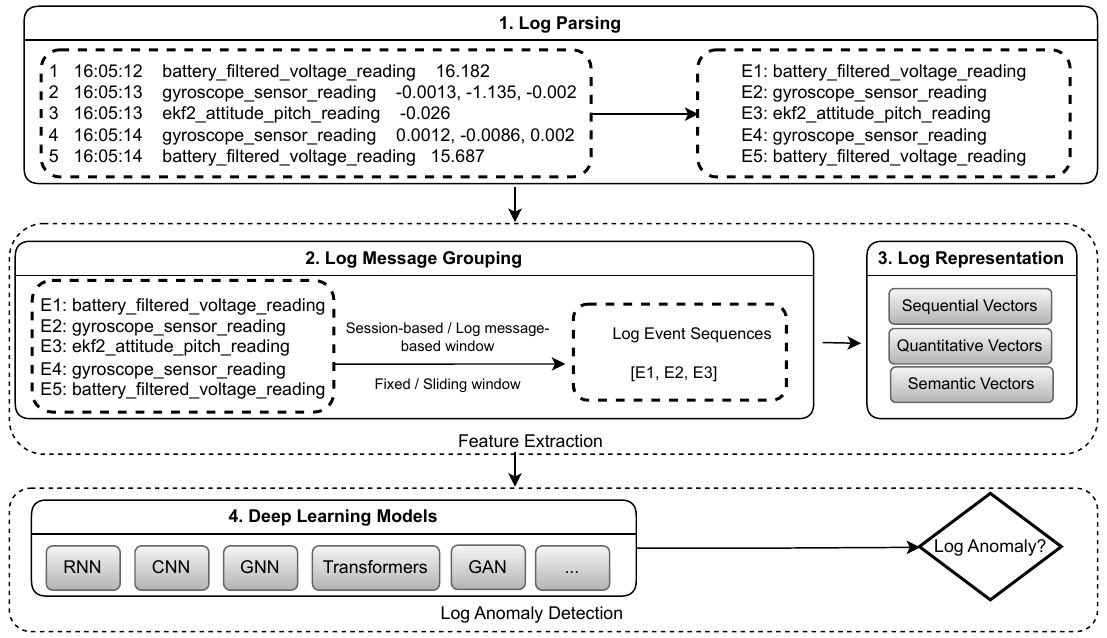}

\caption{\new{Common Workflow of LAD using Deep Learning Models}}
\label{fig:ladWorkflow}
\end{figure}
 
An \emph{\ep} \new{\cite{xie2020attention,le2022logHowFar, chen2021experience, catillo2022autolog, zhang2019robust, du2017deeplog, li2022swisslog, landauer2023deep, wu2023effectiveness, huang2020hitanomaly, lu2018detectingCNN, yu2024deep, xiao2024contexlog}} is the projection with respect to the log event occurrences of the sequence of log entries recorded in the log. 
For instance, let us consider the first three log entries in \figurename~\ref{log}. The execution path obtained from these entries is the sequence of the three corresponding log event occurrences (\textit{battery\_filtered\_voltage\_reading}, \textit{gyroscope\_sensor\_reading}, \textit{ekf2\_attitude\_pitch\_reading}). 
An \ep is called \textit{anomalous} (i.e., containing \ep log anomalies) when the order of its sequence of log event occurrences is unexpected. 

Given a log $\sigma$, we denote by $\sigma(l)$ the log event occurrence recorded at the entry of log $\sigma$ having an ID equal to $l$. For instance, given the log in \figurename~\ref{log}, we have $\sigma(2) = \text{\textit{gyroscope\_sensor\_reading}}$.
We introduce a word-based tokenization function $W$ that, given a log event occurrence as input, returns the sequence of words contained in the log event occurrence. For instance, $W(\sigma(2)) = (\text{\textit{gyroscope}}, \text{\textit{sensor}}, \text{\textit{reading}})$.

\subsection{\new{Common workflow of LAD using \new{Deep Learning (DL)} models}}\label{ladWorkflow}
\new{As shown in \figurename~\ref{fig:ladWorkflow}, the common workflow of LAD using DL models~\cite{le2022logHowFar, wu2023effectiveness, zhang2019robust, wang2022lightlog} includes several steps: log parsing, log message grouping, log representation, and log anomaly detection through appropriate deep learning models.}

\new{
Logs are typically unstructured; they contain ID, timestamp, log event occurrences and the parameter values as shown in \figurename~\ref{log}.
To transform these unstructured logs into a structured format, many of the supervised and semi-supervised LAD techniques (except NeuralLog~\cite{le2021logNeural}, LayerLog~\cite{zhang2023layerlog}, Logfit~\cite{almodovar2023logfit}\new{,} LogGD~\cite{xie2022loggd}, ContexLog~\cite{xiao2024contexlog} and SaRLog~\cite{adeba2024sarlog}) rely on log parsing techniques (e.g., Drain~\cite{he2017drain}) to extract log events from raw log messages. 
For instance, given the first log entry `$1$ $16:05:12$  $battery\_filtered\_voltage\_reading$ $16.182$' in \figurename~\ref{log}, the application of Drain leads to the following log event: `$battery\_filtered\_voltage\_reading$'.}

\new{
The extracted log events can then be grouped into session-based windows (windows that correspond to log event occurrences recorded within a full system execution) or log message-based windows~\cite{huang2020hitanomaly, meng2019loganomaly, yang2021semiPlelog, liu2021lognads, wang2022lightlog, le2021logNeural, le2022logHowFar} (windows determined by a specific number of log messages),
 forming different log event sequences. 
}

\new{
Logs are either collected from i) session-based datasets (see Section ~\ref{sessionBasedDatasets}), where the log event sequences are labeled at the level of the full system execution or ii) log message-based datasets (section~\ref{newDatasets}), in which the labeling process is done at the level of individual log messages, without providing a clear indication on how to group them into log event sequences. Therefore, a log message grouping step~\cite{landauer2023deep} is necessary for such datasets. 
Log event sequences from log message-based datasets are either created using log message-based windows or timestamp-based windows\footnote{\new{Windows delimited by log messages, where the time elapsed between the timestamp of the first and the last log messages within the window is equal to the window size.}}~\cite{qi2023logencoder,guo2021logbert,le2022logHowFar}.
}
\new{Each of these log message-based grouping strategies can be further split into fixed and sliding windows.} 

\new{Once the log event sequences are formed, features need to be extracted to enable the use of ML techniques. This can involve i) encoding log event sequences into vector formats such as sequential vectors (where each component is an index-based encoding of
a single log event within a log event sequence~\cite{du2017deeplog}), quantitative vectors (each component is the occurrences of each log event within a log event sequence~\cite{le2022logHowFar}), or semantic vectors (where each component captures the semantic information from the different
log events within a log event sequence~\cite{zhang2019robust}). These vector representations capture the underlying patterns and semantics of the logs and enable the model to understand differences and similarities between log event sequences, which is crucial for accurately identifying log anomalies (see Section~\ref{encoding}).}
\new{
Finally, the numerical representations are fed into the corresponding DL models, such as
RNN~\cite{du2017deeplog,zhu2020approach,xie2020attention,liu2021lognads,meng2019loganomaly,yang2021semiPlelog,zhang2019robust,qi2023logencoder, xia2021loggan, li2022swisslog,zhang2023layerlog,han2021interpretablesad,gong2024logeta, nguyen2024efficient}, CNN~\cite{lu2018detectingCNN,wang2022lightlog, chen2022tcn,hashemi2021onelog,yin2024semi}, Transformers~\cite{huang2020hitanomaly,le2021logNeural,guo2021logbert,almodovar2023logfit,du2021log,huang2023improving,lee2023heterogeneous,guo2024logformer,zang2024mlad,xiao2024contexlog, adeba2024sarlog}, GNN~\cite{xie2022loggd,li2024graph,wang2024loggt} or GAN~\cite{xia2021loggan,qi2022adanomaly,lin2024fastlogad} to detect log anomalies.
}

\subsection{Traditional ML Techniques}\label{traditional}
We briefly describe three traditional ML techniques 
further used in this study: SVM~\cite{cortes1995support}, RF~\cite{breiman2001randomRF}, and OC-SVM~\cite{scholkopf2001estimating}. 
We selected these techniques since one or several of them have been used as alternatives in the evaluation of previous work on LAD~\cite{huang2020hitanomaly, zhang2019robust, le2021logNeural, liu2021lognads, guo2021logbert, qi2022adanomaly, catillo2022autolog, zhang2023layerlog, li2022swisslog,huang2023improving, chen2021experience, wu2023effectiveness,guo2024logformer,li2024graph,wang2024loggt,xiao2024contexlog}.
Furthermore, an extensive analysis conducted in~\cite{fernandez2014we}, which evaluated $179$ classifiers (including variants of RF, decision tree, and logistic regression) across 121 datasets, showed that RF and SVM tend to be the most accurate classifiers.
 
\subsubsection{Support Vector Machine (SVM)} SVM is a supervised classification ML technique.
The key component of SVM is the kernel function, which significantly affects classification accuracy.
The Radial Basis kernel Function (RBF) is typically the default choice when the problem requires a non-linear model (i.e., non-linearly separable data).
SVM is based on a hyperparameter $\gamma$ that 
controls the distance of influence of a single training data point and a regularization hyperparameter $C$ that is used to add a penalty to misclassified data points. 
SVM was used as an alternative supervised traditional ML technique in the evaluation of some of the state-of-the-art LAD techniques~\cite{huang2020hitanomaly, zhang2019robust, le2021logNeural,guo2024logformer,wang2024loggt,xiao2024contexlog}. Several LAD studies~\cite{zhang2019robust,le2021logNeural,zhang2023layerlog,li2022swisslog,xie2022loggd,huang2023improving,wang2024loggt}
show good detection accuracy for SVM,
when evaluated on 
commonly used benchmark datasets (i.e., \hdfs, \hadoop, \bgl, \tbird and \spirit), which we also consider in this empirical study (see Section~\ref{datasets}). 

\subsubsection{Random Forest (RF)} RF is a supervised classification ML technique. 
Two main hyperparameters can impact its accuracy: 
the number of decision trees $\mathit{dTr}$, a hyperparameter driven by data dimensionality, and the number of randomly selected features $\mathit{sFeat}$, a hyperparameter used in an individual tree. 
RF is used as a supervised traditional ML technique in the evaluation of a few LAD techniques (e.g., LogNads~\cite{liu2021lognads}, AdAnomaly~\cite{qi2022adanomaly}) and showed a better detection accuracy than many other alternative techniques, when evaluated on the \hdfs,  \bgl and \openstack public benchmark datasets. 

\subsubsection{One-class SVM (OC-SVM)} OC-SVM (a variant of SVM) is a semi-supervised classification ML technique. It has the same hyperparameters as SVM.
Anomaly detection using OC-SVM requires building a feature matrix from the normal input.
Unlike the unbounded SVM hyperparameter $C$, the regularization hyperparameter $\nu$ of OC-SVM is lower bounded by the fraction of support vectors (i.e., minimum percentage of data points that can act as support vectors). 
Based on experiments conducted in some recent LAD studies~\cite{catillo2022autolog,zhang2023layerlog}, 
OC-SVM showed to be an accurate semi-supervised technique, when evaluated on the \hdfs, \hadoop and \bgl datasets.
                                    
\subsection{Log-based Deep Learning Techniques}\label{deep}
Over the years, many studies have used deep learning for LAD~\cite{du2017deeplog,zhu2020approach,xie2020attention, huang2020hitanomaly, liu2021lognads, meng2019loganomaly, yang2021semiPlelog, zhang2019robust, lu2018detectingCNN, wang2022lightlog, le2021logNeural, guo2021logbert, qi2023logencoder, chen2022tcn,qi2022adanomaly, catillo2022autolog,zhang2023layerlog,almodovar2023logfit,xia2021loggan,hashemi2021onelog,du2021log,li2022swisslog,xie2022loggd,huang2023improving, han2021interpretablesad,lee2023heterogeneous,chen2021experience,le2022logHowFar,wu2023effectiveness,yu2024deep,li2024graph,xiao2024contexlog,guo2024logformer,zang2024mlad,yin2024semi,lin2024fastlogad,gong2024logeta,wang2024loggt}.
\new{Out of the 42 deep learning techniques listed in Table~\ref{evalLimitations}, the majority of models addressing the LAD problem are based on RNNs, followed by Transformer-based models
with 13 and 11 techniques, respectively.} 
\new{More in detail, many} of the semi-supervised and supervised deep learning techniques in the literature rely on RNNs (e.g.,~\cite{du2017deeplog,meng2019loganomaly,zhang2019robust,zhu2020approach,liu2021lognads,qi2023logencoder,han2021interpretablesad,gong2024logeta}), and more specifically LSTM~\cite{hochreiter1997LSTM}.  
\new{Therefore, in our experiments, we considered the vanilla LSTM as a baseline technique, along with two deep RNN-based ML techniques: DeepLog~\cite{du2017deeplog} and LogRobust~\cite{zhang2019robust}. 
}
\new{
We selected DeepLog because:
\begin{enumerate}[i)]
    \item it is the first method to address the LAD problem using deep learning, establishing a foundational benchmark; 
    \item  it is the most cited technique in the literature~\cite{zhu2020approach, xie2020attention, huang2020hitanomaly, liu2021lognads, meng2019loganomaly, yang2021semiPlelog, qi2023logencoder, wang2022lightlog, guo2021logbert, chen2022tcn, le2022logHowFar,zhang2023layerlog,qi2022adanomaly, almodovar2023logfit, xia2021loggan, hashemi2021onelog,li2022swisslog, han2021interpretablesad,chen2021experience, yang2024try, zang2024mlad, xiao2024contexlog, guo2024logformer, lu2018detectingCNN, lee2023heterogeneous, li2024graph, yin2024semi, lin2024fastlogad, gong2024logeta, wang2024loggt, nguyen2024efficient, landauer2024critical} (referenced in 32 out of the 42 studies listed in Table~\ref{evalLimitations}); and
    \item it achieves an overall high detection accuracy in terms of \emph{F1-score} on the benchmark datasets. 
\end{enumerate}
}
\new{
Similarly, LogRobust is the second mostly cited technique in the literature~\cite{huang2020hitanomaly, wang2022lightlog, le2021logNeural, qi2023logencoder, le2022logHowFar,hashemi2021onelog,li2022swisslog,xie2022loggd, huang2023improving,chen2021experience, yang2024try, guo2024logformer, xiao2024contexlog, zang2024mlad, yu2024deep, wang2024loggt, yang2021semiPlelog, lee2023heterogeneous, du2021log, nguyen2024efficient, adeba2024sarlog}(referenced in 21 out of the 42 studies listed in Table~\ref{evalLimitations}), 
showing an overall high \emph{F1-score} on the benchmark datasets.
}

\new{
Further, among the 11 transformer-based deep ML techniques (NeuralLog~\cite{le2021logNeural}, ContexLog~\cite{xiao2024contexlog}, 
HitAnomaly~\cite{huang2020hitanomaly},
LogBERT~\cite{guo2021logbert}, LogFit~\cite{almodovar2023logfit}, LogAttention~\cite{du2021log}, HilBERT~\cite{huang2023improving}, Hades~\cite{lee2023heterogeneous}, LogFormer~\cite{guo2024logformer}, MLAD~\cite{zang2024mlad} and SaRLog~\cite{adeba2024sarlog}) in Table~\ref{evalLimitations}, 
NeuralLog and LogBERT are the most cited transformer-based ML techniques in the literature, with LogBERT being cited in seven studies~\cite{almodovar2023logfit, qi2023logencoder, huang2023improving, hashemi2021onelog, zang2024mlad, lin2024fastlogad, yin2024semi} and NeuralLog in five studies~\cite{hashemi2021onelog, xie2022loggd, xiao2024contexlog, yu2024deep, adeba2024sarlog}. While LogBERT is cited more frequently than NeuralLog, the latter consistently demonstrates high detection accuracy in terms of \emph{F1-score} across the five studies where it was used, for the majority of the datasets.
 In contrast, LogBERT showed low detection accuracy across some of the benchmark datasets.
NeuralLog was thus chosen as a baseline technique for our study.
}

\new{
Finally, among the three GNN-based deep ML techniques (LogGD~\cite{xie2022loggd}, Logs2Graphs~\cite{li2024graph} and LogGT~\cite{wang2024loggt}), the implementation of Logs2Graphs is the only one made publicly available. We therefore included it as a baseline, reflecting the potential of graph-based models for addressing the LAD problem.
}

In the following, we briefly describe \new{three RNN-based (LSTM~\cite{hochreiter1997LSTM}, DeepLog~\cite{du2017deeplog} and LogRobust~\cite{zhang2019robust}), one transformer-based (NeuralLog~\cite{le2021logNeural}) and one GNN-based (Logs2Graphs~\cite{li2024graph})} deep learning techniques 
that we evaluate in this study.

\subsubsection{LSTM} LSTM~\cite{hochreiter1997LSTM} is a supervised deep learning technique. It is known for its capability to learn long-term dependencies between different sequence inputs. 
LSTM is mainly defined with the following hyperparameters: i) a loss function $\mathit{lF}$; ii) an optimizer $\mathit{opt}$; iii) the number of hidden layers $\mathit{hL}$; iv) the 
amount of training data utilized in a single iteration during the training process (i.e., the batch size) $\mathit{bS}$; and v) a number of epochs $\mathit{epN}$.
\subsubsection{DeepLog} Deeplog~\cite{du2017deeplog} is a semi-supervised technique. It relies on a forecasting-based detection model, i.e., anomalies are detected by predicting the next log event given preceding log events. Since it is based on LSTM, the same aforementioned LSTM hyperparameters apply: loss function $\mathit{lF}$, optimizer $\mathit{opt}$, hidden layers $\mathit{hL}$, batch size $\mathit{bS}$, and epochs $\mathit{epN}$.
DeepLog has been used as an alternative technique in many past studies~\cite{zhu2020approach, xie2020attention, huang2020hitanomaly, liu2021lognads, meng2019loganomaly, yang2021semiPlelog, qi2023logencoder, wang2022lightlog, guo2021logbert, chen2022tcn, le2022logHowFar,zhang2023layerlog,qi2022adanomaly, almodovar2023logfit, xia2021loggan, hashemi2021onelog,li2022swisslog, han2021interpretablesad,chen2021experience, yang2024try, zang2024mlad, xiao2024contexlog, guo2024logformer, lu2018detectingCNN, lee2023heterogeneous, li2024graph, yin2024semi, lin2024fastlogad, gong2024logeta, wang2024loggt, nguyen2024efficient}.
\subsubsection{LogRobust}\label{logRobustBG} LogRobust~\cite{zhang2019robust} is a supervised technique that relies on a classification-based detection model. 
LogRobust detects log anomalies by means of an attention-based Bi-LSTM model, allowing it to capture the contextual semantics across log events within a log event sequence. LogRobust is characterized by the same hyperparameters as LSTM and DeepLog, plus an additional hyperparameter, $\mathit{nEpStop}$, which is used to terminate the model training process if it does not improve after having reached a certain number of epochs.
Further, the attention-based mechanism of LogRobust comes with an attention layer and Bi-LSTM weights that are incrementally updated by means of the 
gradient descent method~\cite{kiefer1952stochastic}.
LogRobust uses the FastText~\cite{joulin2016fasttext} semantics-based embedding technique to encode the log messages from the input logs.
LogRobust has been frequently used in past studies~\cite{huang2020hitanomaly, wang2022lightlog, le2021logNeural, qi2023logencoder, le2022logHowFar,hashemi2021onelog,li2022swisslog,xie2022loggd, huang2023improving,chen2021experience, yang2024try, guo2024logformer, xiao2024contexlog, zang2024mlad, yu2024deep, wang2024loggt, yang2021semiPlelog, lee2023heterogeneous, du2021log, nguyen2024efficient, adeba2024sarlog} and showed an overall high detection accuracy. 

\new{\subsubsection{NeuralLog}\label{neuralLogBG}
NeuralLog~\cite{le2021logNeural} is a transformer-based supervised classification technique that directly identifies log anomalies from unstructured logs without applying any log parsing technique to extract templates from input logs. In addition to the hyperparameters that characterize the RNN-based techniques, NeuralLog is also defined by the number of attention heads $\mathit{attH}$ (parallel attention mechanisms that allow the model to simultaneously focus on different parts of the input sequence, thereby capturing contextual relationships) and the feed-forward network size $\mathit{ffnS}$ (the number of units in the layers that process attention outputs, impacting the learning ability of the model). 
NeuralLog uses the Bert encoder~\cite{devlin2018bert} semantics-based embedding technique to encode log messages from the input logs.
NeuralLog showed an overall high detection accuracy when compared with many deep ML techniques~\cite{hashemi2021onelog, xie2022loggd, xiao2024contexlog, yu2024deep, adeba2024sarlog}.}

\new{\subsubsection{Logs2Graphs}\label{logs2GraphsBG}
Logs2Graphs~\cite{li2024graph} is a recent GNN-based semi-supervised deep ML technique that detects log anomalies by modeling the log data as structured graphs, enabling both anomaly detection and interpretability. This technique first organizes the input logs into graph structures where nodes represent unique log events and directed edges capture the sequential relationships between log events.
In addition to the hyperparameters used in RNN-based techniques, Logs2Graphs is further characterized by the number of convolutional layers ($\mathit{cL}$), which controls the network's depth and sets the number of graph convolutional layers; the proximity parameter ($\mathit{k}$), specifying the order of neighborhood proximity considered within the graph; and the embedding dimensions ($\mathit{embD}$), which determine the size of each node's embedding vector. Logs2Graphs uses Glove~\cite{pennington2014glove} embeddings complemented by TF-IDF~\cite{salton1988term} to encode input logs into semantic vectors.
}
 \section{State of the Art}\label{state}

\citet{le2022logHowFar} conducted an in-depth analysis of representative semi-supervised (DeepLog~\cite{du2017deeplog},  LogAnomaly~\cite{meng2019loganomaly} and PleLog~\cite{yang2021semiPlelog}) and supervised (LogRobust~\cite{zhang2019robust} and CNN~\cite{lu2018detectingCNN}) deep learning techniques, in which several model evaluation criteria (i.e., training data selection strategy, log data grouping, early detection ability, imbalanced class distribution, quality of data and early detection ability) were considered to assess the detection accuracy of these different techniques. 
The study concludes that the detection accuracy of the five deep learning LAD techniques, when taking into account the aforementioned evaluation criteria, is lower than the one reported in the original papers. 
For instance, the training data strategies significantly impact the detection accuracy of semi-supervised deep learning techniques. Further, 
data noise such as mislabelled logs (e.g., logs with errors resulting from the domain expert labelling process) heavily impacts the detection accuracy of supervised deep learning techniques.

Further, depending on the evaluation criteria considered, \citet{le2022logHowFar}'s study leads to different conclusions when comparing detection accuracy between supervised LAD techniques and semi-supervised ones.
Although the semi-supervised techniques DeepLog~\cite{du2017deeplog}, LogAnomaly~\cite{meng2019loganomaly} \new{and PleLog~\cite{yang2021semiPlelog}} are sensitive to training data strategies, \new{DeepLog and LogAnomaly, in particular,} are less sensitive to mislabeled logs than supervised techniques.  
However, supervised deep learning techniques show better detection accuracy than semi-supervised ones when evaluated on a large amount of data (e.g., log event sequences), in spite of their sensitivity to mislabeled logs.

Although many deep learning techniques for LAD have shown high detection accuracy (e.g., \cite{du2017deeplog, meng2019loganomaly, zhang2019robust, huang2020hitanomaly, le2021logNeural}), some of them may not perform well, in terms of training time or prediction time, when compared to traditional ML techniques. For instance, NeuralLog~\cite{le2021logNeural} and HitAnomaly~\cite{huang2020hitanomaly} are slower than traditional ML techniques in terms of training and prediction time, respectively. 
Moreover, traditional ML techniques can be more suitable to detect log anomalies, depending on the application domain and dataset. 
Further, traditional techniques such as SVM~\cite{cortes1995support} are defined with significantly less hyperparameters than deep learning techniques (e.g., LSTM~\cite{hochreiter1997LSTM}), 
thus requiring less computational resources for hyperparameter tuning. 
Therefore, if traditional and deep ML techniques show similar detection accuracy, and if the time performance of traditional ML techniques is significantly better than the one recorded for deep learning techniques, the former are preferable from a practical standpoint. This statement is aligned with the results of a very recent study~\cite{yu2024deep} in which the detection accuracy and the time performance of traditional (K-Nearest Neighbor KNN~\cite{fix1989discriminatory}, Decision Tree~\cite{chen2004failure}) and deep  ML (supervised) techniques (SLFN, CNN~\cite{lu2018detectingCNN}, LogRobust~\cite{zhang2019robust} and NeuralLog~\cite{le2021logNeural}) is compared on five different log-based datasets.

Nevertheless, an ML technique, regardless of its type (either traditional or deep learning), can show i) a high detection accuracy and acceptable time performance, when evaluated on a particular hyperparameter setting, and ii) entirely different results when evaluated on other hyperparameter settings.
From the above discussion, we therefore contend that four evaluation criteria should be systemically considered to assess the overall performance of any ML technique, regardless of the type of learning.
These criteria are i) detection accuracy, ii) time performance, sensitivity of iii) detection accuracy and iv) time performance w.r.t. different hyperparameter settings.

In Table~\ref{evalLimitations}, we list \new{42} studies that use LAD techniques, including the five ones considered in the aforementioned work~\cite{le2022logHowFar}, and summarize their evaluation strategies. We selected the studies that use LAD techniques that i) are either semi-supervised or supervised deep learning techniques and ii) are most cited and used as alternative techniques in the literature.
Column \emph{C.L} indicates, using the symbols Y and N, whether the proposed deep learning LAD technique was compared to at least one traditional ML technique that shares the same model learning type.
We also indicate, for each work, whether the evaluation considered:
 the detection accuracy (column \emph{Acc.}), the time performance (column \emph{Time}),  the sensitivity of the detection accuracy to hyperparameter tuning and different datasets (columns \emph{S.H} and \emph{S.D}, respectively, under the \emph{Sensitivity/Acc.} column), as well as the sensitivity of the time performance to hyperparameter tuning across datasets (columns \emph{S.H} and \emph{S.D} under column \emph{Sensitivity/Time}).
For each of these criteria, we use symbol $+$ to indicate if the evaluation criterion is considered for all the techniques used in the experiments; symbol $\pm$ indicates that the evaluation criterion is only considered for the main technique; symbol $-$ indicates that the evaluation criterion is not measured for any of the techniques considered in the paper.

In addition to datasets obtained from industrial contexts (which are not released for confidentiality reasons), LAD techniques have been mostly evaluated on public benchmark datasets (see Section~\ref{datasets}). In Column \emph{Public Datasets}, we indicate whether or not a public benchmark dataset is used to evaluate the ML techniques in each LAD study, using symbols $\checkmark$ and $\times$, respectively. Moreover, in Column \emph{Impl.}, we indicate whether the implementation of a specific LAD technique is made available in the original paper (using symbols Y and N, respectively). We use the symbol Y$\upharpoonright$ in case the implementation of the LAD technique is provided by third parties. 
Column \emph{Window} indicates whether or not the study assesses the impact of fixed window sizes\footnote{Fixed window sizes are windows that are determined by a specific number of log messages on a dataset.} (using symbols $\checkmark$ and $\times$ respectively) on the detection accuracy of ML techniques, considering log message-based datasets. The latter represent datasets that are labeled at the level of individual log messages and do not provide any indication about how to regroup the different log messages into sequences.

\begin{table}[H]
\setlength\extrarowheight{3pt}
\caption{Comparison of the Evaluation Strategies of Deep Learning Log-based Anomaly Detection Approaches}
\label{evalLimitations}
\begin{NiceTabular}{
    >{\raggedright\arraybackslash}p{2.4cm} *{3}{>{\centering\arraybackslash}p{.15cm}} *{4}{>{\centering\arraybackslash}p{.1cm}} *{7}{>{\raggedright\arraybackslash}p{.2cm}} >{\centering\arraybackslash}p{.25cm} *{1}{>{\centering\arraybackslash}p{.15cm}} }[hvlines-except-borders]
\CodeBefore
\rowcolor{gray!15}{1-3}
\Body
\toprule
\Block{3-1}{\rotatebox{90}{\emph{Study}}} & \Block{3-1}{\rotatebox{90}{\emph{C.L}}} & \Block{3-1}{\rotatebox{90}{\emph{Acc.}}} & \Block{3-1}{\rotatebox{90}{\emph{Time}}} & \Block{1-4}{\emph{Sensitivity}} & & & & \Block{2-7}{\emph{Public Datasets}} & & & & & & & \Block{3-1}{\rotatebox{90}{\emph{Impl.}}} & \Block{3-1}{\rotatebox{90}{Window}} \\
& & & & \Block{1-2}{\emph{Acc.}} & & \Block{1-2}{\emph{Time}} & & & & & & & & \\
& & & & \rotatebox{90}{S.H} & \rotatebox{90}{S.D} & \rotatebox{90}{S.H} & \rotatebox{90}{S.D} & \rotatebox{90}{HD} & \rotatebox{90}{HP} & \rotatebox{90}{OS} & \rotatebox{90}{HA} & \rotatebox{90}{BG} & \rotatebox{90}{TB} & \rotatebox{90}{SP} & & \\
\noalign{\smallskip}
\midrule
\citet{du2017deeplog} (DeepLog) & N  & +  & $\pm$   & $\pm$ & - & -   & - & $\checkmark$ &$\times$ &$\checkmark$ &$\times$ &$\times$ &$\times$ &$\times$ & Y$\upharpoonright$ & $\times$ \\ 
\citet{zhu2020approach} (LogNL) & N & +  & -  & $\pm$ & -  & -  & - & $\checkmark$ &$\times$ &$\checkmark$ &$\times$ &$\times$ &$\times$ &$\times$ & N & $\times$ \\
\citet{xie2020attention} (Att-Gru)  & N  & + & $\pm$ & -  & -  & -  & -  &$\checkmark$ &$\times$ &$\times$ &$\times$ &$\times$ &$\times$ &$\times$ & N & $\times$ \\ 
\citet{huang2020hitanomaly} (HitAnomaly) & Y  & + & + & $\pm$ & -  & - & - & $\diamond\checkmark$ &$\times$ &$\checkmark$ &$\times$ &$\checkmark$ &$\times$ &$\times$ & N & $\times$ \\ 
\citet{liu2021lognads} (LogNads) & Y & +  & $\pm$  & - & - & -    & - & $\diamond\checkmark$ &$\times$ &$\times$ &$\times$ &$\checkmark$ &$\times$ &$\times$ & N & $\checkmark$
\\ 
\citet{meng2019loganomaly} (LogAnomaly) & N & + & - & -  & -   & -    & -  & $\checkmark$ &$\times$ &$\times$ &$\times$ &$\checkmark$ &$\times$ &$\times$ & Y$\upharpoonright$ & $\times$ \\ 
\citet{yang2021semiPlelog} (PleLog) & N  & +  & +  & $\pm$ & $\pm$  & -  & -   & $\checkmark$ &$\times$ &$\times$ &$\times$ &$\checkmark$ &$\times$ &$\times$ & Y$\upharpoonright$ & $\times$ \\ 
\citet{zhang2019robust} (LogRobust) & Y  & +  & -  & -  & - & - & - & $\diamond\checkmark$ &$\times$ &$\times$ &$\times$ &$\times$ &$\times$ &$\times$ & Y$\upharpoonright$ & $\times$ \\ 
\citet{lu2018detectingCNN} (CNN)  & N  & + & -   & +  & - & - & -  & $\checkmark$ &$\times$ &$\times$ &$\times$ &$\times$ &$\times$ &$\times$ & Y$\upharpoonright$ & $\times$ \\ 
\citet{wang2022lightlog} (LightLog)  & N  & +  & + & - & -  & - & - & $\checkmark$ &$\times$ &$\times$ &$\times$ &$\checkmark$ &$\times$ &$\times$ & Y & $\times$ \\ 
\citet{le2021logNeural} (NeuralLog)  & Y  & + & + & $\pm$ & $\pm$ & -    & - & $\checkmark$ &$\times$ &$\times$ &$\times$ &$\checkmark$ &$\checkmark$ &$\checkmark$ & Y$\upharpoonright$ & $\times$ \\ 
\citet{guo2021logbert} (logBert) & Y & +  & - & $\pm$ & - & -  & - & $\checkmark$ &$\times$ &$\times$ &$\times$ &$\checkmark$ &$\checkmark$ &$\times$  & Y & $\times$ \\ 
\citet{qi2023logencoder} (LogEncoder) & N & +  & - & $\pm$ & - & -    & - & $\checkmark$ &$\times$ &$\times$ &$\times$ &$\checkmark$ &$\checkmark$ &$\times$  & N & $\times$ \\ 
\citet{chen2022tcn} (EdgeLog) & N & + & $\pm$  & -     & -     & -    & - & 
$\checkmark$ &$\checkmark$ &$\checkmark$ &$\times$ &$\checkmark$ &$\times$ &$\times$ & N   & $\times$ \\  
\citet{qi2022adanomaly} (AdAnomaly) & N & + & +  & - & - & - & - &
$\checkmark$ &$\times$ &$\checkmark$ &$\times$ &$\checkmark$ &$\times$ &$\times$ & N & $\checkmark$ \\  
\citet{catillo2022autolog} (AutoLog) & Y & + & -  & $\pm$  & + & -    & - & $\times$ &$\checkmark$ &$\times$ &$\times$ &$\checkmark$ &$\times$ &$\times$ & Y & $\times$ \\ 
\bottomrule
\end{NiceTabular}
\end{table}

\addtocounter{table}{-1}
\begin{savenotes}
\begin{table}[H]
\setlength\extrarowheight{3pt}
\captionsetup{list=no}
\caption{\textbf{Continued.} Comparison of the Evaluation Strategies of Deep Learning Log-based Anomaly Detection Approaches}
\begin{NiceTabular}{
    >{\raggedright\arraybackslash}p{2.4cm} *{3}{>{\centering\arraybackslash}p{.25cm}} *{4}{>{\centering\arraybackslash}p{.1cm}} *{7}{>{\raggedright\arraybackslash}p{.2cm}} >{\centering\arraybackslash}p{.15cm} *{1}{>{\centering\arraybackslash}p{.15cm}} }[hvlines-except-borders]
\CodeBefore
\rowcolor{gray!15}{1-3}
\Body
\toprule
\Block{3-1}{\rotatebox{90}{\emph{Study}}} & \Block{3-1}{\rotatebox{90}{\emph{C.L}}} & \Block{3-1}{\rotatebox{90}{\emph{Acc.}}} & \Block{3-1}{\rotatebox{90}{\emph{Time}}} & \Block{1-4}{\emph{Sensitivity}} & & & & \Block{2-7}{\emph{Public Datasets}} & & & & & & & \Block{3-1}{\rotatebox{90}{\emph{Impl.}}} & \Block{3-1}{\rotatebox{90}{Window}} \\
& & & & \Block{1-2}{\emph{Acc.}} & & \Block{1-2}{\emph{Time}} & & & & & & & & \\
& & & & \rotatebox{90}{S.H} & \rotatebox{90}{S.D} & \rotatebox{90}{S.H} & \rotatebox{90}{S.D} & \rotatebox{90}{HD} & \rotatebox{90}{HP} & \rotatebox{90}{OS} & \rotatebox{90}{HA} & \rotatebox{90}{BG} & \rotatebox{90}{TB} & \rotatebox{90}{SP} & & \\
\noalign{\smallskip}
\midrule
\citet{zhang2023layerlog} (LayerLog) & Y & + & -  & -  & -  & - & - & $\checkmark$ &$\times$ &$\times$ &$\times$ &$\checkmark$ &$\times$ &$\times$ & N & $\times$ \\ 
\citet{almodovar2023logfit} (LogFit) & N & + & -  & - & - & - & - & $\checkmark$ &$\times$ &$\times$ &$\times$ &$\checkmark$ &$\checkmark$ &$\times$  & N &$\times$ \\ 
\citet{xia2021loggan} (LogGan) & N & + & -  & $\pm$     & $\pm$ & -    & - & $\checkmark$ &$\times$ &$\times$ &$\times$ &$\checkmark$ &$\times$ &$\times$ & N & $\times$ \\ 
\citet{hashemi2021onelog} (OneLog) & N & + & - & $\pm$     & $\pm$ & -    & - & 
$\checkmark$ &$\checkmark$ &$\times$ &$\times$ &$\checkmark$ &$\checkmark$ &$\checkmark$ & N &$\times$ \\ 
\citet{du2021log} (LogAttention) & Y & + & -  & -     & -     & -   & - & $\checkmark$ &$\times$ &$\times$ &$\times$ &$\checkmark$ &$\times$ &$\times$ & N &$\checkmark$ \\ 
\citet{li2022swisslog} (SwissLog)$\oplus$ & Y & + & +  & - & - & - & - & $\diamond\checkmark$ &$\checkmark$ &$\checkmark$ &$\times$ &$\checkmark$ &$\checkmark$ &$\times$ &Y &$\times$ \\ 
\citet{xie2022loggd} (LogGD) & Y & + & -  & - & - & - & - & $\checkmark$ &$\times$ &$\times$ &$\times$ &$\checkmark$ &$\checkmark$ &$\checkmark$ & N &$\checkmark$ \\ 
\citet{huang2023improving} (HilBert) & Y & + & $\pm$ & - & - & - & - & $\diamond\checkmark$ &$\times$ &$\times$ &$\times$ &$\checkmark$ &$\times$ &$\times$ & N &$\times$ \\ 
\citet{han2021interpretablesad} (InterpretableSAD) & N & + & -  & -  & -     & - & - & $\diamond\checkmark$ &$\times$ &$\times$ &$\times$ &$\diamond\checkmark$ &$\diamond\checkmark$ &$\times$ & Y &$\times$ \\ 
\citet{lee2023heterogeneous} (Hades) & N & + & -  & - & - & - & - & $\times$ &$\times$ &$\times$ & $\checkmark$  &$\times$ &$\times$ &$\times$ & Y &$\times$ \\ 
\citet{chen2021experience} & N~$\dagger$ & + & +  & - & - & - & - & $\checkmark$ &$\times$ &$\times$ &$\times$ &$\checkmark$ &$\times$ &$\times$ & N &$\times$ \\ 
\citet{le2022logHowFar} & N & + & -  & -  & -     & - & - & $\diamond\checkmark$ &$\times$ &$\times$ &$\times$ &$\checkmark$ &$\checkmark$ &$\checkmark$ & Y &$\checkmark$ \\ 
\citet{wu2023effectiveness} & N & + & -  & -  & - & - & - & $\checkmark$ &$\times$ &$\times$ &$\times$ &$\checkmark$ &$\checkmark$ &$\checkmark$ & Y &$\checkmark$ \\ 
\citet{yu2024deep} (LightAD) & Y & + & + & - & - & - & - & $\checkmark$ & $\times$ & $\times$ & $\times$ & $\checkmark$ & $\checkmark$ & $\checkmark$ & Y &$\checkmark$ \\

\new{\citet{li2024graph} (Logs2Graphs)} & \new{Y} & \new{+} & \new{+} & \new{$\pm$} & \new{$\pm$} & \new{-} & \new{+} &
\new{$\checkmark$} & \new{$\checkmark$} & \new{$\times$} &\new{$\times$}  & \new{$\checkmark$} & \new{$\checkmark$} 
& \new{$\checkmark$} 
& \new{Y} & \new{$\times$} \\
\bottomrule
\end{NiceTabular}
\end{table}
\end{savenotes}

\addtocounter{table}{-1}
\begin{savenotes}
\begin{table}[H]
\setlength\extrarowheight{3pt}
\captionsetup{list=no}
\caption{\textbf{Continued.} Comparison of the Evaluation Strategies of Deep Learning Log-based Anomaly Detection Approaches}
\begin{NiceTabular}{
    >{\raggedright\arraybackslash}p{2.4cm} *{3}{>{\centering\arraybackslash}p{.25cm}} *{4}{>{\centering\arraybackslash}p{.1cm}} *{7}{>{\raggedright\arraybackslash}p{.2cm}} >{\centering\arraybackslash}p{.15cm} *{1}{>{\centering\arraybackslash}p{.15cm}} }[hvlines-except-borders]
\CodeBefore
\rowcolor{gray!15}{1-3}
\Body
\toprule
\Block{3-1}{\rotatebox{90}{\emph{Study}}} & \Block{3-1}{\rotatebox{90}{\emph{C.L}}} & \Block{3-1}{\rotatebox{90}{\emph{Acc.}}} & \Block{3-1}{\rotatebox{90}{\emph{Time}}} & \Block{1-4}{\emph{Sensitivity}} & & & & \Block{2-7}{\emph{Public Datasets}} & & & & & & & \Block{3-1}{\rotatebox{90}{\emph{Impl.}}} & \Block{3-1}{\rotatebox{90}{Window}} \\
& & & & \Block{1-2}{\emph{Acc.}} & & \Block{1-2}{\emph{Time}} & & & & & & & & \\
& & & & \rotatebox{90}{S.H} & \rotatebox{90}{S.D} & \rotatebox{90}{S.H} & \rotatebox{90}{S.D} & \rotatebox{90}{HD} & \rotatebox{90}{HP} & \rotatebox{90}{OS} & \rotatebox{90}{HA} & \rotatebox{90}{BG} & \rotatebox{90}{TB} & \rotatebox{90}{SP} & & \\
\noalign{\smallskip}
\midrule
\new{\citet{xiao2024contexlog} (ContexLog)} & \new{Y} & \new{+} & \new{+} & \new{-} & \new{-} & \new{-} & \new{-} & \new{$\checkmark$} & \new{$\times$} & \new{$\times$} &\new{$\times$}  & \new{$\checkmark$} & \new{$\diamond\checkmark$} & \new{$\times$} & \new{N} & \new{$\times$} \\

\new{\citet{guo2024logformer} (LogFormer)} & \new{Y} & \new{+} & \new{+} & \new{-} & \new{-} & \new{-} & \new{-} & \new{$\checkmark$} & \new{$\times$} & \new{$\times$} &\new{$\times$}  & \new{$\checkmark$} & \new{$\checkmark$} & \new{$\times$} & \new{Y} & \new{$\times$} \\

\new{\citet{zang2024mlad} (MLAD)} & \new{N} & \new{+} & \new{-} & \new{$\pm$ } & \new{$\pm$ } & \new{-} & \new{-} 
& \new{$\checkmark$} & \new{$\times$} & \new{$\times$} &\new{$\times$}  & 
\new{$\checkmark$} & \new{$\checkmark$} & \new{$\times$} &
\new{N} & \new{$\times$} \\

\new{\citet{yin2024semi} (BTCNLog)} & \new{N} & \new{+} & \new{+} & \new{-} & \new{-} & \new{-} & \new{-} & \new{$\times$} & \new{$\times$} & \new{$\times$} &\new{$\times$}  & \new{$\checkmark$} & \new{$\checkmark$} & \new{$\checkmark$} &
\new{N} & \new{$\checkmark$} \\
\new{\citet{lin2024fastlogad} (FastLogAD)} & \new{N} & \new{+} & \new{+} & \new{$\pm$} & \new{$\pm$} & \new{-} & \new{-} 
& \new{$\checkmark$} & \new{$\times$} & \new{$\times$} &\new{$\times$}  & \new{$\checkmark$} & \new{$\checkmark$} & \new{$\times$} &
\new{N} & \new{$\times$} \\

\new{\citet{gong2024logeta} (LogETA)} & \new{N} & \new{+} & \new{-} & \new{-} & \new{-} & \new{-} & \new{-} 
& \new{$\times$} & \new{$\times$} & \new{$\times$} &\new{$\times$}  & \new{$\checkmark$} & \new{$\checkmark$} & \new{$\times$} &
\new{N} & \new{$\times$} \\

\new{\citet{wang2024loggt} (LogGT)} & \new{Y} & \new{+} & \new{-} & \new{$\pm$} & \new{$\pm$} & \new{-} & \new{-} 
& \new{$\checkmark$} & \new{$\times$} & \new{$\times$} &\new{$\times$}  & \new{$\checkmark$} & \new{$\checkmark$} & \new{$\times$} &
\new{N} & \new{$\checkmark$} \\
\new{\citet{landauer2024critical} } & \new{Y} & \new{+} & \new{-} & \new{-} & \new{-} & \new{-} & \new{-} 
& \new{$\checkmark$} & \new{$\checkmark$} & \new{$\times$} &\new{$\times$}  & \new{$\checkmark$} & \new{$\checkmark$} & \new{$\times$} &
\new{Y} & \new{$\times$} \\
\new{\citet{yang2024try} (SemPCA)} & \new{N} & \new{+} & \new{+} & \new{-} & \new{-} & \new{-} & \new{-} 
& \new{$\checkmark$} & \new{$\times$} & \new{$\times$} &\new{$\times$}  & \new{$\checkmark$} & \new{$\times$} & \new{$\checkmark$} &
\new{Y} & \new{$\times$} \\
\new{\citet{nguyen2024efficient} (DistilLog)} & \new{N} & \new{+} & \new{+} & \new{$\pm$} & \new{-} & \new{-} & \new{-} 
& \new{$\checkmark$} & \new{$\times$} & \new{$\times$} &\new{$\times$}  & \new{$\checkmark$} & \new{$\times$} & \new{$\times$} &
\new{Y} & \new{$\times$} \\
\new{\citet{adeba2024sarlog} (SaRLog)} & \new{N} & \new{+} & \new{-} & \new{$\pm$} & \new{-} & \new{-} & \new{-} 
& \new{$\times$} & \new{$\times$} & \new{$\times$} &\new{$\times$}  & \new{$\checkmark$} & \new{$\checkmark$} & \new{$\times$} &
\new{N} & \new{$\times$} \\

\textbf{Our study} & \textbf{Y} & \textbf{+}  & \textbf{+} & \multicolumn{1}{c|}{\textbf{+}} & \multicolumn{1}{l|}{\textbf{+}} & \multicolumn{1}{l|}{\textbf{+}}  & \textbf{+} &$\checkmark$ &$\checkmark$ &\new{$\diamond\checkmark$} &$\checkmark$ &$\checkmark$ &$\checkmark$ &$\checkmark$ & \textbf{Y} &$\checkmark$\\

\bottomrule
\end{NiceTabular}
\\ \\
\emph{HD}, \emph{HP}, \emph{OS}, \emph{HA}, \emph{BG}, \emph{TB} and \emph{SP}
refer to \hdfs, \hadoop, \openstack, \hades, \bgl, \tbird and \spirit datasets respectively.  Dataset \emph{HA} (Hades) is named after the technique~\citet{lee2023heterogeneous} (Heterogeneous Anomaly DEtector via Semi-supervised learning), in which the dataset was first used and released.\\
$^\diamond$Authors used the first version of the dataset and\new{/or} a synthetic version of it. \\ 
$\oplus$ Not all datasets are used to evaluate the overall performance of SwissLog. For instance, only \hdfs is used to assess its time performance, whereas \bgl is used to assess the effectiveness of the proposed log parser and the semantic embedding technique used by SwissLog. \\ 
$\dagger$ The study compares different supervised and unsupervised, traditional and deep ML techniques. However, it does not compare any semi-supervised traditional ML technique to a semi-supervised deep ML technique.\\

\end{table}
\end{savenotes}

\paragraph{Comparison among Techniques}
As shown in Table~\ref{evalLimitations}, all empirical studies report the detection accuracy of all the techniques they consider. 
Only a subset of these studies --- focusing on supervised~\cite{huang2020hitanomaly, liu2021lognads, zhang2019robust, le2021logNeural, du2021log, li2022swisslog, xie2022loggd,huang2023improving,zhang2023layerlog, yu2024deep,xiao2024contexlog,gong2024logeta,wang2024loggt} 
and semi-supervised~\cite{guo2021logbert, catillo2022autolog,zhang2023layerlog,li2024graph,zang2024mlad,yin2024semi,lin2024fastlogad} approaches\footnote{LayerLog~\cite{zhang2023layerlog} adopts two log anomaly detection models, supervised and semi-supervised.} --- compare, in terms of detection accuracy, the proposed technique with at least one traditional ML technique. 

\new{
To the best of our knowledge, the most relevant study to our work is an experience report~\cite{chen2021experience}, which systematically evaluates traditional and deep ML techniques in terms of their anomaly detection accuracy, time performance (in terms of model training and prediction time) and robustness (the ability of an ML technique to detect log anomalies in the presence of unseen log events). 
However, the study neither assesses the sensitivity of detection accuracy and time performance to hyperparameter tuning of the different ML techniques across datasets nor investigates the impact of window sizes on detection accuracy.
Further, it does not study the impact of data imbalance—a common characteristic of real-world log-based datasets (e.g., \hdfs, \bgl)—on detection accuracy. 
Additionally, the evaluation of ML techniques in this study is restricted to a very limited number of datasets (\hdfs and \bgl only), thus affecting its generalizability.
In contrast, our work aims to address these limitations by utilizing a broader set of datasets enabling a more comprehensive evaluation of the different ML techniques while  systematically evaluating the impact of data imbalance and window size on detection accuracy, time performance, and sensitivity of both detection accuracy and time performance to hyperparameter tuning.
}\label{stateoftheart:experiencereport}

\paragraph{Datasets}
Most of the LAD techniques~\cite{du2017deeplog, zhu2020approach,xie2020attention,huang2020hitanomaly,liu2021lognads,meng2019loganomaly,yang2021semiPlelog,zhang2019robust,lu2018detectingCNN,wang2022lightlog,guo2021logbert,qi2023logencoder,qi2022adanomaly,catillo2022autolog,almodovar2023logfit,zhang2023layerlog,xia2021loggan,xie2020attention,huang2023improving,han2021interpretablesad,chen2021experience,xiao2024contexlog,guo2024logformer,zang2024mlad,yin2024semi,lin2024fastlogad,gong2024logeta,wang2024loggt, yang2024try, nguyen2024efficient, adeba2024sarlog} have been evaluated on a small set (two to three datasets only) of public benchmark datasets, among which \hdfs and \bgl are the most commonly used ones. Further, even in the case of studies in which LAD techniques are evaluated on a larger set of datasets~\cite{le2021logNeural,chen2022tcn,hashemi2021onelog,li2022swisslog,xie2022loggd,le2022logHowFar,wu2023effectiveness, yu2024deep, li2024graph, landauer2024critical}, they either i) do not report the time performance of the different ML techniques or ii) do not study their sensitivity, in terms of detection accuracy or time performance, to hyperparameter tuning across datasets. 
\paragraph{Hyperparameter Tuning}
Hyperparameter tuning is a time and resource-consuming process that can show a gap i) in the computational time (training time and prediction time) and ii) the resource allocation (e.g., memory, CPU) of a single ML technique and, when evaluated on different hyperparameter settings.
To the best of our knowledge, none of the LAD empirical studies reports the results of the hyperparameter tuning, when applicable.
A common practice across these studies consists of reporting only the exact hyperparameter settings that lead to the best results they report in the 
corresponding research papers.

\begin{table}[tb]
\setlength\extrarowheight{3pt}
\begin{center}
\begin{NiceTabular}
{m{2.5cm} m{2.8cm} m{2cm} m{1.8cm} m{.7cm}}[hvlines-except-borders] 
\CodeBefore
\rowcolor{gray!15}{1-2}
\Body
\toprule
\Block{2-1}{\emph{Technique}} & \Block{1-3}{\emph{Datasets}} & & & \Block{2-1}{\emph{Alt.}} \\ 
& \bgl & \tbird & \spirit \\
Empirical study~\cite{le2022logHowFar} & $[20, 100, 200]$ & $[20, 100, 200]$ & $[20, 100, 200]$ & Y \\
LogNads~\cite{liu2021lognads} & $[10, 20, 30, 40]$ & - & - & N \\
AdAnomaly~\cite{qi2022adanomaly} & $[5, 10, 15, 20, 25, 30]$ & - & - & N \\
LogGD~\cite{xie2022loggd} & $[20, 60, 100]$ & $[20, 60, 100]$ & $[20, 60, 100]$ & Y \\
LogAttention~\cite{du2021log} & $[200, 350, 450, 500]$ & - & - & N \\
Embedding techniques evaluation~\cite{wu2023effectiveness}$\ast$ & - & $[20, 100, 200]$ & -  & N \\
LightAD~\cite{yu2024deep} & [1, 10] & [1, 10] & [1, 10] & N \\
\new{BTCNLog~\cite{yin2024semi}} & \new{[60, 120, 180, 240]} & \new{[60, 120, 180, 240]} & \new{-} & \new{N} \\
\new{LogGT~\cite{wang2024loggt}} & \new{[5, 10, 15, 20, 25, 40]} & \new{[5, 10, 15, 20, 25, 40]} & \new{-} & \new{N} \\
\bottomrule
\end{NiceTabular} \\
\end{center}
$\ast$ The paper studies the impact of different log message-based grouping strategies from \bgl and \spirit datasets on the detection accuracy of different ML techniques, considering different evaluation criteria (e.g., feature aggregation), which fall outside the scope of our paper.
\caption{Existing studies on the impact of fixed window sizes on the detection accuracy of ML techniques}
\label{bestF1_windowSizes}
\end{table}

\paragraph{Impact of Window Size}
As depicted in Column \emph{Window} of Table~\ref{evalLimitations}, only a few studies~\cite{le2022logHowFar, qi2022adanomaly, liu2021lognads, xie2022loggd, du2021log, wu2023effectiveness, yu2024deep,yin2024semi,wang2024loggt} assessed the impact of different fixed window sizes on the detection accuracy of ML techniques. 
More in detail, Table~\ref{bestF1_windowSizes} shows the exact window size values that were used in such studies. We also report (using symbols $Y$ and $N$) whether  these studies assessed the impact of fixed window sizes on the detection accuracy of all the alternative ML techniques (Column \emph{Alt.}) used in their experiments. Only two~\cite{le2022logHowFar, xie2022loggd} out of the \new{nine} aforementioned studies assessed the impact of the fixed window size 
on the detection accuracy of all the alternative techniques.

\paragraph{Motivations for this Work}
Overall, restricting the evaluation of existing LAD studies to reporting the best results (in terms of the \emph{F1-score}) and sharing the exact hyperparameter settings that led to these results does not help external users (e.g., practitioners or researchers) assess the suitability of a specific ML technique to detect log anomalies in a specific context and datasets w.r.t. its i) overall computational time (model training time and prediction time) and ii) sensitivity to hyperparameter tuning. 

Moreover, most studies do not consistently report the execution time of ML techniques; they include either model training time or prediction time. Further, none of these studies provides a systematic evaluation of all the techniques considered in their experimental campaign w.r.t. the four evaluation criteria discussed above.

We therefore believe that conducting large experiments to evaluate ML techniques would be of a great help for practitioners and researchers to better understand what can be expected from different ML techniques and to thus decide what technique(s) they need to apply to address LAD and get the best possible results with the least resources and effort possible. 

Given the aforementioned limitations of existing empirical studies, in this paper, we report on the first comprehensive empirical study, in which we not only evaluate the detection accuracy of existing supervised and semi-supervised, traditional and deep learning techniques applied to LAD, but also assess their time performance as well as the sensitivity of their detection accuracy and their time performance to hyperparameter tuning across datasets.
 \section{Log Representation}\label{encoding}
To use ML techniques for the detection of \ep log anomalies, sequences of log event occurrences need to be first converted into numerical representations that are understandable by such techniques, while preserving their original meaning (e.g., the different words forming each log event occurrence, the relationship between the different log event occurrences forming these sequences). 
 
A recent study~\cite{wu2023effectiveness} has shown that different semantics-based embedding techniques (Word2Vec~\cite{mikolov2013efficient}, FastText~\cite{joulin2016fasttext} and Bert~\cite{devlin2018bert}), when evaluated on different supervised traditional (e.g., SVM and RF) and deep learning (e.g., CNN, LSTM) techniques on four public benchmark datasets (\hdfs, \tbird, \bgl and \spirit), yield similar results in terms of detection accuracy. 
In this study, we apply FastText \new{with the traditional (RF, SVM, OC-SVM) and deep (LSTM, 
 LogRobust~\cite{zhang2019robust}) ML techniques}  since this embedding technique was \new{already} used by \new{LogRobust, along with} previous LAD studies~\cite{le2022logHowFar, yang2021semiPlelog, xie2020attention} and showed good results.
\new{For NeuralLog~\cite{le2021logNeural} and Logs2Graphs~\cite{li2024graph} techniques, we conducted experiments with the embedding methods (Bert~\cite{devlin2018bert} and Glove~\cite{pennington2014glove}, respectively) used in the original papers.}
\new{Regarding FastText,} we use the same log encoding technique adopted by LogRobust~\cite{zhang2019robust}. 
We first pre-process sequences of log event occurrences (e.g., removing non-character tokens, splitting composite tokens into individual ones). We then apply a three-step encoding technique (i.e., word-vectorization, log event occurrence vectorization, sequence vectorization), which we describe next. 

\paragraph{Word Vectorization}
    FastText~\cite{joulin2016fasttext} maps each word $w_i,  1 \le i \le E$, in the sequence of words $ W(\sigma(l))=(w_1,w_2,\dots,w_E)$ extracted from the log event occurrence $\sigma(l)$, to a $d$-dimensional word vector $v_i$ where $ 1 \le i \le E$ and $d=300$\footnote{The choice of $d=300$ dimensions to encode word vectors is motivated by a few LAD techniques (LogRobust~\cite{zhang2019robust}, PleLog~\cite{yang2021semiPlelog}, and LightLog~\cite{wang2022lightlog}) which use the same value for  $d$ when evaluated on \hdfs (one of the datasets considered in our study). For consistency, we adopted the same dimensionality.}
    
    For instance, let us consider the log event occurrences 
    \textit{battery\_filtered\_voltage\_reading} and \textit{gyroscope\_sensor\_reading}, recorded in the first two log entries in \figurename~\ref{log}.
    The corresponding lists of words are $W(\sigma(1))=(\mathit{battery}, \mathit{filtered}, \mathit{voltage}, \mathit{reading})$ and $W(\sigma(2))= (\mathit{gyroscope}, \mathit{sensor}, \mathit{reading})$. 
        By setting the word vector dimension to $d=2$, the different word vectors resulting from FastText and associated to the words $\mathit{battery}$, $\mathit{filtered}$, $\mathit{voltage}$, $\mathit{reading}$, $\mathit{gyroscope}$, and $\mathit{sensor}$ are $v_1= [-0.2759, -0.0023]$, $v_2=[0.2618, 0.1413]$,  $v_3=[-0.4211, 0.4043]$, $v_4=[0.0834, -0.1302]$, $v_5=[0.3276, 0.4368]$  and $v_6=[-0.3419, 0.4418]$, respectively. 
    \paragraph{Log Event Occurrence Vectorization}
    We transform the word list $W(\sigma(l))$ into a word vector list $\mathit{WV}(\sigma(l))$, such that $\mathit{WV}(\sigma(l))=[v_1, v_2, \dots, v_E]$, where $v_j \in \mathbb{R}^{d}$ and $j \in [1,E]$ denotes the word vector.
    $\mathit{WV}(\sigma(l))$ is finally transformed to an aggregated word vector by aggregating all its word vectors using the weighted aggregation technique TF-IDF~\cite{salton1988term}, i.e., a technique that measures the importance of the different words defined in a log event occurrence within a log.
    For instance, the word vector lists associated with the word lists $W(\sigma(1))$ and $W(\sigma(2))$ are  $\mathit{WV}(\sigma(1))=[[-0.3878, -0.0032], [0.3680, 0.1986], [-0.5918, 0.5682], [0.0834, -0.1302]]$ and $\mathit{WV}(\sigma(2))=[[0.4604, 0.6139], [-0.4805, 0.6209], [0.0834, -0.1302]]$, respectively.
    The corresponding aggregated word vectors obtained by means of TF-IDF are $[-0.1321, 0.1583]$ and $[0.0211,  0.3682]$, respectively. 
    \paragraph{Sequence Vectorization}
    Given the aggregated word vectors from the previous step, the latter are further aggregated to form a sequence vector, i.e., a representation of the sequence of log event occurrences. 
    More in detail, the aggregation is done by means of the average operator for each dimension of the aggregated word vectors. 
    For example, if we consider the sequence of log event occurrences obtained from the first two log entries in \figurename~\ref{log}, given the corresponding aggregated word vectors from the previous step ($[-0.1321, 0.1583]$ and $[0.0211,  0.3682]$), the final sequence vector is  $[-0.0555,  0.2633]$.
 \section{Empirical Study Design}
\label{evaluationMehodology}

\subsection{Research Questions}\label{RQs}
The goal of our study is to evaluate alternative ML techniques (described in Section~\ref{background}) when applied to the detection of \ep log anomalies, considering both supervised and semi-supervised, traditional and deep learning techniques. The evaluation is performed based on the four evaluation criteria described in Section~\ref{state}. 
We address the following research questions:
\begin{itemize}
\item RQ1: How do supervised traditional ML and deep learning techniques compare at detecting \ep log anomalies?
\item RQ2: How do supervised traditional ML and deep learning techniques compare in terms of time performance?
\item RQ3: How do semi-supervised traditional ML  and deep learning techniques compare at detecting \ep log anomalies?
\item RQ4: How do semi-supervised traditional ML and deep learning techniques compare in terms of time performance? 
\end{itemize}

These research questions are motivated by the fact that traditional ML techniques are less data hungry and typically less time consuming than deep learning ones when it comes to training the corresponding ML models, and are therefore more practical in many contexts. Therefore, if the loss in detection accuracy is acceptable, assuming there is any, and if the time performance is significantly better than the one recorded for deep ML techniques, traditional ML techniques are preferable.
Similarly, given the scarcity of anomalies in many logs, semi-supervised techniques should be considered in certain contexts.
Further, a ML technique, regardless of its type (traditional or deep), when evaluated on the same dataset, can show wide variation in detection accuracy or time performance from one hyperparameter setting to another. This motivates us to study the sensitivity of such accuracy and performance to hyperparameter tuning.

\subsection{Benchmark Datasets}\label{datasets}

All of the LAD studies illustrated in Table~\ref{evalLimitations} have been evaluated on at least one of the seven public labeled benchmark datasets (\hdfs, \hadoop, \bgl, \tbird, \spirit, \openstack and \hades) listed in Column \emph{Public Datasets}.
These benchmark datasets, except for \spirit and \hades, are published in the LogHub dataset collection~\cite{he2020loghub}.
Most of these datasets are collected from real system executions (\hdfs\cite{xu2009detecting}, \hadoop\cite{hadooplink}, \bgl\cite{oliner2007supercomputersBGL}, \tbird\cite{oliner2007supercomputersBGL}, \spirit\cite{oliner2007supercomputersBGL} and \openstack\cite{du2017deeplog}), whereas one dataset (\hades\cite{lee2023heterogeneous}) is generated from a simulated system.  
Further, different synthetic versions of the first versions of \hdfs, \bgl and \tbird datasets have been proposed in the context of the empirical evaluation of some of the LAD techniques considered in this study. These versions have been obtained by removing, inserting, or shuffling log events within log event sequences to study the impact of log instability on LAD accuracy. 
These synthetic datasets are marked with $\diamond$ symbol in Column \emph{Public Datasets} in the table.
As seen in Table~\ref{evalLimitations}, \hdfs and \bgl are the most commonly used benchmark datasets across LAD studies. \hades has been only used in one LAD study~\cite{lee2023heterogeneous} as it has only been released recently.

In this empirical study, we evaluate ML techniques on datasets that are i) suitable for detecting \ep log anomalies (i.e., datasets containing sequences of log messages), ii) labeled, and iii) publicly available. Public benchmark datasets are either labeled at the level of a single log message (\bgl, \tbird, \spirit, and \hades) or at the level of a session (\hdfs, \hadoop, and \openstack), representing a full system execution. We therefore regroup these datasets into two categories, based on the nature of their original labeling: log message-based or session-based datasets. 

Among the seven public benchmark datasets we identified satisfying our requirements, \openstack is too imbalanced (i.e., anomalies are only injected in four out of 2069 sequences of log event occurrences) and \new{contains a high overlap of 98.5\% between normal and anomalous log event sequences (identical sequences) according to findings reported in a recent study~\cite{landauer2024critical}}, and is thus not suitable for our experiments. 
\new{As an alternative dataset, we used \fdataset~\cite{cotroneo2019bad}, which was recently reported in the experiments of the Semparser~\cite{huo2023semparser} technique.}

\new{
A recent empirical study~\cite{landauer2024critical} recommended the ADFA-LD (Australian Defence Force Academy Log Dataset) dataset~\cite{creech2013generation} for evaluating LAD techniques, as its log anomalies are more complex to detect than those in commonly used benchmark datasets (\hdfs, \hadoop, \bgl and \tbird). 
However, we could not include ADFA-LD in our experiments since only a preprocessed version with numeric identifiers is available, making it unsuitable for our study, where ML techniques (except DeepLog) are fed with semantics encoding of the original log messages (see Section~\ref{encoding}).
}
\new{Overall, we evaluated the ML-based LAD techniques on the seven aforementioned datasets.}

Since all but one of the datasets are unstructured, we used the Drain~\cite{he2017drain} log parsing tool to parse them. We chose Drain since it was already used to parse the logs in the \hades dataset (whose log templates are included in the replication package of the corresponding paper~\cite{lee2023heterogeneous}); moreover, Drain has shown to fare much better than other log parsing tools~\cite{kdbb:icse2022}.
We configured Drain with i) the default settings (similarity threshold = 0.5 and tree depth = 4), that are commonly adopted in LAD studies~\cite{le2022logHowFar,guo2021logbert, li2022swisslog}, and ii) the default regular expressions\footnote{We adopted the regular expressions from Logpai~\cite{logPai} for \hdfs and \hadoop datasets. 
The regular expressions used for the \bgl and \tbird datasets do not cover as many cases as the ones used in LogBert~\cite{guo2021logbert} (e.g., IP address, hexadecimals, and warnings). We therefore adopted the regular expressions shared by the latter. As the regular expressions for the \spirit dataset are not shared, we adopted the ones used for \tbird as both datasets share the same data structure.}.

In the following, we describe in more detail the datasets we used in our empirical study.

\subsubsection{Session-based Datasets}\label{sessionBasedDatasets}
  
The Hadoop Distributed File System (HDFS) dataset was produced
from more than 200 nodes of the Amazon EC2 web service.
HDFS contains \num{\hdfsMessages} log messages collected from \num{\hdfsBlocks} different labeled blocks representing \num{\hdfsGoodBlocks} normal and \num{\hdfsBadBlocks} anomalous program executions.

The Hadoop dataset contains logs collected from a computing cluster running two MapReduce jobs (WordCount and PageRank). Different types of failures (e.g., machine shut-down, network disconnection, full hard disk) were injected in the logs. The dataset contains \num{978} executions; \num{167} logs are normal and the remaining ones (\num{811} logs) are abnormal. 

\new{The \fdataset is a synthesized version of the \openstack dataset that integrates additional failure tests across three subsystems—Cinder, Nova, and Neutron—by injecting 16 distinct types of API error failures. The dataset contains \num{1640} executions; \num{1189} are normal and the remaining ones (\num{451} logs) are abnormal.}

Table~\ref{dataStrategies} shows the main characteristics of the \new{three} session-based datasets used in our experiments.  
Column \emph{\#Temp.} indicates the number of unique templates extracted from the original log messages using the Drain tool.
Columns \emph{\#N} and \emph{\#A} under \emph{\#Seq} indicate the total number of normal and anomalous log event sequences, respectively. Column \emph{IR} represents the 
percentage of log event sequences from the minority class\footnote{Anomalous log event sequences represent the minority class across \new{log message-based} datasets on all window sizes, except for \spirit on window size 300 \new{and session-based datasets except \hadoop}.}.
Columns \emph{Min} and \emph{Max} under \emph{\#Len} denote the minimum and the maximum sequence length, respectively.
We therefore observe that \new{the three session-based datasets} \hdfs, \hadoop, \new{and \fdataset} are imbalanced, where normal \new{sequences represent the majority class on \hdfs and \fdataset and anomalous sequences represent the majority class on \hadoop}. Further, \hdfs is more imbalanced than \new{both} \hadoop \new{and \fdataset}.
The percentage of log event sequences from the minority class in the former represents 2.93\% of the dataset (\num{16838} anomalous sequences out of a total of \num{575061} sequences), while the percentage\new{s} in the other dataset\new{s}
\new{are} 17.08\% \new{for \hadoop} (\num{167} normal sequences out of \num{978} sequences) \new{and 27.5\% \new{for \fdataset} (\num{1189} normal sequences out of a total of \num{1640} sequences)}. 

\subsubsection{Log message-based Datasets}
\label{newDatasets}

The \bgl dataset contains logs collected from
a BlueGene/L supercomputer system at Lawrence Livermore
National Labs (LLNL), California. The dataset contains \num{4747963} labeled log messages among which \num{348460} log messages are anomalous (the remaining \num{4399503} log messages are labeled as normal).

The \tbird dataset contains logs collected from a supercomputer system at Sandia National Labs (SNL). The dataset contains more than \num{200000000} log messages labeled by system engineers. In this study, we selected the first ten million~\footnote{Most of the studies in the literature~\cite{le2022logHowFar, le2021logNeural, almodovar2023logfit, xie2022loggd, yu2024deep} used the same subset of \tbird for their experiments.} log messages from the first version of the \tbird dataset. It contains \num{353794} anomalous log messages while the remaining \num{9646206} are normal.

The \spirit dataset contains aggregated system logs collected from a super computing system at Sandia National Labs. The dataset contains more than \num{172000000} labeled log messages. In this study, we selected the first five million~\footnote{For computation time purpose, we used the same subset from the first version of the \spirit dataset used in the experiments of a recent empirical study~\cite{le2022logHowFar}.} log messages from the first version of the dataset. The selected subset contains \num{4235110} normal log messages while the remaining \num{764890} log messages are labeled as anomalous.

The \hades dataset 
contains logs that were obtained by injecting faults on Apache Spark. It is shared by a recent work~\cite{lee2023heterogeneous} in which a novel semi-supervised ML technique is proposed for large-scale software systems. 
\new{The dataset consists of \SI{37.64}{\mega\byte}  of log files collected over a duration of $95.87$ hours.}
The authors share a structured version of the dataset obtained from Drain. 
\hades contains \num{1048575} labeled log messages, among which only \num{575} log messages are anomalous.  
\begin{table}[tb]
\setlength\extrarowheight{3pt}
\centering
\caption{Characteristics of session-based Benchmark Datasets} \label{dataStrategies}
\begin{NiceTabular}
{m{2.2cm} m{1.4cm} m{1cm} m{1cm} m{1cm} m{1cm} m{1cm}}[hvlines-except-borders] 
\CodeBefore
\rowcolor{gray!15}{1-2}
\Body
\toprule
\Block{2-1}{\emph{Dataset}} & \Block{2-1}{\emph{\# Temp.}} & \Block{1-3}{\emph{\#Seq}} & & & \Block{1-2}{\emph{\#Len}} \\
& & \Block{1-1}{\emph{\#N}} & \Block{1-1}{\emph{\#A}} & \Block{1-1}{\emph{IR}} & \Block{1-1}{\emph{Min}} & \Block{1-1}{\emph{Max}} \\
\hdfs  & 48 & \num{558223} & \num{16838}   & 2.93\% & 1 &  \num{297} \\ 
\hadoop  & \num{340} &  \num{167} & \num{811}  & 17.08\% & 5 & \num{11846} \\
\new{\fdataset}  & \new{\num{97}} &  \new{\num{1189}} & \new{\num{451}}  & \new{\num{27.5}\%} & \new{\num{35}} & \new{\num{1616}} \\
\bottomrule
\end{NiceTabular}
\end{table}
 
\new{Recall that } \new{u}nlike session-based datasets in which sequences are labeled and determined by full executions of a system, log message-based datasets are labeled at the level of individual log messages and do not provide any indication about how to regroup the different log messages into sequences \new{(see Section ~\ref{ladWorkflow})}. Therefore, a log message grouping~\cite{landauer2023deep} step first needs to be applied to such datasets.
\new{More in detail, in some studies log messages are grouped using log message-based windows~\cite{huang2020hitanomaly, meng2019loganomaly, yang2021semiPlelog, liu2021lognads, wang2022lightlog, le2021logNeural, le2022logHowFar} or timestamp-based windows
~\cite{qi2023logencoder,guo2021logbert,le2022logHowFar}. Each of these log message-based grouping strategies can be further split into fixed and sliding windows. 
}

\subsection{Evaluation Metrics}\label{evalMetrics}

In the context of (log-based) anomaly detection, we define the standard concepts of 
\textit{True Positive}, \textit{False Positive}, \textit{True Negative}, and \textit{False Negative} as follows:
 \begin{itemize} 
 \item \textit{TP} (True Positive)\footnote{Note that the positive class in our experiments is always associated with the anomalous log event sequences in every dataset, even when this class is not the minority class in a dataset.} is the number of the abnormal sequences of log event occurrences that are correctly detected by the model. 
\item \textit{FP} (False Positive) is the number of normal sequences of log event occurrences that are wrongly identified as anomalies by the model. 
\item \textit{TN} (True Negative) are normal sequences of log event occurrences that are classified correctly.
\item \textit{FN} (False Negative) is the number of abnormal sequences of log event occurrences that are not detected by the model.
\end{itemize}

In Table~\ref{evalMeasures}, we list the evaluation metrics adopted in the existing studies (already introduced in Section~\ref{state}) to evaluate the corresponding LAD techniques. \emph{Precision} (column \emph{Prec}) indicates the percentage of the \emph{correctly} detected anomalous sequences of log event occurrences over all the anomalous sequences detected by the model; the corresponding formula is $\mathit{Prec}=\frac{\mathit{TP}}{\mathit{TP}+\mathit{FP}}$. 
\emph{Recall} (column \emph{Rec}) is the percentage of sequences of log event occurrences that are \emph{correctly} identified as anomalous over all real anomalous sequences in the dataset;  
it is defined as: $\mathit{Rec}=\frac{\mathit{TP}}{\mathit{TP}+\mathit{FN}}$. 
The \emph{F1-score} (column \emph{F1}) represents the harmonic mean of precision and recall: $\mathit{F1}=\frac{2 * \mathit{Prec} * \mathit{Rec}}{\mathit{Prec} + \mathit{Rec}}$.
\emph{Specificity} (column \emph{Spec}) is the percentage of sequences of log event occurrences that are \emph{correctly} identified as normal over all real normal sequences in the dataset; it is defined as: $\mathit{Spec}=\frac{\mathit{TN}}{\mathit{TN}+\mathit{FP}}$.
\emph{Accuracy} (column \emph{Acc}) is defined as: $\mathit{Acc}=\frac{\mathit{TP} + \mathit{TN}}{\mathit{TP}+\mathit{TN}+\mathit{FN}+\mathit{FP}}$.
\emph{False Positive Rate} (column \emph{FPR}) is defined as: $\mathit{FPR}=\frac{\mathit{FP}}{\mathit{FP}+\mathit{TN}}$.
The corresponding formula for the \emph{Area Under Curve} (column \emph{AUC}) is :
$\mathit{AUC}=\frac{\mathit{Rec} + (1 - \mathit{FPR})}{\mathit{2}}$. \\

\begin{table}[H]
\setlength\extrarowheight{3pt}
\caption{Evaluation metrics considered in existing studies}
\label{evalMeasures}
\begin{NiceTabular}{
    >{\centering\arraybackslash}p{4.5cm} *{7}{>{\centering\arraybackslash}p{0.6cm}}
}[hvlines-except-borders]
\CodeBefore
\rowcolor{gray!15}{1-1}
\Body
\toprule
\Block{1-1}{\emph{Study}} & \Block{1-1}{\emph{Prec}} & \Block{1-1}{\emph{Rec}} & \Block{1-1}{\emph{F1}} & \Block{1-1}{\emph{Acc}}& \Block{1-1}{\emph{Spec}} & \Block{1-1}{\emph{FPR}} & \Block{1-1}{\emph{AUC}} \\
\midrule
\citet{du2017deeplog} (DeepLog) &$\checkmark$&$\checkmark$&$\checkmark$  &$\times$ & $\times$ & $\times$ & $\times$ \\ 
\citet{zhu2020approach} (LogNL)&$\checkmark$&$\checkmark$&$\checkmark$  &$\times$ & $\times$ & $\times$ & $\times$\\
\citet{xie2020attention} (Att-Gru)&$\checkmark$&$\checkmark$&$\checkmark$  &$\checkmark$ & $\times$ & $\times$ & $\times$\\
\citet{huang2020hitanomaly} (HitAnomaly)&$\checkmark$&$\checkmark$&$\checkmark$  &$\times$ & $\times$ & $\times$ & $\times$\\ 
\citet{liu2021lognads} (LogNads)&$\checkmark$&$\checkmark$&$\checkmark$  &$\checkmark$ & $\times$ &$\checkmark$ &$\checkmark$\\
\citet{meng2019loganomaly} (LogAnomaly)&$\checkmark$&$\checkmark$&$\checkmark$  &$\times$ & $\times$ & $\times$ & $\times$\\ 
\citet{yang2021semiPlelog} (PleLog)&$\checkmark$&$\checkmark$&$\checkmark$  &$\times$ & $\times$ & $\times$ & $\times$\\ 
\citet{zhang2019robust} (LogRobust)&$\checkmark$&$\checkmark$&$\checkmark$  &$\times$ & $\times$ & $\times$ & $\times$\\ 
\citet{lu2018detectingCNN}&$\checkmark$&$\checkmark$&$\checkmark$  &$\times$ & $\times$ & $\times$ & $\times$\\ 
\citet{wang2022lightlog} (LightLog)&$\checkmark$&$\checkmark$&$\checkmark$  &$\times$ & $\times$ & $\times$ & $\times$\\ 
\citet{le2021logNeural} (NeuralLog)&$\checkmark$&$\checkmark$&$\checkmark$  &$\times$ & $\times$ & $\times$ & $\times$\\ 
\citet{guo2021logbert} (logBert)&$\checkmark$&$\checkmark$&$\checkmark$  &$\times$ & $\times$ & $\times$ & $\times$\\  
\citet{qi2023logencoder} (LogEncoder)&$\checkmark$&$\checkmark$&$\checkmark$  &$\times$ & $\times$ & $\times$ & $\times$\\ 
\citet{chen2022tcn} (EdgeLog)&$\checkmark$&$\checkmark$&$\checkmark$  &$\times$ & $\times$ & $\times$ & $\times$\\  
\citet{qi2022adanomaly} (AdAnomaly)&$\checkmark$&$\checkmark$&$\checkmark$  &$\times$ & $\times$ & $\times$ & $\times$\\ 
\citet{catillo2022autolog} (AutoLog)&$\checkmark$&$\checkmark$&$\checkmark$  &$\times$ & $\times$ & $\times$ & $\times$\\ 
\citet{zhang2023layerlog} (LayerLog)&$\checkmark$&$\checkmark$&$\checkmark$  &$\times$ & $\times$ & $\times$ & $\times$\\ 
\citet{almodovar2023logfit} (LogFit)&$\checkmark$&$\checkmark$&$\checkmark$  &$\times$ & $\checkmark$ & $\times$ & $\times$\\  
\citet{xia2021loggan} (LogGan)&$\checkmark$&$\checkmark$&$\checkmark$  &$\times$ & $\times$ & $\times$ & $\times$\\ 
\citet{hashemi2021onelog} (OneLog)&$\checkmark$&$\checkmark$&$\checkmark$  &$\times$ & $\times$ & $\times$ & $\times$\\ 
\citet{du2021log} (LogAttention)&$\checkmark$&$\checkmark$&$\checkmark$  &$\times$ & $\times$ & $\times$ & $\times$\\ 
\citet{li2022swisslog} (SwissLog)&$\checkmark$&$\checkmark$&$\checkmark$  &$\times$ & $\times$ & $\times$ & $\times$\\  
\citet{xie2022loggd} (LogGD)&$\checkmark$&$\checkmark$&$\checkmark$  &$\times$ & $\times$ & $\times$ & $\times$\\  
\citet{huang2023improving} (HilBert)&$\checkmark$&$\checkmark$&$\checkmark$  &$\times$ & $\times$ & $\times$ & $\times$\\ 
\citet{han2021interpretablesad} (InterpretableSAD)&$\checkmark$&$\checkmark$&$\checkmark$  &$\times$ & $\times$ & $\times$ & $\times$\\ 
\citet{lee2023heterogeneous} (Hades)&$\checkmark$&$\checkmark$&$\checkmark$  &$\times$ & $\times$ & $\times$ & $\times$\\ 
\citet{chen2021experience}&$\checkmark$&$\checkmark$&$\checkmark$  &$\times$ & $\times$ & $\times$ & $\times$\\ 
\citet{le2022logHowFar}&$\checkmark$&$\checkmark$&$\checkmark$  &$\times$ & $\checkmark$ & $\times$ & $\times$\\ 
\citet{wu2023effectiveness}&$\checkmark$&$\checkmark$&$\checkmark$  &$\times$ & $\times$ & $\times$ & $\times$\\
\citet{yu2024deep}&$\checkmark$&$\checkmark$&$\checkmark$  &$\times$ & $\times$ & $\times$ & $\times$\\
\bottomrule
\end{NiceTabular}
\end{table}

\addtocounter{table}{-1}
\begin{savenotes}
\begin{table}[H]
\setlength\extrarowheight{3pt}
\captionsetup{list=no}
\caption{\textbf{Continued.} Evaluation metrics considered in existing studies}
\label{evalMeasures2}
\begin{NiceTabular}{
    >{\centering\arraybackslash}p{4.5cm} *{7}{>{\centering\arraybackslash}p{0.6cm}}
}[hvlines-except-borders]
\CodeBefore
\rowcolor{gray!15}{1-1}
\Body
\toprule
\Block{1-1}{\emph{Study}} & \Block{1-1}{\emph{Prec}} & \Block{1-1}{\emph{Rec}} & \Block{1-1}{\emph{F1}} & \Block{1-1}{\emph{Acc}}& \Block{1-1}{\emph{Spec}} & \Block{1-1}{\emph{FPR}} & \Block{1-1}{\emph{AUC}} \\
\midrule

\new{\citet{li2024graph} (Logs2Graphs)} &  \new{$\checkmark$} & \new{$\times$} &  \new{$\times$} & \new{$\times$} & \new{$\times$} & \new{$\times$} & \new{$\checkmark$} \\ 
\new{\citet{xiao2024contexlog}(ContexLog)} & \new{$\checkmark$} & \new{$\checkmark$} & \new{$\checkmark$} & \new{$\times$} & \new{$\times$} & \new{$\times$} & \new{$\times$} \\
\new{\citet{guo2024logformer}(LogFormer)} 
& \new{$\checkmark$} & \new{$\checkmark$} & \new{$\checkmark$} & \new{$\times$} & \new{$\times$} & \new{$\times$} & \new{$\times$} \\
\new{\citet{zang2024mlad} (MLAD)} & \new{$\checkmark$} & \new{$\checkmark$} & \new{$\checkmark$} & \new{$\times$} & \new{$\times$} & \new{$\times$} & \new{$\times$} \\
\new{\citet{yin2024semi} (BTCNLog)} & \new{$\checkmark$} & \new{$\checkmark$} & \new{$\checkmark$} & \new{$\times$} & \new{$\checkmark$} & \new{$\times$} & \new{$\times$} \\
\new{\citet{lin2024fastlogad} (FastLogAD)} & \new{$\checkmark$} & \new{$\checkmark$} & \new{$\checkmark$} & \new{$\times$} & \new{$\times$} & \new{$\times$} & \new{$\times$} \\
\new{\citet{gong2024logeta} (LogETA)} & \new{$\checkmark$} & \new{$\checkmark$} & \new{$\checkmark$} & \new{$\times$} & \new{$\times$} & \new{$\times$} & \new{$\checkmark$} \\
\new{\citet{wang2024loggt} (LogGT)} & \new{$\checkmark$} & \new{$\checkmark$} & \new{$\checkmark$} & \new{$\times$} & \new{$\times$} & \new{$\times$} & \new{$\checkmark$} \\
\new{\citet{landauer2024critical}} & \new{$\checkmark$} & \new{$\checkmark$} & \new{$\checkmark$} & \new{$\times$} & \new{$\checkmark$} & \new{$\times$} & \new{$\times$} \\
\new{\citet{yang2024try} (SemPCA)} & \new{$\checkmark$} & \new{$\checkmark$} & \new{$\checkmark$} & \new{$\times$} & \new{$\times$} & \new{$\times$} & \new{$\times$} \\
\new{\citet{nguyen2024efficient} (DistilLog)} & \new{$\checkmark$} & \new{$\checkmark$} & \new{$\checkmark$} & \new{$\times$} & \new{$\times$} & \new{$\times$} & \new{$\times$} \\
\new{\citet{adeba2024sarlog} (SaRLog)} & \new{$\checkmark$} & \new{$\checkmark$} & \new{$\checkmark$} & \new{$\times$} & \new{$\times$} & \new{$\times$} & \new{$\times$} \\
\bottomrule

\end{NiceTabular}
\end{table}
\end{savenotes}
 
We indicate whether or not an evaluation metric is used to evaluate the ML techniques in each LAD study, using symbols $\checkmark$ and $\times$, respectively. As shown in Table~\ref{evalMeasures}, most of the studies (\new{41} out of \new{42}) evaluated the different LAD techniques by means of \emph{Prec}, \emph{Rec} and \emph{F1}. 
\new{This is because most of the log-based datasets (see Section ~\ref{datasets}) are highly imbalanced, with normal log event sequences representing the majority class. This imbalance makes evaluation metrics such as the \emph{F1-score}, which prioritize the accurate detection of the minority (anomalous log event sequences) class, particularly valuable for assessing the detection accuracy of log anomalies. 
}

\new{
In contrast, evaluation metrics such as accuracy (\emph{Acc}) can be misleading in such contexts, as they are skewed by the majority class and, therefore, unreliable for evaluating LAD techniques~\cite{yao2021impact}. Similarly, while AUC measures the ability of a model to distinguish between normal and anomalous log event sequences across various thresholds, it does not provide detailed insights into precision or false positive rates—key factors in imbalanced scenarios where the majority class heavily influences the detection accuracy~\cite{hancock2023evaluating}. Further, \emph{FPR}, which quantifies the proportion of normal log event sequences incorrectly classified as anomalous can be problematic in the context of imbalanced log-based datasets. This is because the false positives become obfuscated by the large number of normal log event sequences (the negative class). Since the denominator in the definition of FPR is the size of the negative class (the total number of FP and TN), which is considerably larger in such datasets, even notable changes in the number of false positives may appear negligible. This limitation makes FPR an unsuitable evaluation metric for effectively evaluating LAD techniques in scenarios where minimizing false alarms is critical~\cite{hancock2023evaluating}.} For \new{these} reason\new{s}, we adopt \emph{Prec}, \emph{Rec} and \emph{F1} to assess the detection accuracy of the different ML techniques considered in our study.

\new{Further, although specificity is not commonly reported in the literature (it was used in only four studies), we select this evaluation metric because i) it is relevant for assessing the ability of ML models to recognize normal log event sequences (the majority class in most benchmark datasets) and ii) its usage  was strongly recommended in a recent empirical study~\cite{le2022logHowFar}, in which deep ML techniques show a low specificity (below 0.5), revealing that the corresponding models perform poorly by classifying many normal log event sequences as anomalies, causing many false alarms.}

\subsection{Experimental Setup} \label{setup}
In this empirical study, as discussed in Section~\ref{background}, we consider \new{nine} alternative ML techniques. Three of them are traditional: SVM, RF (supervised) and OC-SVM (semi-supervised); see Section~\ref{traditional}. 
The others are deep learning-based:
LogRobust~\cite{zhang2019robust}, LSTM~\cite{hochreiter1997LSTM}, \new{NeuralLog~\cite{le2021logNeural}\footnote{~\label{neuralLog}\new{NeuralLog was intentionally designed without incorporating any template extraction technique (see Section~\ref{deep}). We therefore consider two versions of this ML technique in our experiments: one version (called ``NeuralLog1'') that we trained using raw log messages from all the seven datasets (see Section~\ref{datasets}) 
and another version (called ``NeuralLog2''), where the ML model is fed with the templates extracted by means of Drain; such templates are also fed to the remaining ML techniques for consistency  (see Section~\ref{experimentalMethodology}).}}}(supervised), 
DeepLog~\cite{du2017deeplog}, \new{and Logs2Graphs~\cite{li2024graph}} (semi-supervised); see Section~\ref{deep}.

\subsubsection{Hyperparameter Settings}\label{hyperParameterSettings}
Each of the \new{nine} alternative ML techniques considered in our study requires hyperparameter tuning before models can be trained. 
In the following, we provide the hyperparameter settings associated with each of the techniques considered in this study.

\begin{itemize} 
\item \textbf{SVM.} 
We used the RBF kernel function, set
the values of $\mathit{C}$ to \{1, 10, 100, 1000\}
and  $\gamma$  to \{0.0001, 0.001, 0.01, 0.1\}.
These values of $\gamma$ and $\mathit{C}$ are within the range of values that were recommended in a study~\cite{probst2019tunability} in which hyperparameter tuning was conducted to assess the impact of different hyperparameter settings on the detection accuracy of SVM on 38 datasets.
Setting the hyperparameters of SVM to the above values leads to 16 different hyperparameter settings (i.e., combinations of hyperparameter values).

\item \textbf{RF.}
We set the number of decision trees $\mathit{dTr}$ to values ranging from 10 to 100 in steps of 10 based on the findings reported in the past studies~\cite{oshiro2012many, probst2018tune} which thoroughly investigated the impact of the number of decision trees on the detection accuracy of RF using a large number of datasets. The findings suggest that RF can achieve the highest detection accuracy using 100 trees. Additionally, considering that computational time (training and prediction time) increases linearly with the number of trees~\cite{probst2019hyperparameters}, we aimed to strike a balance between the detection accuracy and the computational time. Consequently, we opted for $\mathit{dTr}$ values ranging from 10 to 100 in steps of 10. We set the number of features $\mathit{sFeat}$ in a single node of each decision tree to the square root\footnote{The square root of the total number of features is a common practice when applying RF~\cite{bernard2009influence, genuer2008random}.} of the total number of features (i.e., the total features represent the $\mathit{d}=300$ dimensions of the encoded sequence of log event occurrences as defined in Section~\ref{encoding}),  
leading to 10 hyperparameter settings.
\item \textbf{OC-SVM.}
We used the RBF kernel function and set 
the values of $\nu$ from 0.1 to 0.9 in steps of 0.1. The selection of $\nu$ values aligns with the recommendations from a previous study~\cite{yu2019clustering} in which they studied the impact of $\nu$ hyperparameter on the performance of OC-SVM, considering different values of $\nu$ ranging within the interval bounded by $0.02$ and $1$
on ten benchmark datasets. For $\gamma$ hyperparameter,
--- similarly to the SVM settings --- we selected values in \{0.0001, 0.001, 0.01, 0.1\}, leading to 36 different hyperparameter settings.
\item \textbf{LSTM, LogRobust, DeepLog, \new{NeuralLog, and Logs2Graphs.}}
To train these deep learning-based techniques\footnote{
For DeepLog, we 
set the top log event candidates, i.e., log events that are likely to occur given a history of previously seen log events, to 9.
}, we set the loss function $\mathit{lF}$ to the binary cross entropy, the optimizer $\mathit{opt}$ to the three commonly used optimizers (\texttt{adam}, \texttt{rmsprop}, and \texttt{adadelta}). 
According to \citet{perin2021influence}, \texttt{adam} and \texttt{rmsprop} are more suitable on small neural networks (e.g., a small number of hidden layers and a small number of neurons), whereas \texttt{adadelta} is more suitable for larger neural networks. Further, another study~\cite{okewu2019experimental} suggests that the three selected optimizers (\texttt{adadelta}, \texttt{adam} and \texttt{rmsprop}) lead to a high detection accuracy of deep learning ML techniques based on CNN. 
We therefore selected these three optimizers to conduct our experiments. We set the batch size $\mathit{bS}$ to three different values (32, 64 and 128) specifically in multiples of 32. We remark that a batch size of 32 was recommended as a default value by~\citet{bengio2012practical}. 
We also set the number of hidden layers $\mathit{hL}$ to 2
and the number of epochs\footnote{Due to the high computational cost of the experiments conducted in the paper, we set the maximum number of epochs for all the deep learning techniques to 150.} $\mathit{epN}$ to \{10, 50, 100, 150\}, leading to 36 different hyperparameter settings for each of these techniques. As LogRobust is defined with an additional hyperparameter  $\mathit{nEpStop}$, we set the latter to 10, as adopted by a previous empirical study~\cite{logadempirical}.
\new{For the hyperparameters that are restricted to the definition of transformer-based ($\mathit{attH}$ and $\mathit{ffnS}$) and GNN-based models ($\mathit{cL}$, $\mathit{k}$ and $\mathit{embD}$) and do not apply to RNN-based models, we set the corresponding values to the ones used in the original papers (see Section~\ref{deep}).}
\end{itemize}

\subsection{Experimental Methodology}\label{experimentalMethodology}
In this section, we discuss the experimental methodology we adopted to answer the four research questions.  
More in detail, we first present the grouping strategy we follow to group log messages in log message-based datasets. We then explain how we perform the hyperparameter tuning and evaluate the different ML techniques across session-based and log message-based datasets.

\subsubsection{Log message-based Grouping Strategy}\label{logMessageGroupingStrategy}
Due to the inconsistent use of fixed window sizes across studies and the lack of coverage of all alternative techniques and common benchmark datasets (\bgl, \tbird
and \spirit) in existing studies, we assess the impact of the size of fixed log message-based windows on the detection accuracy of the traditional and deep, supervised and semi-supervised ML techniques, considering nine window sizes ($\mathit{ws}$) ranging from $10$ to $300$.
\\
\begin{table}[!htb]
\setlength\extrarowheight{8.5pt}
\caption{Characteristics of log message-based Benchmark Datasets} \label{newDatasetsStrategies}
\hskip-1.8cm
\begin{NiceTabular}
{m{.7cm} m{1.1cm} m{.5cm} m{.01cm} m{.01cm} m{.01cm} m{.01cm} m{.01cm} m{.01cm} m{.01cm} m{.01cm} m{.8cm}}[hvlines-except-borders] 
\CodeBefore
\rowcolor{gray!15}{1-2}
\Body
\toprule
\Block{2-1}{\emph{Data.}} & \Block{2-1}{\emph{\#Temp.}} & \Block{2-1}{\emph{Seq.}} & \Block{1-9}{\emph{Window Size}}  & & & & & & & & \\ 
& & & 10 & 15 & 20 & 50 & 100 & 150 & 200 &250  & 300 \\ 

\Block{3-1}{\rotatebox[origin=c]{90}{\hades}} & \Block{3-1}{\num{117}} & 
\#N & \multicolumn{1}{c}{\num{104718}} & \multicolumn{1}{l}{\num{69776}} & \multicolumn{1}{l}{\num{52314}} & \multicolumn{1}{l}{\num{20887}}  & \multicolumn{1}{l}{\num{10410}}  & \multicolumn{1}{l}{\num{6921}} & \multicolumn{1}{l}{\num{5184}} & \multicolumn{1}{l}{\num{4132}} & \num{3439}  \\ 
& & \#A  & \multicolumn{1}{c}{\num{139}}   & \multicolumn{1}{l}{\num{128}}  & \multicolumn{1}{l}{\num{114} } & \multicolumn{1}{l}{\num{84} }  & \multicolumn{1}{l}{\num{75}}     & \multicolumn{1}{l}{\num{69}}     & \multicolumn{1}{l}{\num{58} }     & \multicolumn{1}{l}{\num{62} } & \num{56} \\
& & \emph{IR}  & \multicolumn{1}{c}{0.13\%}   & \multicolumn{1}{l}{0.18\%}  & \multicolumn{1}{l}{0.22\%} & \multicolumn{1}{l}{0.40\%} & \multicolumn{1}{l}{0.72\%} & \multicolumn{1}{l}{0.99\%}     & \multicolumn{1}{l}{1.11\%}     & \multicolumn{1}{l}{1.48\%} & 1.60\%  \\ 

\Block{3-1}{\rotatebox[origin=c]{90}{\bgl}} & \Block{3-1}{\num{1425}} & \#N  & \multicolumn{1}{c}{\num{432326}} & \multicolumn{1}{l}{\num{287671}} & \multicolumn{1}{l}{\num{215418}}  & \multicolumn{1}{l}{\num{85465}} & \multicolumn{1}{l}{\num{42310}} & \multicolumn{1}{l}{\num{28004}} & \multicolumn{1}{l}{\num{20885}}  & \multicolumn{1}{l}{\num{16621}}  & \num{13775} \\
& & \#A & \multicolumn{1}{c}{\num{39023}}   & \multicolumn{1}{l}{\num{26561}}   & \multicolumn{1}{l}{\num{20256}}   & \multicolumn{1}{l}{\num{8804}}   & \multicolumn{1}{l}{\num{4824}}  & \multicolumn{1}{l}{\num{3419}}  & \multicolumn{1}{l}{\num{2682}}  & \multicolumn{1}{l}{\num{2232}}  & \num{1936}  \\ 
& & \emph{IR} & \multicolumn{1}{c}{8.28\%}   & \multicolumn{1}{l}{8.45\%}   & \multicolumn{1}{l}{8.59\%}   & \multicolumn{1}{l}{9.34\%}   & \multicolumn{1}{l}{10.23\%}  & \multicolumn{1}{l}{10.88\%}  & \multicolumn{1}{l}{11.38\%}  & \multicolumn{1}{l}{11.84\%} & 12.32\%  \\ 

\Block{3-1}{\rotatebox[origin=c]{90}{\tbird}} & \Block{3-1}{\num{4265}} & \#N & \multicolumn{1}{c}{\num{832313}}           & \multicolumn{1}{l}{\num{525125}}           & \multicolumn{1}{l}{\num{377278}}           & \multicolumn{1}{l}{\num{129342}}          & \multicolumn{1}{l}{\num{61737}}           & \multicolumn{1}{l}{\num{40501}}           & \multicolumn{1}{l}{\num{30096}}           & \multicolumn{1}{l}{\num{23925}}           & \num{19829}           \\
& & \#A  & \multicolumn{1}{l}{\num{167686} } & \multicolumn{1}{l}{\num{141541}} & \multicolumn{1}{l}{\num{122721} } & \multicolumn{1}{l}{\num{70657}} & \multicolumn{1}{l}{\num{38262} } & \multicolumn{1}{l}{\num{26165} } & \multicolumn{1}{l}{\num{19903} } & \multicolumn{1}{l}{\num{16074}} & \num{13504}  \\
& & \emph{IR} & \multicolumn{1}{l}{16.77\%} & \multicolumn{1}{l}{21.23\%} & \multicolumn{1}{l}{24.54\%} & \multicolumn{1}{l}{35.33\%} & \multicolumn{1}{l}{38.26\%} & \multicolumn{1}{l}{39.25\%} & \multicolumn{1}{l}{39.81\%} & \multicolumn{1}{l}{40.19\%} & 40.51\%\\ 

\Block{3-1}{\rotatebox[origin=c]{90}{\spirit}} & \Block{3-1}{\num{15487}} & \#N & \multicolumn{1}{c}{\num{353642}} & \multicolumn{1}{l}{\num{230075}} & \multicolumn{1}{l}{\num{169976}} & \multicolumn{1}{l}{\num{64611}}  & \multicolumn{1}{l}{\num{30270}} & \multicolumn{1}{l}{\num{18987}}   & \multicolumn{1}{l}{\num{13399}} & \multicolumn{1}{l}{\num{10153}}  & \num{8131}  \\
& & \#A  & \multicolumn{1}{l}{\num{146356} } & \multicolumn{1}{l}{\num{103257}} & \multicolumn{1}{l}{\num{80023}} & \multicolumn{1}{l}{\num{35388}} & \multicolumn{1}{l}{\num{19729}} & \multicolumn{1}{l}{\num{14346}} & \multicolumn{1}{l}{\num{11600}} & \multicolumn{1}{l}{\num{9846}} & \num{8535}  \\
& & \emph{IR} & \multicolumn{1}{l}{29.27\%} & \multicolumn{1}{l}{30.98\%} & \multicolumn{1}{l}{32.01\%} & \multicolumn{1}{l}{35.39\%} & \multicolumn{1}{l}{39.46\%} & \multicolumn{1}{l}{43.04\%} & \multicolumn{1}{l}{46.40\%} & \multicolumn{1}{l}{49.23\%} &48.79\%\\
\bottomrule
\end{NiceTabular}
\end{table}
 Table~\ref{newDatasetsStrategies} describes the characteristics of the four log message-based datasets based on the nine window sizes we considered in our study (Column \emph{Window size}). For each dataset, we indicate: the number of unique templates extracted from the original log messages (Column \emph{\#Temp.}); the total number of normal and anomalous sequences (Column \emph{\#N} and Column \emph{\#A} under \emph{Seq.}, respectively); the percentage of log event sequences from the minority class (Column \emph{IR} under \emph{Seq.}) computed for each window size in the different log message-based datasets.
As shown in Table~\ref{newDatasetsStrategies}, log message-based datasets become less imbalanced with the increase of window size. In other words, the percentage of log event sequences from the minority class (Column \emph{IR}) increases from small to large window sizes, across datasets. 
For instance, we observe that \hades is the most imbalanced dataset, in which the percentage of log event sequences from the minority class varies between \hadesMinIR on $\mathit{ws}=\hadesMinWS$ and \hadesMaxIR on $\mathit{ws}=\hadesMaxWS$.
\spirit is one of the two less imbalanced log message-based datasets. The percentage of log event sequences from the minority class ranges between \spiritMinIR and \spiritMaxIR on $\mathit{ws}=\spiritMinWS$ and $\mathit{ws}=\spiritMaxWS$, respectively.

\subsubsection{Hyperparameter Tuning Phase}\label{hyperParameterTuning}
Table~\ref{datasetsSetUp} summarizes the strategy we followed to divide the benchmark datasets so as to enable training. 
Symbol $C_1$ denotes the majority class in each dataset, whereas $C_2$ denotes the minority class\footnote{Recall that the majority class is ``normal'' for all the datasets, except \spirit on the largest window size ($\mathit{ws}=300$) and \hadoop.}.
Column \emph{Learning} indicates the learning type, semi-supervised or supervised.
We divided each dataset used in our experiments into training, validation\footnote{\new{The validation set is used to determine the hyperparameter settings that lead to the highest detection accuracy for semi-supervised and supervised ML techniques by testing the model on unseen data.}}\label{validationSet}, and testing sets and assigned different proportions for these sets depending on the learning type of each technique as follows:
\begin{itemize}
    \item \emph{Semi-supervised.} Models are trained on 70\%
    of the majority class, 
        validated on 10\% of each class and tested on the remaining set (20\%~$C_1$ and 90\%~$C_2$).
    \item \emph{Supervised.} Models are trained on 70\% of each class, validated on 10\% of each class, and tested on the remaining set (20\%~$C_1$ and 20\%~$C_2$).
\end{itemize}

\begin{footnotesize} 
\begin{table}[tb]
\centering
\caption{Set up of benchmark datasets} \label{datasetsSetUp} 
\begin{NiceTabular}
{m{3cm} m{2cm} m{2cm} m{2cm}}[hvlines-except-borders] 
\CodeBefore
\rowcolor{gray!15}{1-1}
\Body
\toprule
\emph{Learning} & \emph{Training} & \emph{Validation} & \emph{Test} \\

\emph{Semi-supervised} &
70\% $C_1$ & \begin{tabular}[c]{@{}l@{}}10\% $C_1$\\ 
10\% $C_2$ \end{tabular} & \begin{tabular}[c]{@{}l@{}}20\% $C_1$ \\ 90\% $C_2$\end{tabular} \\
\emph{Supervised} & \begin{tabular}[c]{@{}l@{}}
70\% $C_1$ \\ 70\% $C_2$
\end{tabular} & \begin{tabular}[c]{@{}l@{}}
10\% $C_1$\\ 10\% 
$C_2$\end{tabular} & \begin{tabular}[c]{@{}l@{}}
20\% $C_1$\\ 20\% $C_2$ \end{tabular} \\
\bottomrule
\end{NiceTabular}  \\
$C_1$ ($C_2$) is the majority (minority) class in each dataset
\end{table}
\end{footnotesize} 

It is typically challenging to specify what hyperparameter values to use for a specific ML technique, on a particular dataset. Therefore, for each learning algorithm, we carried out hyperparameter tuning, using a grid search~\cite{bergstra2012random}, which is one of the commonly used strategies.

To perform our experiments and answer all the research questions (see Section~\ref{RQs}), we first trained the different ML techniques with features extracted from the \new{seven} benchmark datasets used in this study (see Section~\ref{datasets}). We then test the different ML models on these datasets, considering different combinations of hyperparameter settings per technique (see Section~\ref{hyperParameterSettings}). 
At the end of this step, we collected i) the different \emph{F1-score} values and ii) the different \emph{training} time values from both supervised and semi-supervised techniques to study their sensitivity to hyperparameter tuning.

For each hyperparameter setting, we trained the ML technique on the training set and validated it on the validation set. To avoid biased results and assess the stability of the detection accuracy of each technique, we repeated this process (training and validation) five times, computed \emph{precision}, \emph{recall}, \emph{F1-score}, and \emph{Specificity}, and recorded the computational time needed for the training phase (training time and validation time) for each iteration; we reported the average values from the five iterations.

Given that there are 16, 10 and 36 hyperparameter settings for  the traditional techniques considered in this study (respectively, SVM, RF and OC-SVM) and 36 hyperparameter settings for each of the \new{five} deep learning techniques (LSTM, DeepLog, LogRobust, \new{both versions of NeuralLog, and Logs2Graphs}), the total number of hyperparameter settings considered in this study during hyperparameter tuning is \new{278}.
Concurrently executing\footnote{All the experiments were conducted on
cloud computing platforms provided by the Digital
Research Alliance of Canada~\cite{computecanada}: a) the Narval cluster with a total of 636 NVIDIA A100 GPUs with 8 to 40 GB of memory, b) the Cedar cluster with a total of 1352 NVIDIA P100 Pascal GPUs with 8 to 64 GB of memory, \new{c) the Beluga cluster with a total of 688 NVIDIA NVidia V100SXM2 GPUs with 4 to 16 GB of memory, and d) the Graham cluster with a total of 520 NVIDIA (P100 Pascal, V100 Volta and T4 Turing) GPUs with 8 to 32 GB of memory.}} each algorithm five times for all the \new{278} hyperparameter settings i) on \new{three} session-based datasets leads to 
$5 \times \new{278} \times 3 = \new{4170}$ executions and ii) on four log message-based datasets with nine different window sizes leads to $ 5 \times \new{278} \times 4 \times 9 = \new{\num{50040}}$ executions. 
The total number of executions is therefore set to \new{$\num{4170}$} + \new{$\num{50040}$} = \new{$\num{54210}$}, leading to \new{$\num{1933}$} days ($\approx$ \new{5.30} years) of computation time.

We collected the average \emph{F1-score} for each hyperparameter setting of a ML technique, across datasets, to analyze its sensitivity in terms of detection accuracy.
Similarly, we collected the average computational time needed for the training phase for each hyperparameter setting  to assess the time performance sensitivity of each technique, considering each dataset separately.
\\
\paragraph{Best Hyperparameter Settings}
Table~\ref{rq1BestSettings} shows the hyperparameter settings that led to the highest detection accuracy \new{on the validation set} for each ML technique, on each benchmark dataset. 
Recall that unlike session-based datasets (\hdfs, \hadoop, \new{and \fdataset}), log message-based datasets (\hades, \bgl, \tbird and \spirit) are labeled at the level of individual log messages. After extracting log events from the raw log messages, we generated sequences of log event occurrences from such datasets using nine fixed window sizes (see Section~\ref{logMessageGroupingStrategy}). We therefore evaluated each ML technique on all window sizes and reported the results associated with the one which yields the highest detection accuracy in terms of \emph{F1-score}.
\subsubsection{Testing Phase}

\begin{table}[tb]
\setlength\extrarowheight{4pt}
\centering
\caption{Best Hyperparameter Settings} \label{rq1BestSettings} 
\begin{NiceTabular}
{m{1.8cm} m{.9cm} m{1.22cm} m{1.22cm} m{1.4cm} m{1.22cm} m{1.22cm} m{1.6cm} m{1.22cm}}[hvlines-except-borders] 
\CodeBefore
\rowcolor{gray!15}{1-2}
\Body
\toprule
\Block{2-1}{\emph{Technique}} & \Block{2-1}{\emph{Hyper.}} & \Block{1-7}{\emph{Dataset}} & \\
& & \Block{1-1}{\small{\hdfs}} & \Block{1-1}{\small{\hadoop}}  & \Block{1-1}{\new{\small{\fdataset}}} & \Block{1-1}{\small{\hades}} & \Block{1-1}{\small{\bgl}} & \Block{1-1}{\small{\tbird}} & \Block{1-1}{\small{\spirit}}\\

\Block{2-1}{SVM} & $C$ & $1$ & \num{1000} & \new{\num{1000}} & \num{1000} &\num{1000} & $10$ & \num{1000} \\ 
& $\gamma$ & $0.1$ & $0.0001$ & \new{\num{0.1}} & $0.001$ & $0.001$ & $0.1$ & $0.01$ \\ 
\Block{1-1}{RF} & $\mathit{dTr}$ & $60$ & $50$ & \new{\num{100}} & $80$ & $80$ & $80$ & $100$ \\
\Block{3-1}{LSTM} & 
$\mathit{opt}$ &  \texttt{adam} & \texttt{adam} & \new{\texttt{adam}} & \texttt{adam} & \texttt{rmsprop} & \texttt{adam} & \texttt{adam} \\ 
& $\mathit{epN}$ & $150$ & $10$ & \new{$150$} & $100$ & $10$ & $150$ & $100$\\ 
& $\mathit{bS}$ & $64$ & $32$ & \new{$32$} & $64$ & $128$ & $32$ & $32$ \\ 
\Block{3-1}{LogRobust} & $\mathit{opt}$ &  \texttt{rmsprop} & \texttt{rmsprop} & \new{\texttt{rmsprop}} & \texttt{rmsprop} & \texttt{adam} & \texttt{rmsprop} & \texttt{adam} \\ 
& $\mathit{epN}$ & $100$ & $150$ & \new{$150$} & $100$ & $100$ & $10$ & $100$ \\ 
& $\mathit{bS}$ & $64$ & $128$ & \new{$128$} & $32$ & $128$ & $64$ & $64$ \\
\Block{3-1}{\new{NeuralLog1}} & 
$\new{\mathit{opt}}$ &  \new{\texttt{adadelta}} & \new{\texttt{rmsprop}} & \new{\texttt{rmsprop}} & \new{\texttt{adam}} & \new{\texttt{adam}} & \new{\texttt{rmsprop}} & \new{\texttt{adam}} \\ 
& $\new{\mathit{epN}}$ & $\new{150}$ & $\new{150}$ & $\new{150}$ & $\new{150}$ & $\new{150}$ & $\new{150}$ & $\new{100}$ \\ 
& $\new{\mathit{bS}}$ & $\new{32}$ & $\new{128}$ &$\new{128}$ & $\new{128}$ & $\new{128}$ & $\new{32}$ & $\new{128}$ \\

\Block{3-1}{\new{NeuralLog2}} & 
$\new{\mathit{opt}}$ &  \new{\texttt{rmsprop}} & \new{\texttt{rmsprop}} & \new{\texttt{rmsprop}} & \new{\texttt{adam}} & \new{\texttt{rmsprop}} & \new{\texttt{rmsprop}} & \new{\texttt{adam}} \\ 

& $\new{\mathit{epN}}$ & $\new{50}$ & $\new{150}$ & $\new{150}$ & $\new{100}$ & $\new{100}$ & $\new{100}$ & $\new{50}$ \\

& $\new{\mathit{bS}}$ & $\new{128}$ & $\new{128}$ &$\new{128}$ & $\new{64}$ & $\new{128}$ & $\new{64}$ & $\new{128}$ \\

\Block{2-1}{OC-SVM} & $\nu$ & $0.2$ & $0.1$ & \new{$0.3$} & $0.1$ & $0.1$ & $0.9$ & $0.4$ \\ 
& $\gamma$ & $0.0001$ & $0.01$ & \new{$0.0001$} & $0.1$ & $0.1$ & $0.0001$ & $0.1$ \\

\Block{3-1}{DeepLog} & $\mathit{opt}$ &  \texttt{rmsprop} & \texttt{rmsprop} & \new{\texttt{rmsprop}} & \texttt{adam} & \texttt{adam} & \texttt{rmsprop} & \texttt{adam} \\ 
& $\mathit{epN}$ & $10$ & $150$ & \new{$100$} & $50$ & $100$ & $150$ & $150$\\ 
& $\mathit{bS}$ & $64$ & $32$ &\new{$32$} & $32$ & $64$ & $32$ & $64$\\
\Block{3-1}{\new{Logs2Graphs}} & 
$\new{\mathit{opt}}$ &  \new{\texttt{rmsprop}} & \new{\texttt{rmsprop}} & \new{\texttt{adam}} & \new{\texttt{adadelta}} & \new{\texttt{adadelta}} & \new{\texttt{rmsprop}} & \new{\texttt{adadelta}} \\ 
& $\new{\mathit{epN}}$ & $\new{100}$ & $\new{150}$ & $\new{150}$ & $\new{50}$ & $\new{50}$ & $\new{10}$ & $\new{150}$ \\ 
& $\new{\mathit{bS}}$ & $\new{32}$ & $\new{128}$ &$\new{128}$ & $\new{64}$ & $\new{32}$ & $\new{32}$ & $\new{128}$ \\
 
\bottomrule
\end{NiceTabular}

\end{table}

We selected the best hyperparameter setting for each ML technique on each dataset obtained from the previous step to i) re-train the different ML models on the training and validation sets and ii) evaluate them on the test set. We repeated the process five times for each ML technique, on each dataset, and then computed  \emph{precision}, \emph{recall}, and \emph{F1-score}, as well as \emph{re-train time} and 
\emph{test time} per iteration~\footnote{We computed the time performance metrics (model training, re-training and prediction time)  by means of Python \texttt{time} function~\cite{python-time}.}. We finally computed and reported the average \emph{F1-score} and the average \emph{re-train time} from the five iterations associated with the best hyperparameter setting for each ML technique, evaluated on each dataset, to reflect the best possible detection accuracy and time performance of that technique.
More in details, each of the \new{nine} ML techniques was concurrently executed five times for the best hyperparameter setting on each of i) the \new{three} session-based datasets (\hdfs\new{,} \hadoop \new{and \fdataset}), leading to $\new{9} \times 5 \times \new{3} = \new{135}$ executions and ii) the four log message-based datasets (\bgl, \tbird, \spirit and \hades) with \new{nine window sizes }
leading to $\new{9} \times 5 \times 4 \new{\times 9} = \new{\num{1620}}$ executions. 
The total number of executions during the testing phase is therefore set to $\new{135}+ \new{\num{1620}} = \new{\num{1755}}$ leading to \new{96} days ($\approx$ \new{three months}) of computation time. 

We remark that research questions RQ1 and RQ2 are both dedicated to supervised ML techniques, whereas RQ3 and RQ4 concern semi-supervised ones. We therefore used the same hyperparameter settings for RQ1 and RQ2. Similarly, RQ3 and RQ4 share the same settings.

\subsubsection{Statistical Analysis of the Results}\label{sec:stat-analysis}
To assess the significance of the difference among the semi-supervised and supervised, traditional and deep ML techniques used in this study, we applied the non-parametric statistical Kruskal-Wallis test~\cite{kruskal1952kruskal} on the results obtained from answering our research questions.
We selected the Kruskal-Wallis test because it is i) suitable for non-normally distributed data and ii) commonly used to evaluate the performance of ML techniques on multiple datasets. \new{ This test was chosen as it does not require assumptions about the underlying data distribution, making it particularly well-suited for comparing multiple independent groups, especially when dealing with datasets of varying sizes and distributions.}

More in detail, we conducted five statistical tests, each associated with one of the evaluation criteria: a) detection accuracy, b) sensitivity of the detection accuracy to hyperparameter tuning, c1) time performance - re-training time; c2) time performance - prediction time, and d) the sensitivity of the time performance (training time) to hyperparameter tuning.  
We provided as input (score) to these tests i) the highest F1-score for each ML technique on each dataset; ii) the range of F1-score (i.e., the difference between the minimum and maximum F1-score) across hyperparameter settings from the sensitivity analysis; iii) the model re-training time and iv) the prediction time (both associated with the best F1-score reported for each ML technique on each dataset); v) the range of the training time (i.e., the difference between the minimum and maximum training time) across hyperparameter settings from the sensitivity analysis. 
\new{More in detail, we performed the five statistical tests on the detection accuracy (in terms of \emph{F1-score}) and the time performance of nine alternative ML techniques (see Sections~\ref{traditional} and ~\ref{deep}) across seven datasets (see Section~\ref{datasets}), leading to a sample size of 9 $\times 7$ = 63.}
For each of the five statistical tests, we set the null hypothesis to: ``There is no significant difference among ML techniques across datasets''. 
We considered a confidence level of $95\%$, setting the significance level value to $0.05$. We then calculated the test statistic and the corresponding \emph{p-value}.
We rejected the null hypothesis when the \emph{p-value} was below that selected significance level ($\text{p-value} < 0.05$).

Further, we conducted a post-hoc analysis on the 
results associated with each of the evaluation criteria in which the null hypothesis was rejected. 
To do so, we applied the non-parametric pairwise post-hoc statistical Dunn's test~\cite{dunn1964multiple} to compare all the different pairs of ML techniques in terms of the sensitivity of the \emph{F1-score} to hyperparameter tuning.

 \section{Results}
\label{results}

\begin{table}[tb]
\setlength\extrarowheight{3pt}
\caption{Window sizes associated with the highest detection accuracy for supervised and semi-supervised, traditional and deep ML techniques}
\label{bestWSFourdatasets}
\centering
\begin{NiceTabular}
{m{2.5cm} m{1.9cm} m{1.5cm} m{1.5cm} m{1.9cm} m{1.5cm}}[hvlines-except-borders] 
\CodeBefore
\rowcolor{gray!15}{1-2}
\Body
\toprule
\Block{2-1}{\emph{Learning Type}} & \Block{2-1}{\emph{Technique}} & \Block{1-4}{\emph{Log message-based dataset}} & & & \\
& & \hades & \bgl & \tbird & \spirit \\

\Block{6-1}{\emph{Supervised}} & SVM  & 15 & 15  & 50 & 20 \\ 
& RF & 10 & 15 & 10 & 15 \\ 
& LSTM & 15 & 15 & 10 & 20 \\ 
& LogRobust & 10 & 10 & 20 & 10 \\ 
& \new{NeuralLog1} & \new{10} & \new{15} & \new{15} & \new{10} \\ 
& \new{NeuralLog2} & \new{10} & \new{15} & \new{15} & \new{50} \\ 
\Block{3-1}{\emph{Semi-supervised}} 
& OC-SVM  & 300 & 50 & 250 & 10 \\ 
& DeepLog & 10 & 10 & 200 & 15 \\
& \new{Logs2Graphs} & \new{10} & \new{200} & \new{15} & \new{10} \\
\bottomrule
\end{NiceTabular}
\end{table}

\subsection{RQ1 - Detection accuracy of supervised traditional and deep ML techniques}\label{RQ1results} 
\subsubsection{Detection Accuracy}\label{RQ1detectionaccuracy}

As shown in Table~\ref{rq1Accuracy}, both supervised traditional (SVM, RF) and deep (LSTM, LogRobust\new{, NeuralLog1 and NeuralLog2}) ML techniques show a high detection accuracy (\emph{F1-score}) when evaluated on \new{the} session-based datasets \hdfs and \hadoop, with better results on \hdfs than \hadoop.
\new{On \fdataset, traditional ML techniques by far outperform deep ML techniques with the highest \emph{F1-score} of $96.58$ achieved by RF and the lowest \emph{F1-score} ($0.00$) recorded for LogRobust and both versions of NeuralLog (NeuralLog1 and NeuralLog2).
The low detection accuracy of the latter techniques is likely due to the small number of anomalous log event sequences in \fdataset relative to the other datasets, which makes it difficult for complex models like the attention-based RNN model (LogRobust) and the Transformer-based model (NeuralLog) to effectively learn the minority class features.
We remark that the specificity of all supervised ML techniques is high and similar across session-based datasets except for \hadoop, due to the fact that the majority class of this dataset corresponds to anomalous log event sequences, making it challenging for the different supervised ML techniques to recognize the normal log event sequences.
}

When evaluated on three log message-based datasets (\bgl, \tbird, and \spirit), supervised ML techniques yield a high detection accuracy \new{(in terms of \emph{F1-score})}, with slightly better results on \tbird and \spirit than \bgl.
Except for SVM, the remaining supervised ML techniques show a large decrease in \new{\emph{F1-score}} when evaluated on \hades. More specifically, while SVM achieves an \emph{F1-score} of $93.88$, the \emph{F1-score} of the remaining techniques ranges from \new{$49.26$} for \new{NeuralLog2} to \new{$84.00$} for \new{NeuralLog1}, with $72.73$ for RF.
\new{The higher \emph{F1-score} of SVM on the most imbalanced dataset \hades, compared to that of the remaining supervised ML techniques, can be attributed to hyperparameters $\mathit{C}$, which penalizes the misclassification of the minority class (anomalous log event sequences), and $\mathit{\gamma}$, which makes the decision boundary more flexible to effectively differentiate between normal and anomalous log event sequences.}

\new{We remark that all supervised, traditional and deep ML techniques show very high and similar detection accuracy on \hdfs, \hadoop, \bgl and \tbird. This is due to the nature of the datasets, where a recent study~\cite{landauer2024critical} shows that simple, non-ML detection techniques, like counting sequence lengths, can also effectively detect log anomalies and achieve high accuracy.}
\new{This is because log anomalies typically manifest themselves through new log event types, variations in log event frequencies and, to a lesser extent, changes in sequence lengths. Overall, the study suggests that a majority of the anomalies are straightforward to identify and the relation between log anomalies and sequential patterns is less pronounced than expected within these commonly used benchmark datasets.}

\new{The specificity (\emph{Spec}) of all supervised traditional and deep ML techniques is high on all log message-based datasets, as all these datasets contain a high number of normal log event sequences.}
\new{We also observe that NeuralLog1 outperforms NeuralLog2 in terms of \emph{F1-score} and \emph{Spec} on all datasets. This is expected given that NeuralLog~\cite{le2021logNeural} is designed to detect log anomalies directly from raw logs rather than from log templates extracted by means of a log parsing technique.}

\begin{table}[!htbp]
\setlength\extrarowheight{3pt}
\caption{Comparison of the detection accuracy of supervised traditional and deep ML techniques on all datasets}
\label{rq1Accuracy}
\centering
\begin{NiceTabular}
{m{.15cm} m{1.8cm} m{.9cm} m{.85cm} m{.85cm} m{.85cm} m{1.35cm} m{1.45cm} m{1.4cm}}[hvlines-except-borders] 
\CodeBefore
\rowcolor{gray!15}{1-2}
\Body
\toprule
\Block{2-2}{\emph{Dataset}} & & \Block{2-1}{\emph{Metric}} & \Block{1-6}{\emph{Technique}} & & & \\
& & & \small{SVM} & \small{RF} & \small{LSTM} & \small{LogRobust} & \small{\new{NeuralLog1}} & \small{\new{NeuralLog2}} \\
\Block{12-1}{\rotatebox[origin=c]{90}{\emph{Session}}} & \Block{4-1}{\hdfs} & \emph{Prec}   & 99.20  & 99.64 & 98.41 & 100.00 & \new{99.32}  & \new{99.64}    \\ 
& & \emph{Rec} & 99.91 & 99.91 & 98.69 & 99.48  & \new{97.34}  & \new{92.56}   \\
&  & \emph{F1} & \textbf{99.56}  & \textbf{99.78} & \textbf{98.53} & \textbf{99.74} & \new{\textbf{98.31}}  & \new{\textbf{95.60}}     \\ 
&  & \emph{\new{Spec}} & \new{99.98}  & \new{99.99} & \new{99.95} & \new{100.00} & \new{100.00}  & \new{100.00}  \\ 
& \Block{4-1}{\hadoop}  & \emph{Prec}  & 82.74  & 82.87 & 82.74 & 82.74 & \new{83.33}  & \new{83.33}\\ 
& & \emph{Rec}  & 100.00 & 97.91 & 100.00 & 100.00  & \new{100.00}  & \new{100.00}  \\  
& & \emph{F1} & \textbf{90.56}  & \textbf{89.76} & \textbf{90.56} & \textbf{90.56} & \new{\textbf{90.91}}  & \new{\textbf{90.91}} \\
&  & \emph{\new{Spec}} & \new{0.00}  & \new{2.94} & \new{0.00} & \new{0.00} & \new{0.00}  & \new{0.00}  \\ 

& \Block{4-1}{\new{\fdataset}}  & 
\emph{\new{Prec}}  & \new{98.68}  & \new{93.97} & \new{100.00} & \new{0.00} & \new{\textbf{0.00}} & \new{\textbf{0.00}} \\ 
& & 
\emph{\new{Rec}}  & \new{82.42} & \new{99.34} & \new{61.76} & \new{0.00} & \new{0.00} & \new{0.00}   \\  
& & 
\emph{\new{F1}} & \new{\textbf{89.82}}  & \new{\textbf{96.58}} & \new{\textbf{76.25}} & \new{\textbf{0.00}}  & \new{\textbf{0.00}} & \new{\textbf{0.00}}  \\ 
&  & 
\emph{\new{Spec}} & \new{99.58}  & \new{97.56} & \new{100.00} & \new{99.50} & \new{100.00}  & \new{100.00}  \\ 
\Block{16-1}{\rotatebox[origin=c]{90}{\emph{Log message}}}

& \Block{4-1}{\hades}  & \emph{Prec}  & 100.00  & 100.00 & 95.05 & 95.00 & \new{100.00}  & \new{60.00}\\   
& & \emph{Rec}  & 88.46 & 57.14 & 73.85 & 53.57 & \new{73.33}  & \new{42.86}\\
& & \emph{F1}  & \textbf{93.88} & \textbf{72.73} & \textbf{83.11} & \textbf{68.50} & \new{\textbf{84.00}}  & \new{\textbf{49.26}}\\
&  & \emph{\new{Spec}} & \new{100.00}  & \new{100.00} & \new{99.99} & \new{100.00} & \new{100.00}  & \new{100.00}  \\ 
& \Block{4-1}{\bgl}  & \emph{Prec}  & 97.52  & 93.29 & 97.49 & 99.79 & \new{99.97}  & \new{99.66}\\ 
& & \emph{Rec}  & 92.38 & 78.96 & 84.27 & 95.50 & \new{99.91}  & \new{98.76}\\
& & \emph{F1}  & \textbf{94.88}  & \textbf{85.46} & \textbf{90.39} & \textbf{97.59} & \new{\textbf{99.94}}  & \new{\textbf{99.21}}\\ 
&  & \emph{\new{Spec}} & \new{99.78}  & \new{99.46} & \new{99.80} & \new{99.98} & \new{100.00}  & \new{99.97}  \\ 
& \Block{4-1}{\tbird}  & \emph{Prec}  & 99.99  & 99.49 & 99.38 & 99.98 & \new{99.98}  & \new{99.99}\\   
& & \emph{Rec}  & 98.22 & 97.01 & 98.55 & 99.84 & \new{99.97}  & \new{99.96}\\
& & \emph{F1}  & \textbf{99.10} & \textbf{98.24} & \textbf{98.96} & \textbf{99.91} & \new{\textbf{99.98}}  & \new{\textbf{99.97}}\\ 
&  & \emph{\new{Spec}} & \new{99.99}  & \new{99.90} & \new{99.88} & \new{99.99} & \new{100.00}  & \new{100.00}  \\ 
& \Block{4-1}{\spirit}  & \emph{Prec}  & 97.87  & 98.60 & 97.98 & 100.00 & \new{99.98}  & \new{99.96}\\   
& & \emph{Rec}  & 97.53 & 85.33 & 93.31 & 95.76 & \new{99.97}  & \new{99.77} \\
& & \emph{F1}  & \textbf{97.70} & \textbf{91.49} & \textbf{95.57} & \textbf{97.83} & \new{\textbf{99.98}}  & \new{\textbf{99.86}}  \\
&  & \emph{\new{Spec}} & \new{99.00}  & \new{99.46} & \new{99.08} & \new{100.00} & \new{100.00}  & \new{99.98}  \\ 
\bottomrule
\end{NiceTabular}
\end{table}

\begin{figure}[H]
\centering
\begin{subfigure}{1\linewidth}
  \centering
  \includegraphics[height=3.75cm, width=12cm]{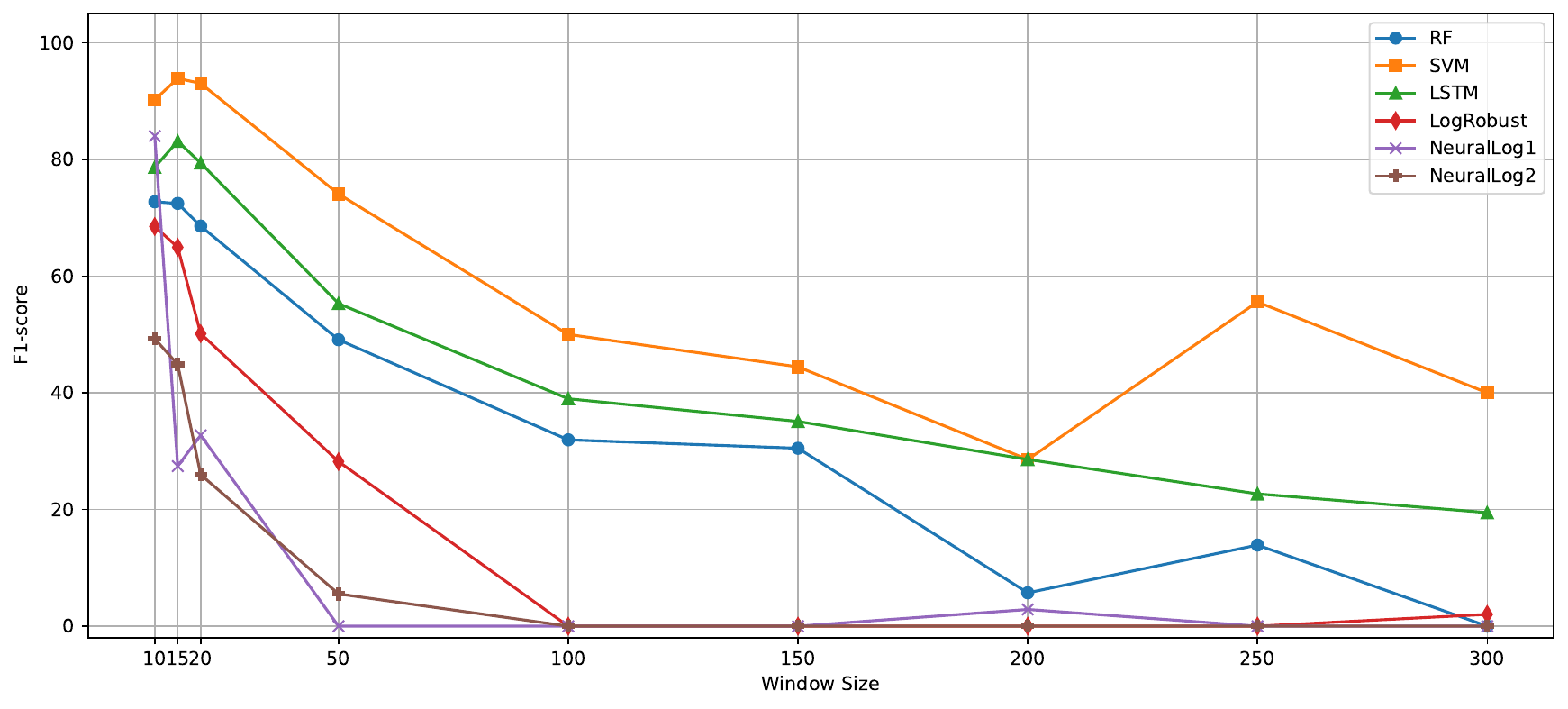}
  \caption{\new{Impact of window size when using the \hades dataset}}
  \label{figRQ1det:hadesAllW}
\end{subfigure}

\begin{subfigure}{1\textwidth}
  \centering
  \includegraphics[height=3.75cm, width=12cm]{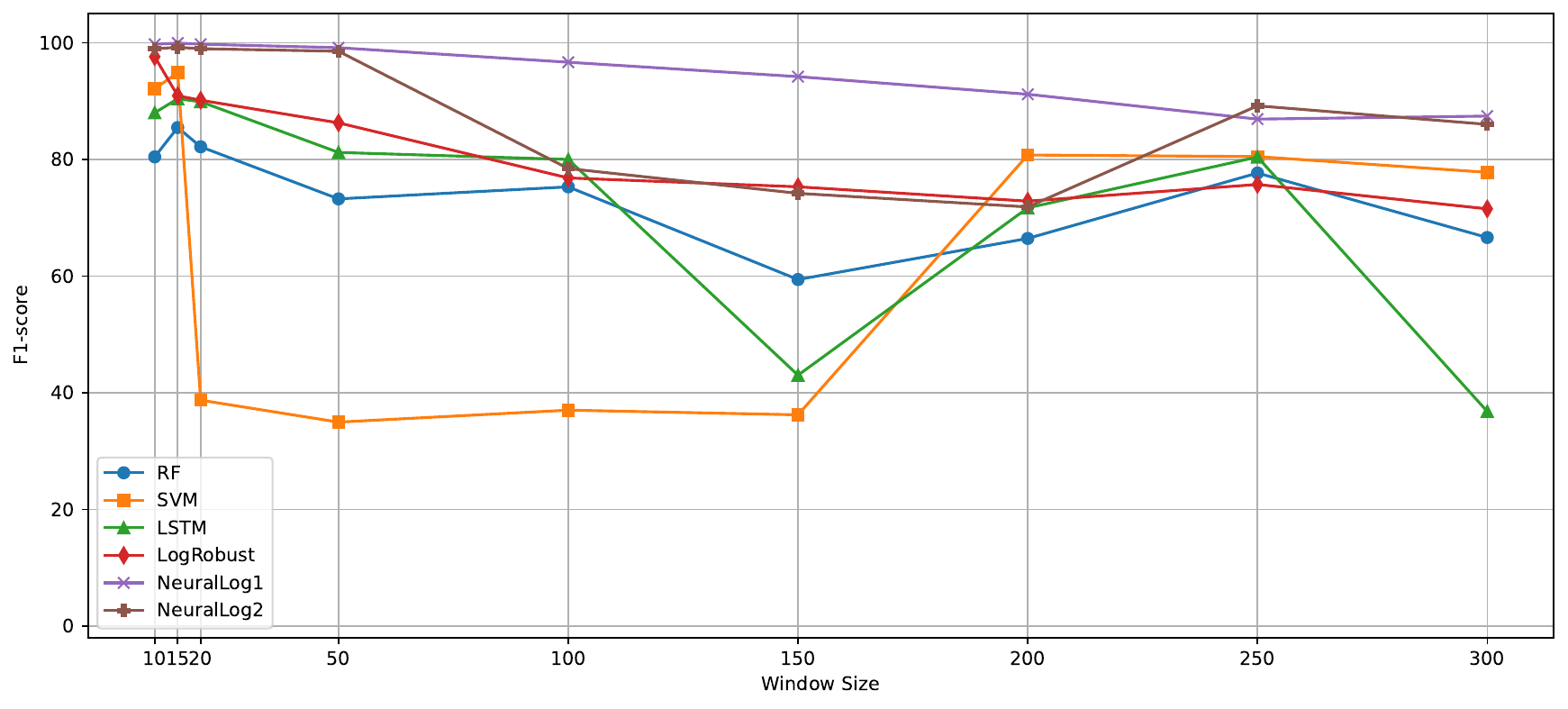}
  \caption{\new{Impact of window size when using the \bgl dataset}}
  \label{figRQ1det:bglAllW}
\end{subfigure}
\begin{subfigure}{1\textwidth}
  \centering
  \includegraphics[height=3.75cm, width=12cm]{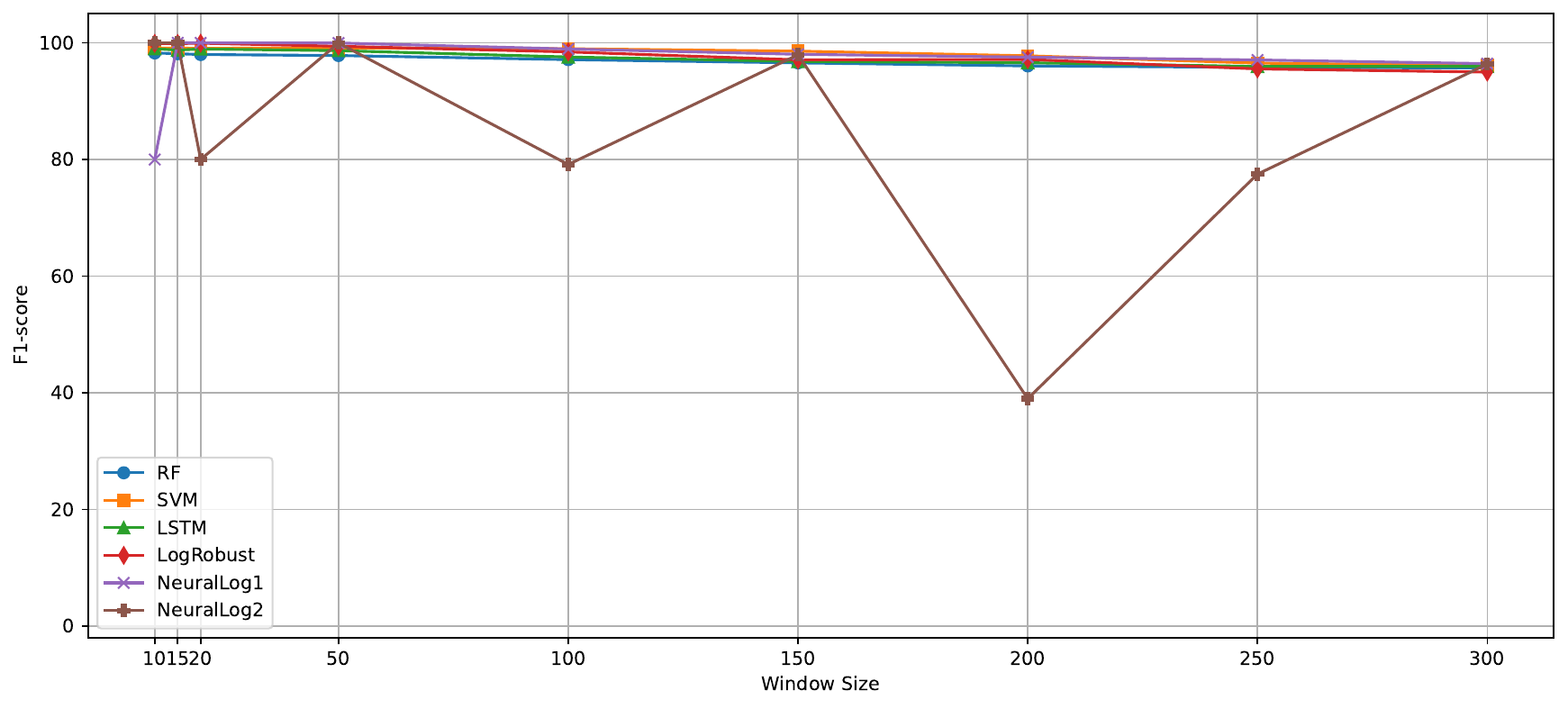}
  \caption{\new{Impact of window size when using the \tbird dataset}}
  \label{figRQ1det:tbirdAllW}
\end{subfigure}

\begin{subfigure}{1\textwidth}
  \centering
  \includegraphics[height=3.75cm, width=12cm]{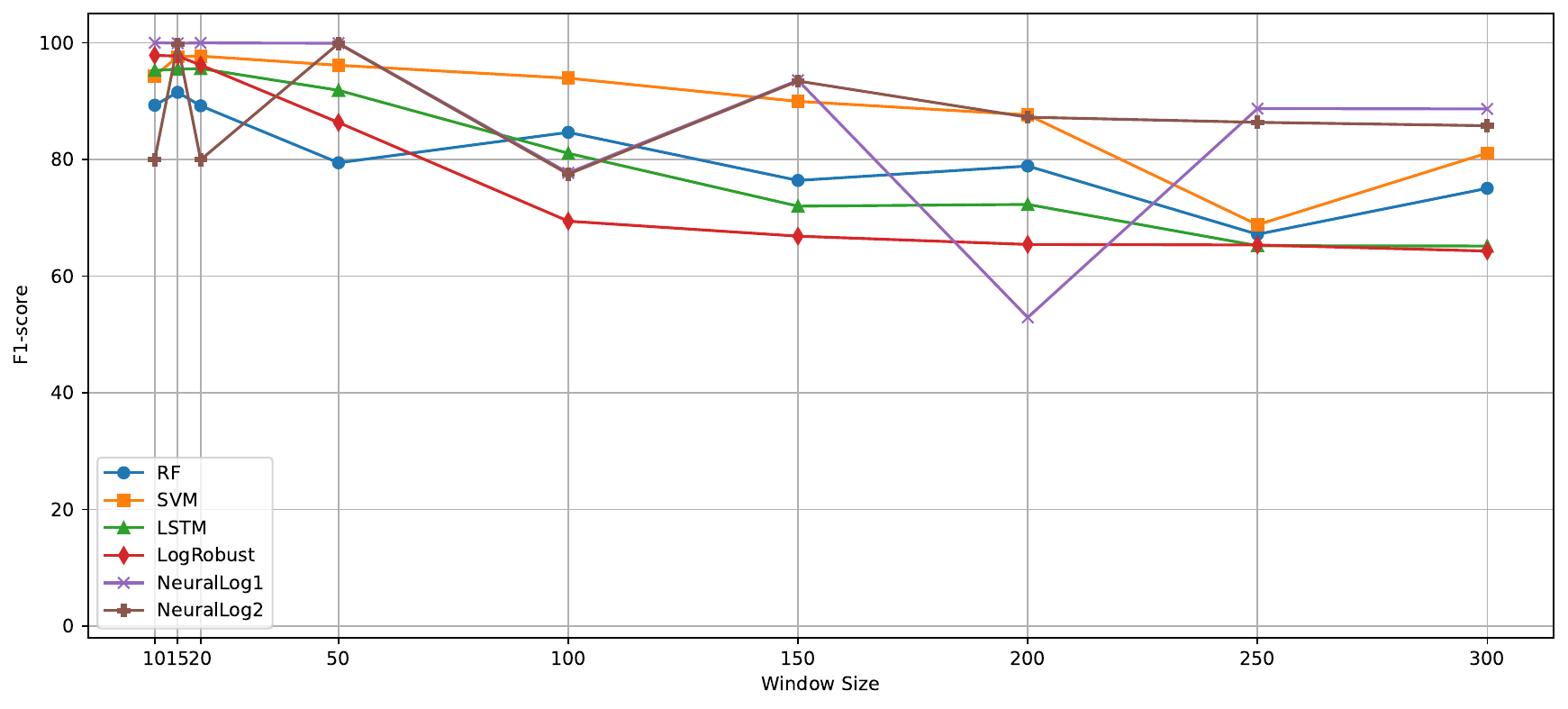}
  \caption{\new{Impact of window size when using the \spirit dataset}}
  \label{figRQ1det:spiritAllW}
\end{subfigure}
\caption{Impact of window size on the detection accuracy of supervised traditional and deep ML techniques on log message-based datasets}
\label{fig:sup-detaccAllW}
\end{figure}

\figurename~\ref{fig:sup-detaccAllW} shows the impact of different window sizes on the detection accuracy of supervised traditional and deep ML techniques on log message-based datasets.
\begin{itemize}
       \item \textit{Small window sizes $ \{10, 15, 20\}$.}
       As depicted in Table~\ref{bestWSFourdatasets}, supervised ML techniques yielded their highest detection accuracy (in terms of \emph{F1-score}) on smaller window sizes across the log message-based datasets. 
For instance, on \hades, RF and LogRobust obtained their highest detection accuracy  with $\mathit{ws}=10$. This may be expected given that small window sizes lead to more sequences to train supervised ML models.

   \item \textit{Large window sizes $\{50, 100, 150, 200, 250, 300\}$.} 
All the supervised ML techniques showed a decrease in detection accuracy when evaluated on large window sizes across all the log message-based datasets. The \new{overall} decrease in detection accuracy is higher on more imbalanced datasets (\hades, \bgl) than on  less imbalanced datasets (\tbird, \spirit). For instance, on \hades, RF yielded an \emph{F1-score} that decreased from $72.73$ with $\mathit{ws}=10$ to $0.0$ with $\mathit{ws}=300$; on \tbird, RF shows a detection accuracy ranging from $98.24$ with $\mathit{ws}=10$ to $95.67$ with $\mathit{ws}=300$. 
This confirms that larger window sizes often lead to lower detection accuracy (especially on highly imbalanced datasets), indicating potential challenges for the supervised ML techniques in capturing log patterns effectively.
\end{itemize}

Statistical analysis (see \S~\ref{sec:stat-analysis}) yields a \emph{p-value} of 
\new{$0.88$}, suggesting the detection accuracy of the different supervised traditional and deep ML techniques is not significantly different.
Therefore accuracy is not a distinguishing factor among techniques on these datasets.

\subsubsection{Sensitivity of Detection Accuracy}\label{RQ1SensDetectionAccuracy}
\begin{figure}[tb]
\begin{subfigure}{{.5\textwidth}}
  \includegraphics[width=6cm]{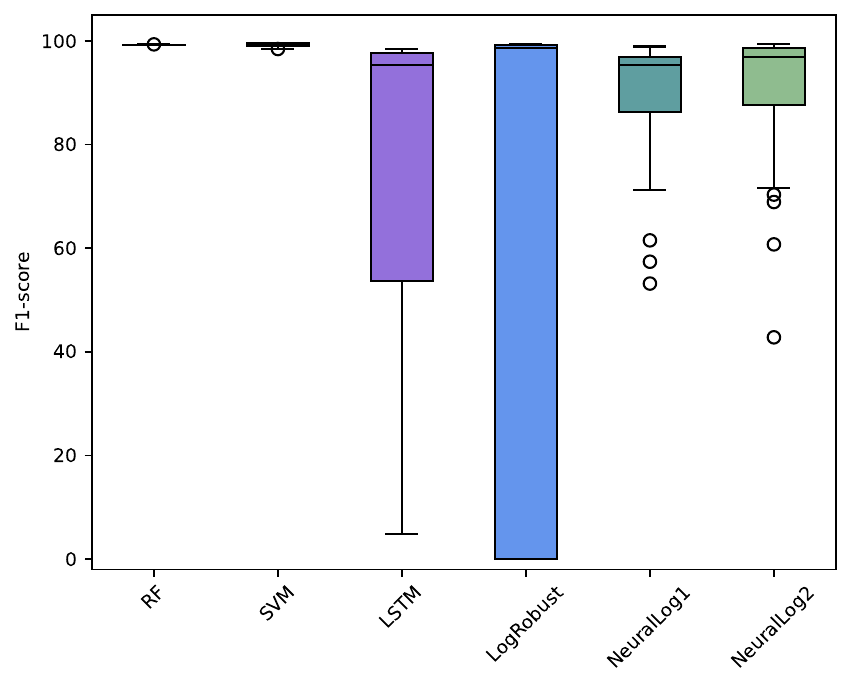}
  \caption{\new{\hdfs dataset}}
  \label{rq1VarAcchdfs}
\end{subfigure}\begin{subfigure}{{.5\textwidth}}
  \includegraphics[width=6cm]{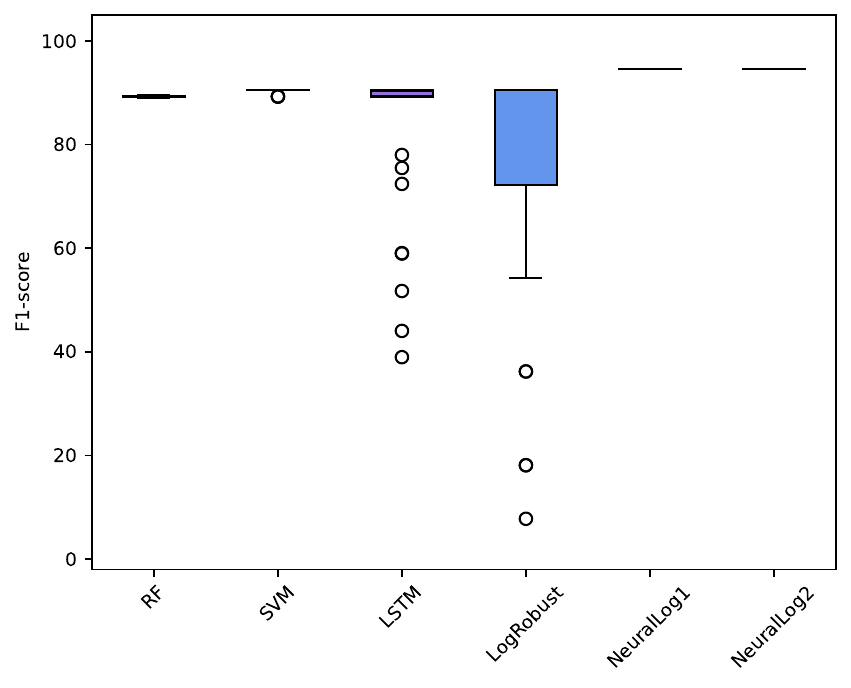}
  \caption{\new{\hadoop dataset}}
  \label{rq1VarAcchadoop}
\end{subfigure}

\begin{subfigure}{{\textwidth}}
    \centering
  \includegraphics[width=6cm]{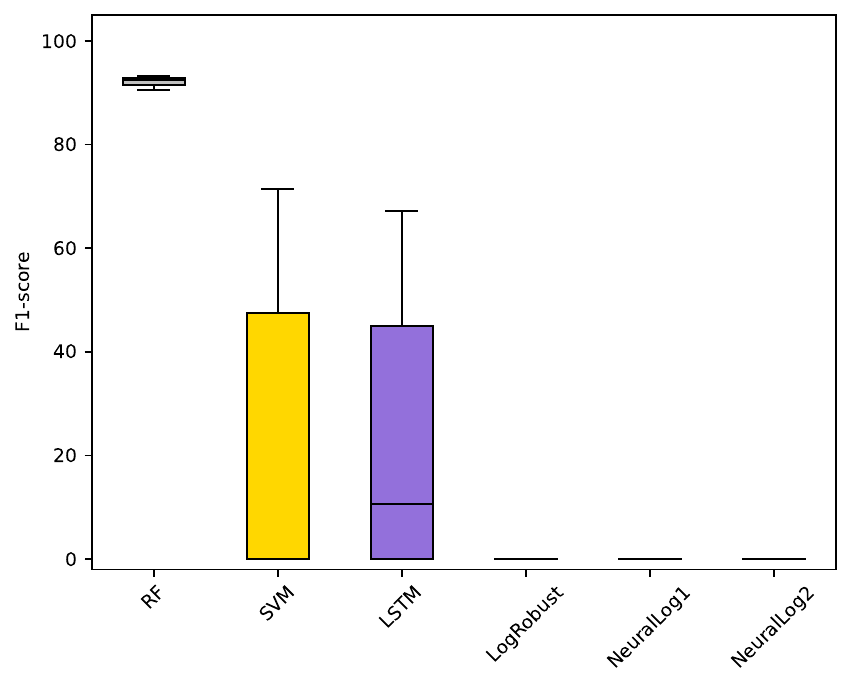}
  \caption{\new{\fdataset}}
  \label{rq1VarAccfdataset}
\end{subfigure}

\caption{Sensitivity of the detection accuracy of supervised traditional and deep ML techniques on session-based datasets}
\label{rq1VarAcc}
\end{figure}
 As depicted in \figurename~\ref{rq1VarAcc}, 
\new{the overall sensitivity of RF on the three session-based datasets \hdfs, \hadoop and \fdataset (plots in~\ref{rq1VarAcchdfs}, ~\ref{rq1VarAcchadoop} and ~\ref{rq1VarAccfdataset}, respectively) is far lower than that of the remaining traditional (SVM) and deep (LSTM, LogRobust, NeuralLog1 and NeuralLog2) ML techniques.
More in detail, RF shows a detection accuracy ranging from i) $99.27$ to $99.33$ (avg $\approx 99.29$, stdDev $\approx 0.02$) on \hdfs, ii) $89.07$ to $89.47$ (avg $\approx 89.28$, stdDev $\approx 0.13$) on \hadoop and iii) $90.44$ to $93.18$ (avg $\approx 92.15$, stdDev $\approx 0.83$) on \fdataset.}  

\new{In contrast, on \hdfs, the detection accuracy of LogRobust ranges from $0.00$ to $99.34$ (avg $\approx 65.87$, stdDev $\approx 46.58$), while for NeuralLog1, it ranges from $53.17$ to $98.92$ (avg $\approx 89.16$, stdDev $\approx 12.26$). On \hadoop, the detection accuracy of LogRobust ranges from $7.77$ to $90.50$ (avg $\approx 76.62$, stdDev $\approx 24.28$), and on the \fdataset, the detection accuracy of SVM ranges from $0.00$ to $71.43$ (avg $\approx 22.29$, stdDev $\approx 29.70$). These results show high sensitivity in terms of \emph{F1-score} across the session-based datasets.}
\new{LSTM is the only ML technique that shows a high sensitivity to hyperparameter tuning across all the session-based datasets. Its detection accuracy ranges from $4.84$ to $98.51$ (avg $\approx 80.17$, stdDev $\approx 25.40$) on \hdfs, from $38.96$ to $90.50$ (avg $\approx 83.45$, stdDev $\approx 14.18$) on \hadoop and from $0.00$ to $67.15$ (avg $\approx 21.50$, stdDev $\approx 24.36$) on \fdataset.}
\new{Although NeuralLog1 and NeuralLog2 show a very small sensitivity to hyperparameter tuning on \hadoop, their detection accuracy on the \fdataset remains consistently 0 across all hyperparameter settings, indicating that the model is not learning. The same observation applies to LogRobust on the same dataset.}

\begin{figure}[p]
\centering
\begin{subfigure}{1\textwidth}
  \centering
  \includegraphics[height=3.9cm, width=14cm]{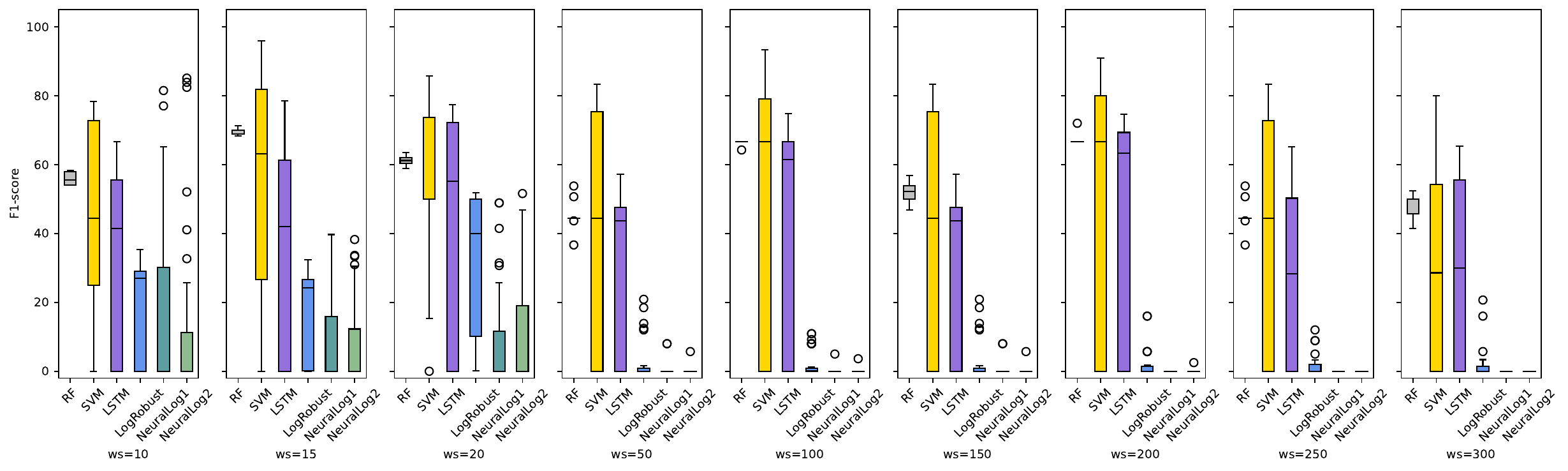}
  \caption{\new{Sensitivity on the \hades dataset}}
  \label{figRQ1detAcc:hades}
\end{subfigure}

\begin{subfigure}{1\textwidth}
  \centering
  \includegraphics[height=3.9cm, width=14cm]{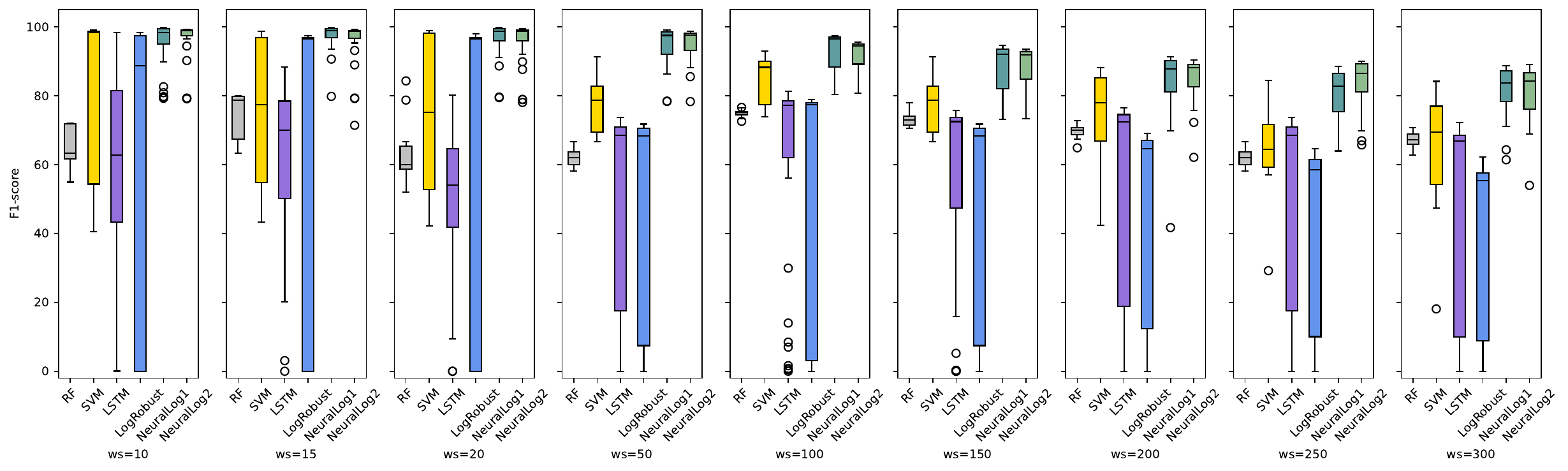}
  \caption{\new{Sensitivity on the \bgl dataset}}
  \label{figRQ1detAcc:bgl}
\end{subfigure}

\begin{subfigure}{1\textwidth}
  \centering
  \includegraphics[height=3.9cm, width=14cm]{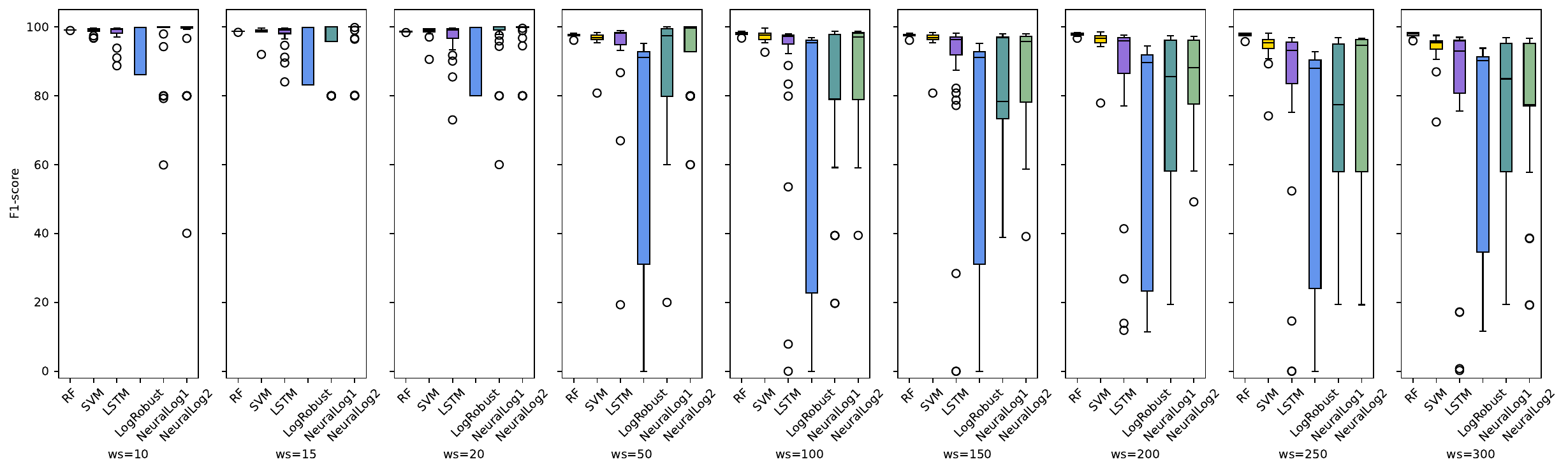}
  \caption{\new{Sensitivity on the \tbird dataset}}
  \label{figRQ1detAcc:tbird}
\end{subfigure}

\begin{subfigure}{1\textwidth}
  \centering
  \includegraphics[height=3.9cm, width=14cm]{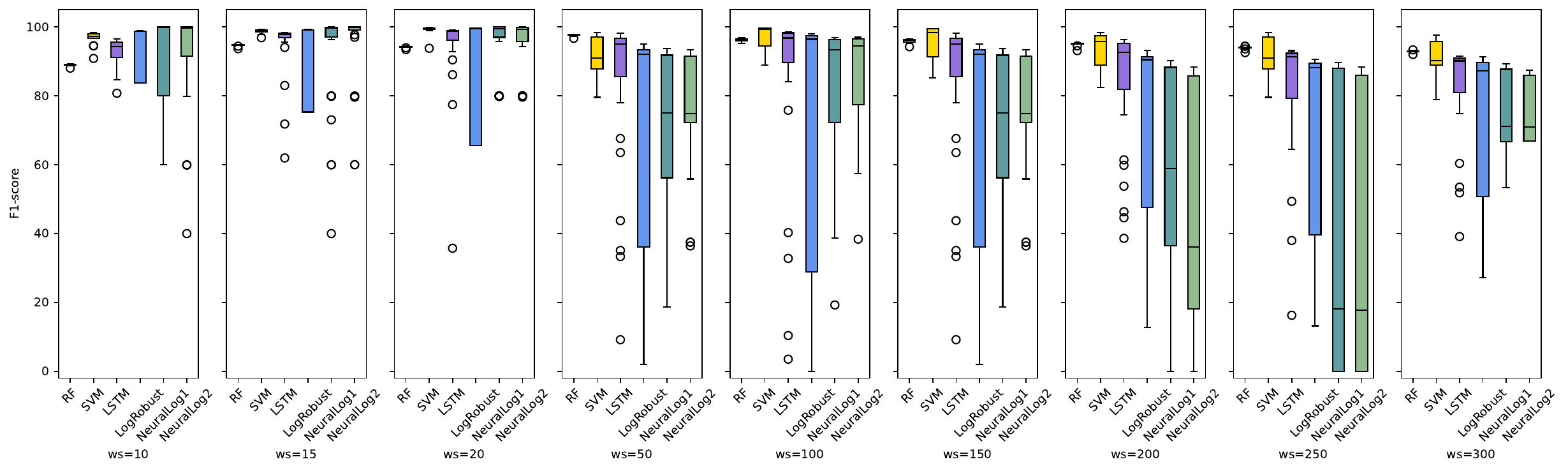}
  \caption{\new{Sensitivity on the \spirit dataset}}
  \label{figRQ1detAcc:spirit}
\end{subfigure}
\caption{Sensitivity of the detection accuracy of supervised traditional and deep ML techniques on log message-based datasets}
\label{fig:sens-sup-detAcc}
\end{figure}

\figurename~\ref{fig:sens-sup-detAcc} shows the sensitivity of the detection accuracy of supervised traditional and deep ML techniques across different window sizes on log message-based datasets.

\begin{itemize}
    \item \textit{Small window sizes $ \{10, 15, 20\}$.}  
    On small window sizes, \new{the  supervised ML techniques (except NeuralLog1 and NeuralLog2 on \hades)} showed limited sensitivity in terms of detection accuracy to hyperparameter tuning on most of the log message-based datasets, in particular \tbird and \spirit. For instance, on \spirit, with $\mathit{ws}=10$, the detection accuracy of RF is far less sensitive (\emph{F1-score} avg $\approx 88.93$, stdDev $\approx 0.35$) than that of all the remaining supervised ML techniques.
   As for deep ML techniques, the \emph{F1-score} observed for LSTM ranges from $80.71$ to $96.52$ (avg $\approx 93.05$, stdDev $\approx 3.51$); \new{NeuralLog2} is the most sensitive deep ML technique showing an \emph{F1-score} ranging from \new{$39.96$} to \new{$99.97$} (avg \new{$\approx 92.24$}, stdDev \new{$\approx 14.08$)}.
    
    \item \textit{Large window sizes $ \{50, 100, 150, 200, 250, 300\}$.} The overall detection accuracy of all supervised ML techniques is more sensitive to hyperparameter tuning across most of the log message-based datasets and large window sizes. 
    For instance, the detection accuracy of LogRobust on \spirit (see \figurename~\ref{figRQ1detAcc:spirit})
    with $\mathit{ws}=100$ ranges from $0.00$ to $97.95$ (avg $\approx 70.93$, stdDev $\approx 37.50$); RF is the least sensitive ML technique to hyperparameter tuning (\emph{F1-score} avg $\approx 96.19$, stdDev $\approx 0.49$).
\end{itemize}

Overall, RF is the least sensitive supervised \new{traditional}  ML technique to hyperparameter tuning in terms of detection accuracy across datasets. \new{One possible reason of the stability of its detection accuracy across different datasets is its decision tree ensemble, which effectively averages out individual tree errors and mitigates overfitting, allowing it to maintain consistent detection accuracy.}
Except for \hades (see \figurename~\ref{figRQ1detAcc:hades}), on which SVM is the most sensitive supervised ML technique\footnote{The choice of hyperparameter $\mathit{C}$ highly impacts the detection accuracy of SVM, especially when evaluated on a highly imbalanced dataset like \hades: the larger the hyperparameter value, the bigger the misclassification penalty, leading to reduced bias towards the majority class in SVM.} \new{and both NeuralLog1 and NeuralLog2 are not learning on larger window sizes, showing a near 0 \emph{F1-score} across hyperparameter settings\footnote{\new{NeuralLog1 and NeuralLog2 rely on the transformer architecture, which requires large amounts of data. They therefore struggle when the training data is limited, and class imbalance further hinders their ability to differentiate between normal and anomalous log event sequences.}}}, supervised deep ML techniques 
are more sensitive to hyperparameter tuning than supervised traditional ML techniques on the remaining datasets (\bgl, \tbird, and \spirit) across window sizes. \new{LogRobust is particularly sensitive on \bgl, \tbird, and \spirit, and NeuralLog1 and NeuralLog2 show increased sensitivity on \tbird and \spirit.}

Statistical analysis (see \S~\ref{sec:stat-analysis}) 
\new{yields a \emph{p-value} of $0.052$, suggesting the sensitivity of detection accuracy across supervised traditional and deep ML techniques is not a distinguishing factor on the seven datasets.}

The answer to RQ1 is that \new{the overall detection accuracy (\emph{F1-score}) of} supervised traditional (RF and SVM) \new{and deep (LSTM, LogRobust, NeuralLog1 and NeuralLog2)} ML techniques yield\new{s}  similar results \new{ on all the benchmark datasets except \fdataset, on which all deep ML models struggle to learn, resulting in poor predictions of log anomalies, whereas traditional ML techniques continue to perform well.
In terms of specificity (\emph{Spec}), all supervised traditional and deep ML techniques show high and similar values across most of the datasets (except for \hadoop), showing that the corresponding models accurately identify normal log event sequences. The low specificity of all supervised ML techniques on \hadoop is explained by the majority class consisting of anomalous log event sequences, making it difficult for these models to recognize the normal log event sequences.}

Further, traditional ML techniques (especially RF) show much less sensitivity, in terms of detection accuracy, to hyperparameter tuning, compared to deep learning techniques on most of the datasets. 
Specifically, RF is the least sensitive on all datasets, followed by SVM which, in spite of being the most sensitive technique on \hades \new{and \fdataset}, is less sensitive than deep ML techniques on the remaining datasets, across window sizes. 
\new{Overall, deep ML techniques} are the most sensitive techniques to hyperparameter tuning, with LSTM \new{(followed by NeuralLog1 and NeuralLog2)}, showing more outliers across datasets.

All the studied traditional and deep ML techniques show their best detection accuracy (\emph{F1-score}) on window sizes ranging from 10 to 50, across log message-based datasets (Table~\ref{bestWSFourdatasets}). 
As expected, we also observed that data imbalance has a negative impact on the detection accuracy of all supervised ML techniques across log message-based datasets: detection accuracy improves from more imbalanced datasets (\hades and \bgl) to less imbalanced ones (\tbird and \spirit).

\subsection{RQ2 - Time performance of supervised traditional and deep ML techniques}\label{RQ2results} 
\subsubsection{Time performance}\label{RQ2timeperformance}
In Table~\ref{timePerformRQ2sup}, we report the model re-training time\footnote{Re-training refers to training the model again using the best hyperparameter settings. The data used to re-train the model consists of both the training set and the validation set from the training phase.} and the prediction time (rows \emph{Re-train.} and \emph{Pred.}, respectively) of the supervised traditional and deep ML techniques on session-based (\hdfs, \hadoop, \new{\fdataset}) and log message-based (\hades, \bgl, \tbird and \spirit) datasets.
One important result is that the overall model re-training time of traditional ML techniques is about \emph{one order of magnitude shorter} than that of deep learning techniques on \new{all} session-based datasets (\hdfs, \hadoop \new{and \fdataset}).

We further study the impact of different window sizes on the time performance of supervised traditional and deep ML techniques across log message-based datasets.

\begin{itemize}
    \item \textit{Small window sizes $ \{10, 15, 20\}$.} 
    Supervised traditional ML techniques are faster (in terms of model re-training time) than deep ML techniques on \hades and \tbird datasets across small window sizes (see Table~\ref{bestWSFourdatasets} for the window sizes associated with the highest detection accuracy and Table~\ref{timePerformRQ2sup} for the model re-training time of the \new{six supervised} ML techniques considered in this study).
    For instance, on \hades, with $\mathit{ws}=10$, RF takes $\SI{66.14}{\second}$, whereas \new{NeuralLog1}  shows the highest model re-training time ($\SI{2479.34}{\second}$) among all the supervised ML techniques on the same dataset and window size. On \tbird, with $\mathit{ws}=10$, RF takes $\SI{847.18}{\second}$ ($\approx \SI{14}{\minute}$), whereas LSTM shows \new{a much higher} model re-training time of $\SI{16135.18}{\second}$ ($\approx \SI{269}{\minute}$). On the other hand, although LSTM shows the lowest model re-training time on \bgl with $\mathit{ws}=15$ ($\SI{60.57}{\second}$), the time taken by RF to re-train the model is relatively close ($\SI{101.38}{\second}$). On \spirit, the model re-training time taken by RF ($\SI{407.28}{\second}$) is by far lower than the time taken by all the remaining supervised techniques to re-train the corresponding models. 
    Recall that small window sizes lead to more sequences to train the supervised ML models (see Section ~\ref{RQ1detectionaccuracy}). This explains the longer model re-training time taken by all the supervised deep ML techniques, when trained on small window sizes.

    \item \textit{Large window sizes $ \{50, 100, 150, 200, 250, 300\}$.}
    All the supervised ML techniques show a short model re-training time across the log message-based datasets on large window sizes. For instance, on \tbird, with $\mathit{ws}=50$, SVM takes $\SI{475.41}{\second}$, whereas re-training the same model takes  longer ($\SI{4928.68}{\second}$) with $\mathit{ws}=10$. This shows that larger window sizes with fewer log event sequences fed into the ML models lead to shorter re-training time. 
\end{itemize}

The prediction time computed for \new{most of} the supervised traditional and deep learning techniques \new{(except for NeuralLog1 on \tbird and \spirit, and both NeuralLog1 and NeuralLog2 on \hdfs and \bgl)} is similar on \new{\hadoop, \fdataset and \hades}, with no practically significant differences. More in detail, the prediction time is less than one minute for all the supervised techniques, ranging from $\SI{0.01}{\second}$ for SVM, RF and LogRobust on \hadoop dataset to $\SI{45.74}{\second}$ for SVM on \spirit dataset. \new{However, the prediction time of NeuralLog2 on \hdfs takes $\SI{180.64}{\second}$ and that of NeuralLog1 on \spirit takes $\SI{190.62}{\second}$.
Therefore, prediction time is generally not a distinguishing factor among most of the techniques, except for the transformer-based models (NeuralLog1 and NeuralLog2), which tend to have significantly longer prediction times due to their complex architecture. Indeed, the number of transformer layers $\mathit{ffnS}$ and the number of attention heads $\mathit{attH}$ control the learning ability of the transformer-based model to capture complex patterns and dependencies in log messages (see Section~\ref{background}).}

Statistical analysis (see \S~\ref{sec:stat-analysis}) indicate that the time performance of the supervised traditional and deep ML techniques, in terms of model re-training \new{time is not significantly different, showing a \emph{p-value} of $0.0826$. While the prediction time of all the supervised ML techniques is significantly different ($\emph{p-value} = 0.047$),} no significant pairwise difference using the post-hoc statistical Dunn’s test is observed (no pair of ML techniques shows a \emph{p-value} smaller than $0.05$).

\begin{table}[tb]
\setlength\extrarowheight{3pt}
\centering
\caption{Time performance (in seconds) of supervised traditional and deep ML techniques on all  datasets} \label{timePerformRQ2sup}
\begin{NiceTabular}
{m{.1cm} m{1.8cm} m{.9cm} m{1cm}m{1.1cm} m{1.2cm} m{1.35cm} m{1.45cm} m{1.4cm}}[hvlines-except-borders] 
\CodeBefore
\rowcolor{gray!15}{1-2}
\Body
\toprule
\Block{2-2}{\emph{Dataset}} & & \Block{2-1}{\emph{Metric}} & \Block{1-6}{\emph{Technique}} & & & \\
& & & \small{SVM} & \small{RF} & \small{LSTM} & \small{LogRobust} & \new{\small{NeuralLog1}} & \new{\small{NeuralLog2}} \\
\Block{6-1}{\rotatebox[origin=c]{90}{Session}} & \Block{2-1}
{\hdfs} & \emph{Re-train.} & 397.64 & 96.01 & \num{2222.09} & \num{1135.53} & \new{\num{57497.17}} & \new{\num{13718.53}} \\ 
& & \emph{Pred.} & 23.88 & 0.89 & 3.01 & 3.53 & \new{\num{70.10}} & \new{180.64} \\

& \Block{2-1}{\hadoop} & 
\emph{Re-train.} & 0.04 & 0.16 & 2.62 & 2.26 & \new{\num{15.95}} & \new{53.96}\\ 
& & \emph{Pred.} & 0.01 & 0.01 & 0.33 & 0.01 & \new{\num{0.62}} & \new{1.54} \\

& \Block{2-1}{\new{\fdataset}} & 
\emph{\new{Re-train.}} & \new{0.07} & \new{2.11} & \new{29.49} & \new{2.12} & \new{\num{13.87}} & \new{13.87} \\ 
& & \emph{\new{Pred.}} & \new{0.03} & \new{0.01} & \new{0.49} & \new{0.01} & \new{\num{0.48}} & \new{0.48}\\

\Block{8-1}{\rotatebox[origin=c]{90}{Log message}} & \Block{2-1}{\hades} & \emph{Re-train.} & 4.45 &  66.14 & 178.03 & 327.60 & \new{\num{2479.34}} & \new{1974.00}\\ 
& & \emph{Pred.} & 0.65 & 0.28 & 0.61 &  0.69 & \new{13.37} & \new{12.22}\\ 
& \Block{2-1}{\bgl} & \emph{Re-train.} & 246.29 & 101.38 & 60.57 & 813.34 & \new{6377.34} & \new{\num{8387.62}}\\ 
& & \emph{Pred.} & 14.55 & 0.72 & 1.96 & 2.89 & \new{95.29} & \new{87.42} \\ 
& \Block{2-1}{\tbird} & \emph{Re-train.} & 475.41 & 847.18 & \num{16135.18} & 856.60  & \new{\num{25067.90}}  & \new{\num{3329.26}}\\  
& & \emph{Pred.} & 32.60 & 3.61 & 11.71 & 3.38 & \new{127.37} & \new{42.21} \\ 
& \Block{2-1}{\spirit} & \emph{Re-train.} & \num{2317.00} & 407.28 & \num{1416.07} & \num{1184.84} & \new{\num{18582.84}} & \new{\num{2138.44}} \\ 
& & \emph{Pred.} & 45.74 & 1.12 & 1.61 & 3.05 & \new{190.62} & \new{14.99} \\ 
\bottomrule
\end{NiceTabular}
\end{table}
 
\subsubsection{Sensitivity of Training Time}\label{RQ2SensTimePerformance}
\begin{figure}[tb]
\begin{subfigure}{{.5\textwidth}}
  \includegraphics[width=6cm]{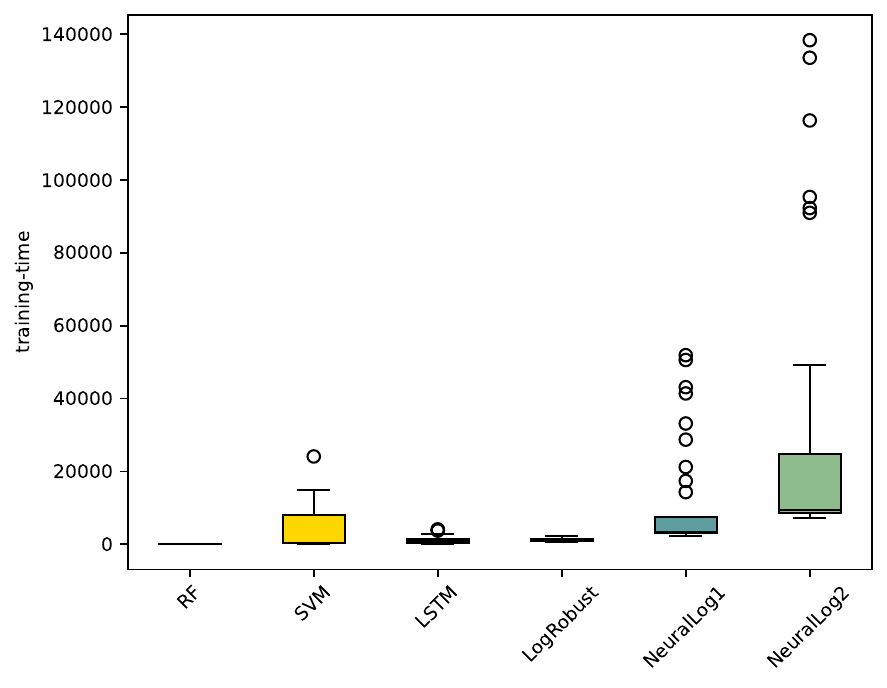}
  \caption{\new{\hdfs dataset}}
  \label{rq2Vartimehdfs}
\end{subfigure}\begin{subfigure}{{.5\textwidth}}
  \includegraphics[width=6cm]{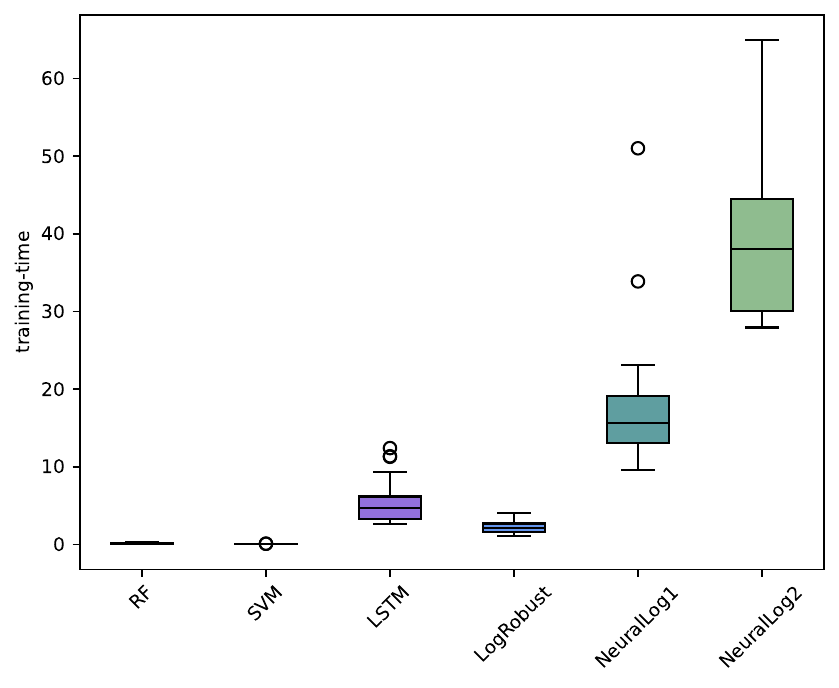}
  \caption{\new{\hadoop dataset}}
  \label{rq2Vartimehadoop}
\end{subfigure}

\begin{subfigure}{{\textwidth}}
    \centering
  \includegraphics[width=6cm]{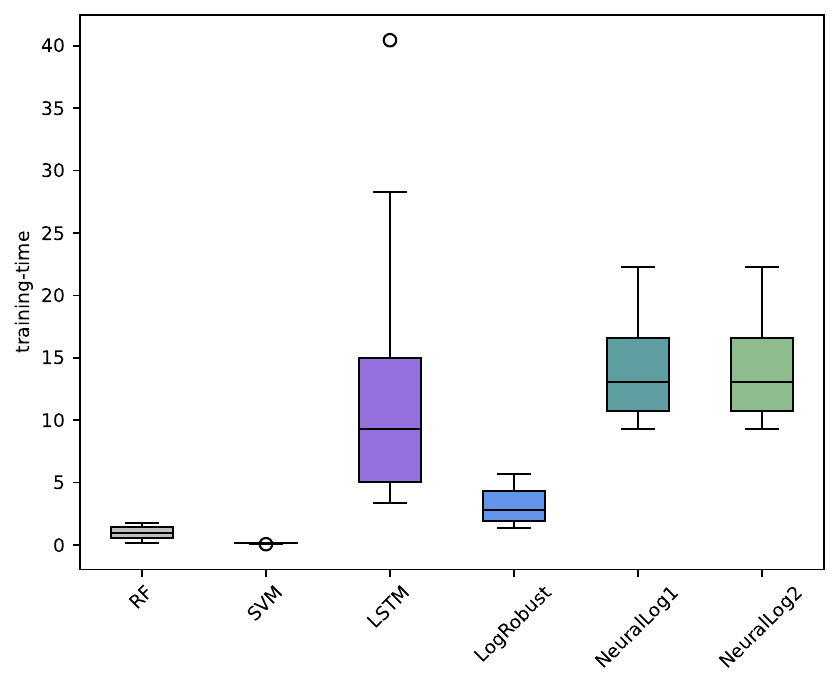}
  \caption{\new{\fdataset dataset}}
  \label{rq2Vartimefdataset}
\end{subfigure}

\caption{Sensitivity of the time performance (in seconds) of supervised traditional and deep ML techniques on session-based datasets (The difference in the y-axis scale of the \new{three} plots is due to the difference in the training size of the datasets)}
\label{rq2supVarTime}
\end{figure}
 As depicted in~\figurename~\ref{rq2supVarTime}, 
the \new{overall model training time of traditional ML techniques (RF and SVM) }is much less sensitive to hyperparameter tuning than that of deep ML techniques on \new{the three} session-based datasets. 
\new{For instance, RF takes from $\SI{13.24}{\second}$ to $\SI{130.64}{\second}$ to train the model (avg $\approx \SI{71.98}{\second}$, stdDev $\approx \SI{37.43}{\second}$) on \hdfs (box plot in \figurename~\ref{rq2Vartimehdfs}) and SVM takes from $\SI{0.029}{\second}$ to $\SI{0.092}{\second}$ to train the model (avg $\approx \SI{0.042}{\second}$, stdDev $\approx \SI{0.019}{\second}$) on \hadoop (box plot in \figurename~\ref{rq2Vartimehadoop}).}

Regarding deep ML techniques, the model training time of LogRobust is less sensitive to hyperparameter tuning than that of \new{the remaining techniques on all the} session-based datasets. For instance, on \hdfs, LogRobust takes from $\SI{632.88}{\second}$ to $\SI{2334.76}{\second}$ (avg $\approx \SI{1285.13}{\second}$, stdDev $\approx \SI{490.36}{\second}$), whereas the model training time of \new{NeuralLog2} ranges from $\SI{7090.97}{\second}$ to $\SI{138370.41}{\second}$ (avg $\approx \SI{29868.56}{\second}$, stdDev $\approx \SI{38664.96}{\second}$) to train the corresponding models. \new{This is expected, given the complex architecture of transformer-based models.}

\begin{figure}[htbp]
\centering
\begin{subfigure}{1\textwidth}
  \centering
  \includegraphics[height=3.9cm, width=14cm]{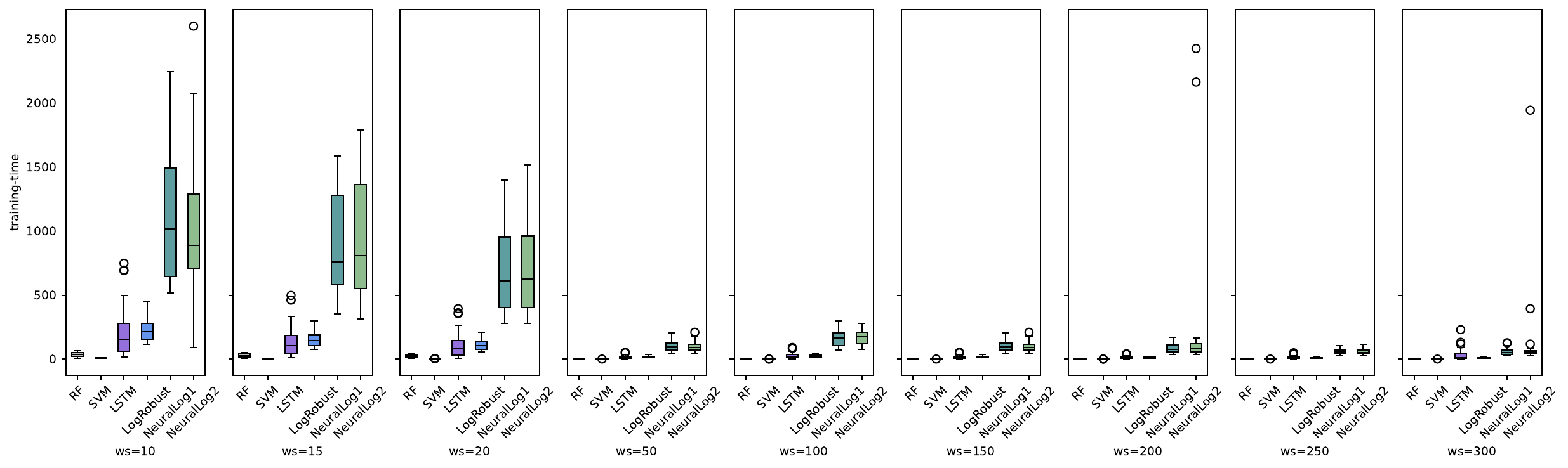}
  \caption{\new{Sensitivity on \hades dataset}}
  \label{figRQ2time:hades}
\end{subfigure}

\begin{subfigure}{1\textwidth}
  \centering
  \includegraphics[height=3.9cm, width=14cm]{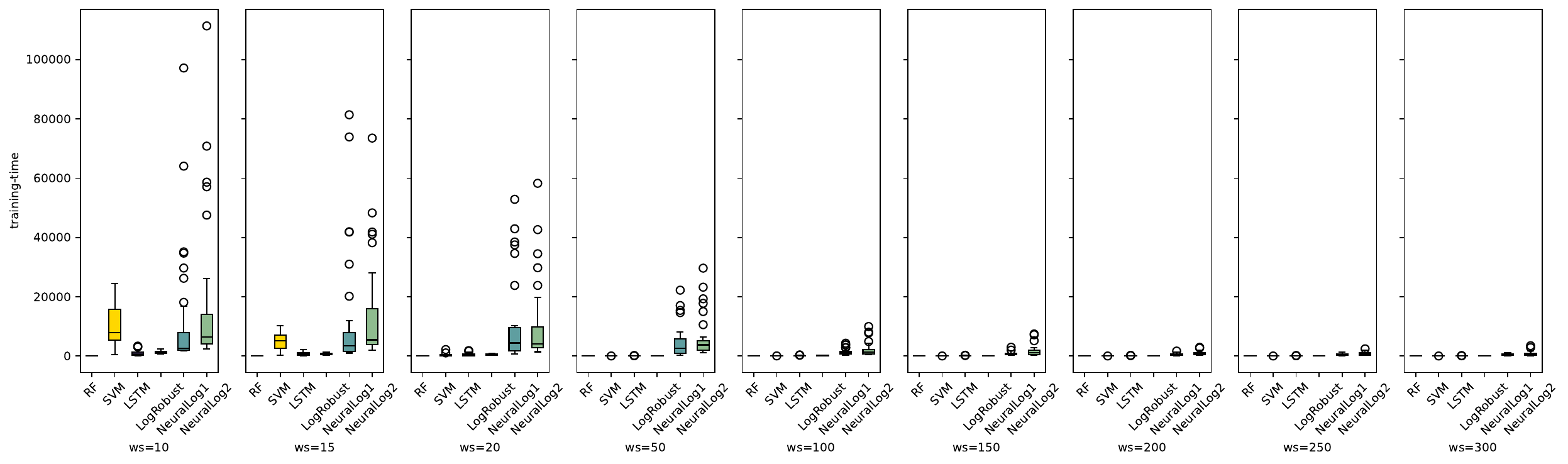}
  \caption{\new{Sensitivity on \bgl dataset}}
  \label{figRQ2time:bgl}
\end{subfigure}
\begin{subfigure}{1\textwidth}
  \centering
  \includegraphics[height=3.9cm, width=14cm]{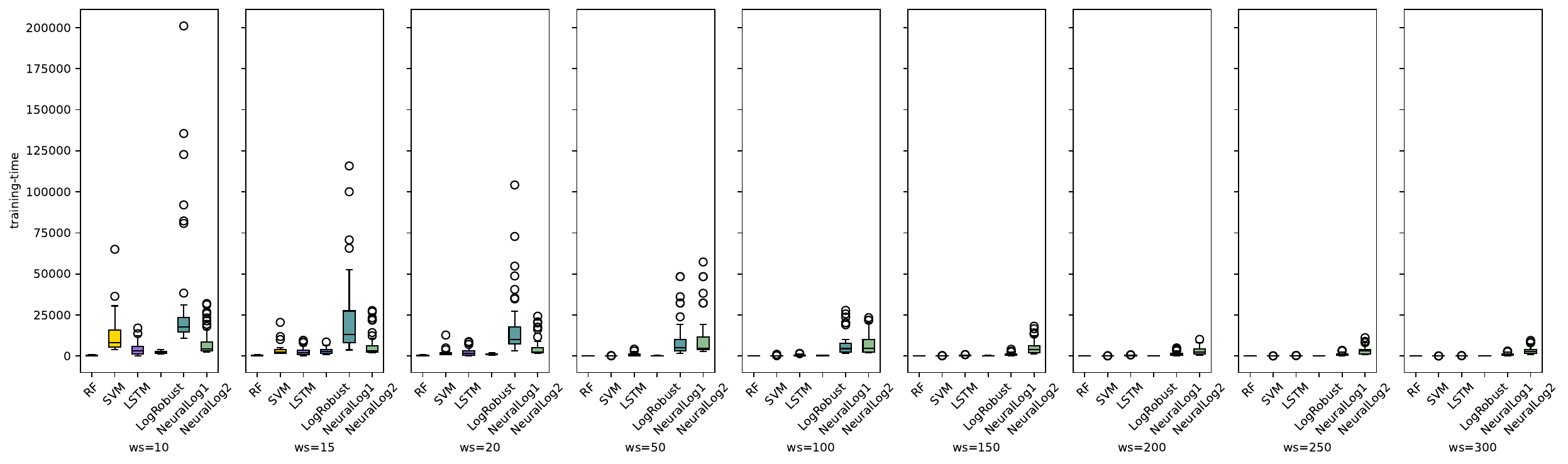}
  \caption{\new{Sensitivity on \tbird dataset}}
  \label{figRQ2time:tbird}
\end{subfigure}

\begin{subfigure}{1\textwidth}
  \centering
  \includegraphics[height=3.9cm, width=14cm]{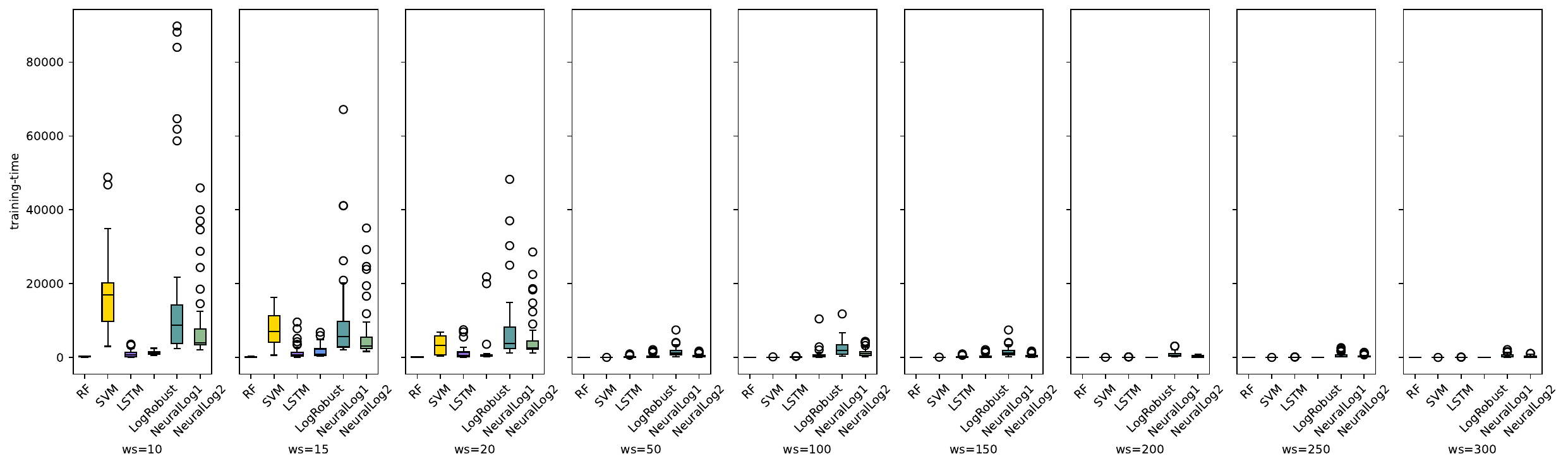}
  \caption{\new{Sensitivity on \spirit dataset}}
  \label{figRQ2time:spirit}
\end{subfigure}
\caption{Sensitivity of the time performance (in seconds) of supervised traditional and deep ML techniques on four log message-based datasets}
\label{fig:sens-sup-time}
\end{figure}

\figurename~\ref{fig:sens-sup-time} shows the sensitivity of training time of supervised traditional and deep ML techniques across different window sizes on log message-based datasets.
\begin{itemize}
    \item \textit{Small window sizes $ \{10, 15, 20\}$.}  
    The overall model training of RF is less sensitive to hyperparameter tuning (with no outliers) than that of the remaining supervised deep ML techniques across most of the datasets and window sizes. 
    The only exception we observe is on window sizes ranging from 10 to 50 on \hades, in which RF is slightly more sensitive to hyperparameter tuning than SVM. Nevertheless, the difference in model training between RF and SVM on these window sizes is negligible. 
    For instance, on \hades, with $\mathit{ws}=10$, RF takes from $\SI{6.43}{\second}$ to $\SI{65.45}{\second}$ (avg $\approx \SI{36.41}{\second}$, stdDev $\approx \SI{19.03}{\second}$), whereas SVM takes from $\SI{4.71}{\second}$ to $\SI{12.31}{\second}$ (avg $\approx \SI{7.33}{\second}$, stdDev $\approx \SI{2.90}{\second}$) to train the corresponding model.
    \new{The two versions of NeuralLog (NeuralLog1 and NeuralLog2) show the highest sensitivity to hyperparameter tuning on small window sizes on all log message-based datasets with many outliers.}
    \new{For instance, on \hades, with $\mathit{ws}=10$, NeuralLog1 takes from $\SI{516.72}{\second}$ to $\SI{2247.86}{\second}$ (avg $\approx \SI{1127.87}{\second}$, stdDev $\approx \SI{516.89}{\second}$) to train the corresponding model.} 
    \new{On \spirit, on the same window size, the model training time of the latter technique takes from 
    $\SI{2430.81}{\second}$ to $\SI{89768.24}{\second}$ (avg $\approx \SI{18852.15}{\second}$, stdDev $\approx \SI{25835.39}{\second}$)}    
    \item \textit{Large window sizes $ \{50, 100, 150, 200, 250, 300\}$.}
    Overall, the model training time of all supervised ML techniques is much less sensitive to hyperparameter tuning across large window sizes than that observed on small window sizes. For instance, on \bgl, SVM takes from $\SI{7.04}{\second}$ to $\SI{38.01}{\second}$ (avg $\approx \SI{14.18}{\second}$, stdDev $\approx \SI{7.02}{\second}$) with $\mathit{ws}=100$. 
    This implies that larger window sizes, with fewer log event sequences fed to the ML models, result in a lower sensitivity to hyperparameter tuning of the model training time for all the supervised ML techniques. 
\end{itemize}

Statistical analysis (see \S~\ref{sec:stat-analysis}) indicates that the sensitivity of the training time of the different supervised traditional and deep ML techniques across hyperparameter settings is not significantly different (\emph{p-value}=\new{$0.11$}). 

The answer to RQ2 is that\new{, except for the transformer-based ML techniques NeuralLog1 and NeuralLog2, the remaining} supervised traditional (RF and SVM) and deep (LSTM and LogRobust) ML techniques show similar model prediction time with no practically significant differences across the different session-based (\hdfs\new{,} \hadoop \new{and \fdataset}) and log message-based (\hades, \bgl, \tbird and \spirit) datasets.

The model re-training time taken by traditional ML techniques (especially RF) is, however, significantly lower than the time taken by deep ML techniques on most of the datasets, notably \hdfs, \hades and \tbird.

Overall, supervised traditional and deep ML techniques generally i) take less time for model training and ii) are less sensitive to hyperparameter tuning when using large window sizes compared to small ones. This trend holds across both more imbalanced datasets like \hades and \bgl and less imbalanced ones such as \tbird and \spirit. 
Notably, the model training time of RF shows less sensitivity to hyperparameter tuning on \new{all} session-based and log message-based datasets 
compared to deep ML techniques across window sizes.

\subsection{RQ3 - Detection accuracy of semi-supervised traditional and deep ML techniques}\label{RQ3results}

\subsubsection{Detection Accuracy}\label{RQ3detectionaccuracy}
\begin{table}[!htbp]
\setlength\extrarowheight{3pt}
\caption{Comparison of the detection accuracy of semi-supervised traditional and deep ML techniques on all datasets}
\label{rq3Accuracy}
\centering
\begin{NiceTabular}
{m{1cm} m{1.9cm} m{1cm} m{1.5cm} m{1.3cm} m{1.8cm}}[hvlines-except-borders] 
\CodeBefore
\rowcolor{gray!15}{1-2}
\Body
\toprule
\Block{2-2}{\emph{Dataset}} & & \Block{2-1}{\emph{Metric}} & \Block{1-3}{\emph{Technique}} & \\
& & & OC-SVM & DeepLog & \new{Logs2Graphs}\\
\Block{12-1}{\rotatebox[origin=c]{90}{\emph{Session}}} & \Block{4-1}{\hdfs} & \emph{Prec}  & 46.32  & 93.86 & \new{95.27}  \\  
& & \emph{Rec} & 67.94  & 72.19 & \new{44.88} \\ 
& & \emph{F1} & \textbf{55.09} & \textbf{81.16} & \new{\textbf{61.02}}  \\
& & \new{\emph{Spec}} & \new{89.31} & \new{96.82} & \new{99.70} \\
& \Block{4-1}{\hadoop}  & 
\emph{Prec}  & 50.36  & 51.01 & \new{50.68} \\ 
& & \emph{Rec}  & 86.50 & 90.80 & \new{92.02} \\
& & \emph{F1}  & \textbf{63.66} & \textbf{64.54} & \new{\textbf{65.36}}\\ 
& & \new{\emph{Spec}} & \new{7.95} & \new{9.17} & \new{3.31} \\

& \Block{4-1}{\new{\fdataset}}  & 
\emph{\new{Prec}} & \new{63.04}  & \new{61.90} & \new{0.00} \\ 
&& \emph{\new{Rec}}  & \new{100.00}  & \new{95.37} & \new{0.00} \\
&& \emph{\new{F1}}  & \new{\textbf{77.33}} & \new{\textbf{75.06}} & \new{\textbf{0.00}} \\ 
& & \new{\emph{Spec}} & \new{0.00} & \new{0.00} & \new{100.00} \\
\Block{16-1}{\rotatebox[origin=c]{90}{\emph{Log message}}} & 
\Block{4-1}{\hades}  & \emph{Prec}  & 89.19 & 66.92 & \new{99.02}\\
& & \emph{Rec}  & 64.71 & 48.85 & \new{80.16}\\ 
& & \emph{F1}  & \textbf{75.00} & \textbf{56.46} & \new{\textbf{80.60}}\\ 
& & \new{\emph{Spec}} & \new{99.42} & \new{99.38} & \new{100.0} \\

& \Block{4-1}{\bgl}  & \emph{Prec}  & 43.65 &  70.86  & \new{96.86}\\
& & \emph{Rec} & 96.80 &  64.70 & \new{84.35} \\
& & \emph{F1}  & \textbf{60.16} & \textbf{67.61} & \new{\textbf{90.17}}\\ 
& & \new{\emph{Spec}} & \new{42.06} & \new{90.48} & \new{92.86} \\
& \Block{4-1}{\tbird}  & \emph{Prec}  & 74.09 & 76.73 & \new{87.44}\\ 
& & \emph{Rec}  & 81.95 & 63.11 & \new{17.22}\\ 
& & \emph{F1}  & \textbf{77.82} & \textbf{67.98} & \new{\textbf{90.55}}\\ 
& & \new{\emph{Spec}} & \new{13.33} & \new{42.11} & \new{96.82} \\
& \Block{4-1}{\spirit}  & \emph{Prec}  & 74.15 & 90.89 & \new{99.87} \\ 
& & \emph{Rec}  & 86.65 & 77.87 & \new{91.18}\\ 
& & \emph{F1}  & \textbf{79.91} & \textbf{83.87} & \new{\textbf{95.33}}\\  
& & \new{\emph{Spec}} & \new{43.73} & \new{78.57} & \new{99.71} \\
\bottomrule
\end{NiceTabular}
\end{table}

 As shown in Table~\ref{rq3Accuracy}, DeepLog far outperforms OC-SVM \new{and Logs2Graphs} on \hdfs in terms of detection accuracy with a notable difference  of \SI{26.07}{\pp} \new{and \SI{20.14}{\pp}} (\si{\pp} = percentage points)\new{, respectively}.
\new{
 A recent study in log-based datasets~\cite{landauer2024critical} shows a high redundancy in log event sequences within \hdfs (a total of \num{575061} sequences can be reduced to \num{26814} sequences only). So many nearly identical event sequences make the index-based encoding technique DeepLog more effective than semantics-based encoding techniques (OC-SVM and Logs2Graphs) on that dataset. 
 More in detail, index-based encoding preserves the order of log event occurrences within sequences and handles the high redundancy of the dataset, whereas semantics-based encoding struggles to capture the differences in order, leading to reduced detection accuracy.}

However, the detection accuracy of \new{the semi-supervised ML techniques is very similar on \hadoop and \fdataset (with a difference in detection accuracy of only \SI{0.88}{\pp} \new{and \SI{2.27}{\pp}, respectively}), with the exception of Logs2Graphs on the latter dataset (\emph{F1-score}=0.00), suggesting that the corresponding ML model is not able to detect anomalous log event sequences on that dataset.}
\new{In terms of specificity, all the semi-supervised ML techniques show a high value on \hdfs, indicating their ability to detect normal log event sequences. They, however, do not perform well on \hadoop and \fdataset. The only exception on \fdataset is for Logs2Graphs (\emph{Spec}=100.00), which perfectly detects normal log event sequences.
}
According to a recent study~\cite{landauer2024critical}, there is a high overlap in the \hadoop \minor{dataset, in the sense that 83.2\% of normal log event sequences contain at least one log event sequence that also appears in anomalous log event sequences. Additionally, 75.5
\% of anomalous log event sequences are identical to normal ones. 
} 
 Therefore, this overlap makes it difficult for the semi-supervised ML models to effectively distinguish between normal and anomalous log event sequences, resulting in poor detection accuracy (the \emph{F1-score} ranges from $63.66$ for OC-SVM to $65.36$ for Logs2Graphs) and very low specificity values, ranging from $3.31$ for Logs2Graphs to $9.17$ for DeepLog.

\new{Logs2Graphs far outperforms OC-SVM and DeepLog in terms of \emph{F1-score} and \emph{Spec} on all log message-based datasets. This suggests that the GNN-based semi-supervised approach is more effective at detecting log anomalies compared to the traditional OC-SVM and the RNN-based DeepLog, demonstrating a superior ability to differentiate between normal and anomalous log event sequences.
}
\new{For instance, on \tbird, Logs2Graphs achieves an \emph{F-score} of $90.55$ and a \emph{Spec} of $96.82$, by far outperforming DeepLog with an \emph{F-score} of $67.98$ and a \emph{Spec} of $42.11$, and OC-SVM with an \emph{F-score} of $77.82$ and a \emph{Spec} of $13.33$.
We also observe that the specificity of Logs2Graphs is not impacted by the imbalance ratio (IR, see Table~\ref{newDatasetsStrategies}). For instance, the highest specificity (\emph{Spec}=$100.00\%$) of the latter is recorded on \hades, with an IR of 0.13\% at $ws=10$, while its lowest specificity (\emph{Spec}=$92.86\%$) is recorded on \bgl, with an IR of 11.38\% at $ws=200$.
}
Although DeepLog outperforms OC-SVM on the two log message-based datasets \bgl and \spirit, the difference in detection accuracy \new{(\emph{F1-score})} is relatively small when compared to the one observed on \hdfs: 
the difference in \emph{F1-score} value is \SI{7.45}{\pp} and \SI{3.96}{\pp} on \bgl and \spirit respectively.
OC-SVM, however, outperforms DeepLog on the remaining two log message-based datasets (\hades and \tbird). 

\new{Specificity, however, is a distinguishing factor for OC-SVM and DeepLog on log message-based datasets, except for \hades.}
\new{For instance, the difference in specificity is \SI{28.78}{\pp} on \tbird and \SI{48.42}{\pp} on \bgl.}

\new{Overall, both OC-SVM and DeepLog show a decreasing specificity from more imbalanced (\hades and \bgl) to less imbalanced (\tbird and \spirit) datasets, given that the imbalance ratio (IR) on the former datasets is much less than that on the latter datasets (see Table~\ref{newDatasetsStrategies}).}

\new{For instance, DeepLog achieves its highest specificity (\emph{Spec}=$99.38\%$) on \hades (the most imbalanced dataset), with an IR of 0.13\% at $ws=10$, while its lowest specificity (\emph{Spec}=$42.11\%$) is recorded 
on \tbird on $ws=200$, with an IR=$39.81\%$.
Similarly, the highest specificity of OC-SVM is recorded on \hades, with an IR of 1.6\% at $ws=300$, while its lowest specificity is recorded on \tbird, with an IR of 40.19\% at $ws=250$.}
\new{This trend in terms of specificity reflects the ability of all traditional (OC-SVM) and deep (DeepLog and Logs2Graphs) semi-supervised ML techniques to better distinguish normal from anomalous log event sequences in datasets with lower $IR$ (\hades and \bgl), reflecting that the identification of normal log event sequences decreases with the increase of the imbalance ratio.}

The difference in detection accuracy \new{(F1-score)} between OC-SVM and DeepLog is higher
on the \hades and \tbird datasets than on the \bgl and \spirit ones, showing a higher ability of DeepLog at detecting anomalous log event sequences on these datasets. More in detail, the difference between the detection accuracy of both semi-supervised ML techniques on \hades is \SI{18.54}{\pp}, whereas it is \SI{9.84}{\pp} on \tbird. 
\new{In terms of specificity, DeepLog shows a better ability at avoiding false positives than  OC-SVM on \bgl, \tbird and \spirit, with a difference in specificity values of \SI{48.42}{\pp}, \SI{28.78}{\pp} and \SI{34.84}{\pp}, respectively.}

\begin{figure}[htbp]
\centering
\begin{subfigure}{1\textwidth}
  \centering
  \includegraphics[height=3.75cm, width=12cm]{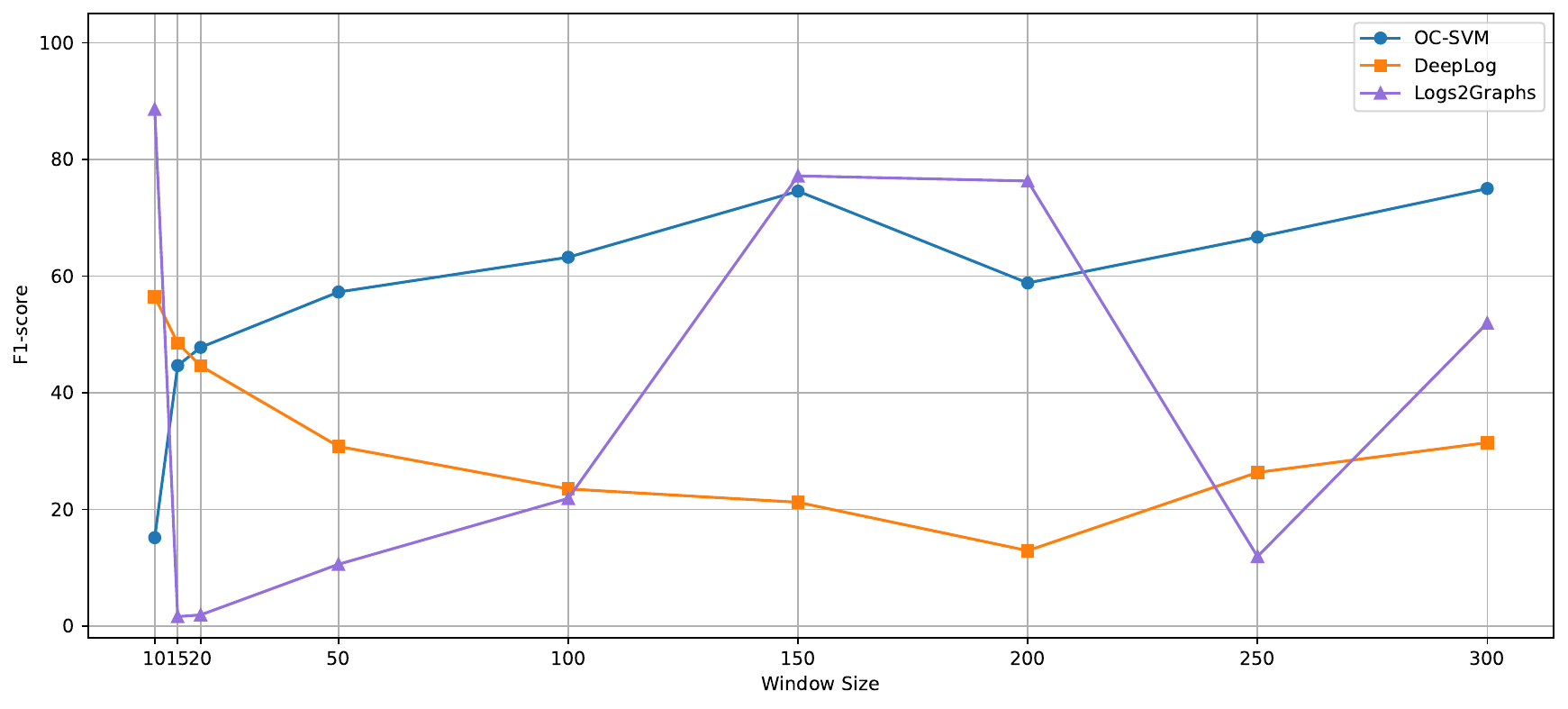}
  \caption{\new{Impact of window size when using the \hades dataset}}
  \label{figRQ3det:hadesAllW}
\end{subfigure}

\begin{subfigure}{1\textwidth}
  \centering
  \includegraphics[height=3.75cm, width=12cm]{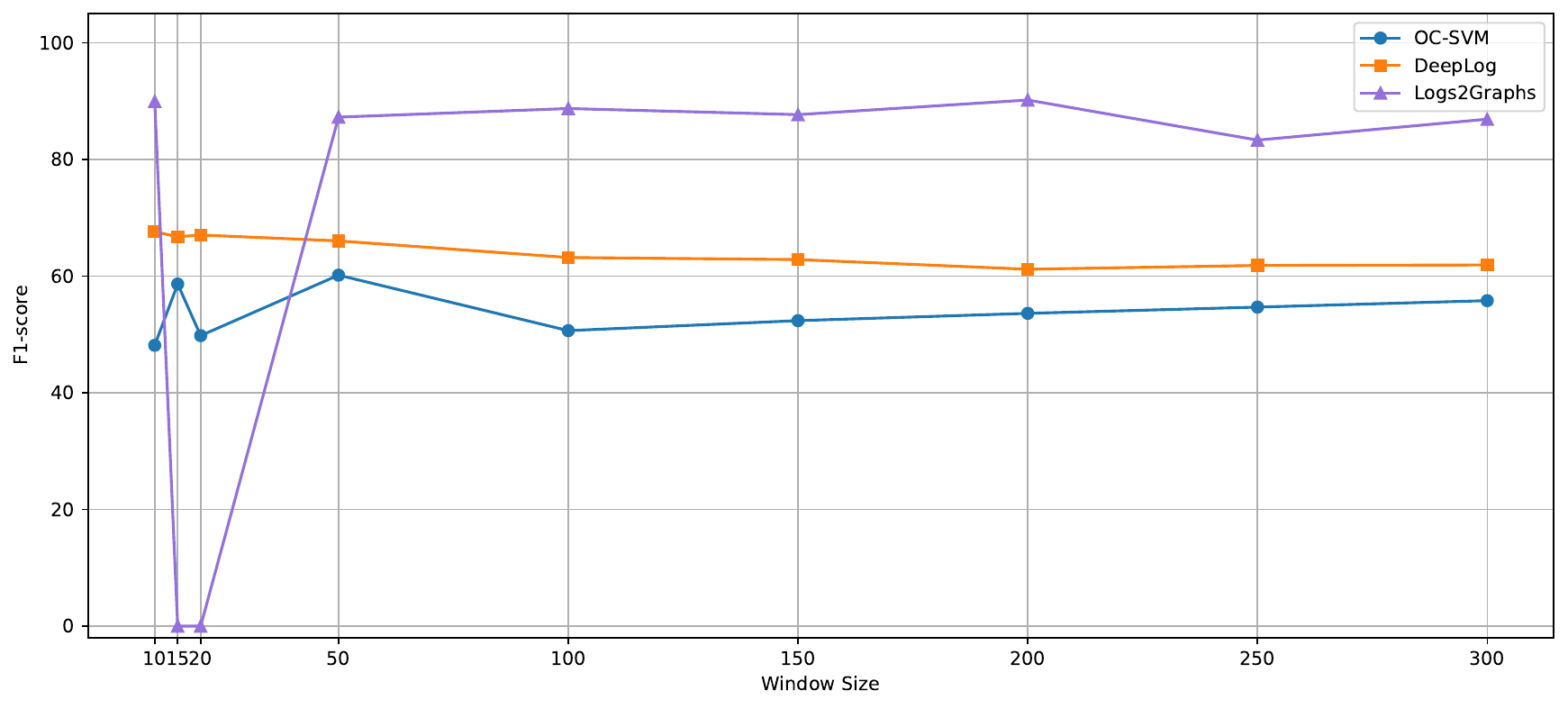}
  \caption{\new{Impact of window size when using the \bgl dataset}}
  \label{figRQ3det:bglAllW}
\end{subfigure}
\begin{subfigure}{1\textwidth}
  \centering
  \includegraphics[height=3.75cm, width=12cm]{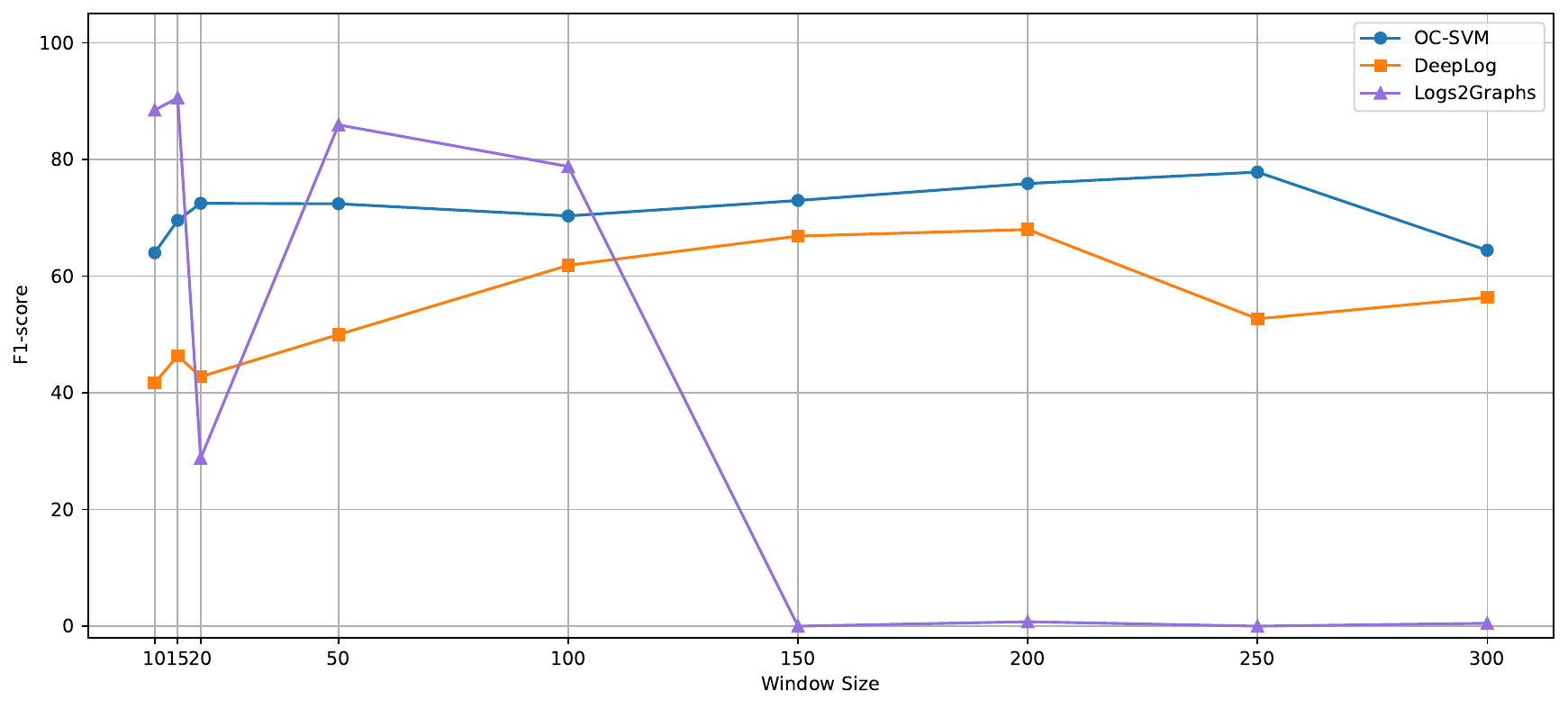}
  \caption{\new{Impact of window size when using the \tbird dataset}}
  \label{figRQ3det:tbirdAllW}
\end{subfigure}

\begin{subfigure}{1\textwidth}
  \centering
  \includegraphics[height=3.75cm, width=12cm]{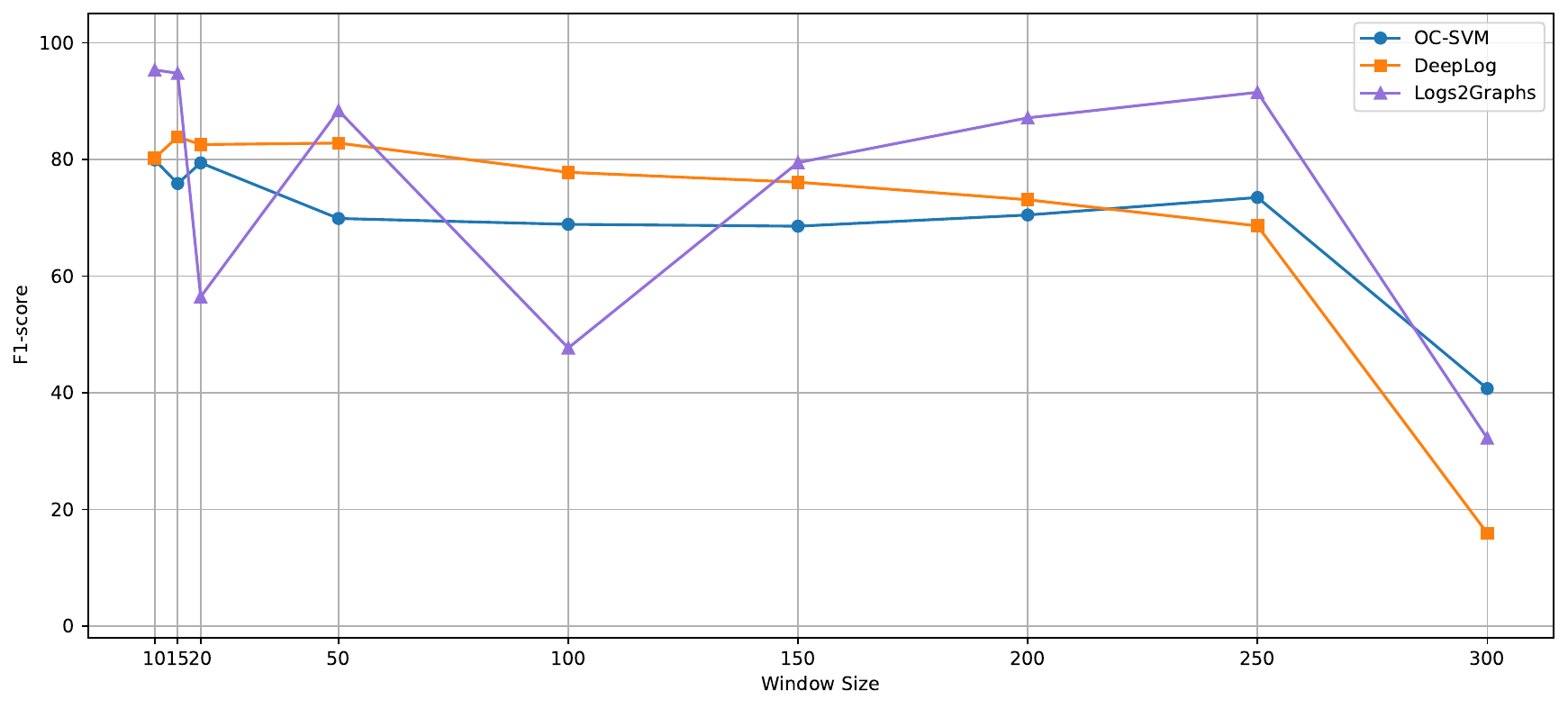}
  \caption{\new{Impact of window size when using the \spirit dataset}}
  \label{figRQ3det:spiritAllW}
\end{subfigure}
\caption{Impact of window size on the detection accuracy of semi-supervised traditional and deep ML techniques on log message-based datasets}
\label{fig:semi-detaccAllW}
\end{figure}

\figurename~\ref{fig:semi-detaccAllW} shows the impact of different window sizes on the detection accuracy of semi-supervised traditional and deep ML techniques on log message-based datasets.

\begin{itemize}
       \item \textit{Small window sizes $ \{10, 15, 20\}$.}
        As depicted in Table~\ref{bestWSFourdatasets}, \new{both deep ML techniques achieve their highest detection accuracy with smaller window sizes on three out of four datasets, with DeepLog showing its highest detection accuracy in terms of \emph{F1-score} on \hades, \bgl and \spirit, and Logs2Graphs on \hades, \tbird and \spirit.} 

        \item \textit{Large window sizes $\{50, 100, 150, 200, 250, 300\}$.}
        As shown in Table~\ref{bestWSFourdatasets}, OC-SVM yields its highest detection accuracy on larger window sizes on three \new{(\hades, \bgl and \tbird)} out of the four log message-based datasets.
        Overall, large window sizes are deemed more suitable for OC-SVM in detecting execution path log anomalies on log message-based datasets. 
\end{itemize}

To conclude, the detection accuracy of semi-supervised traditional \new{(OC-SVM)} and deep ML \new{(DeepLog and Logs2Graphs)} techniques varies across different window sizes. 
Our findings related to DeepLog are consistent with a recent empirical study~\cite{le2022logHowFar}, which also reported similar variations in detection accuracy across different window sizes for semi-supervised deep ML techniques, including DeepLog.

Statistical analysis (see \S~\ref{sec:stat-analysis}) yields a \emph{p-value} of \new{$0.56$}, suggesting the detection accuracy of semi-supervised traditional and deep ML techniques is not significantly different.

\subsubsection{Sensitivity of Detection Accuracy}\label{RQ3SensDetectionAccuracy}
\begin{figure}[tb]
\begin{subfigure}{{.5\textwidth}}
  \includegraphics[width=6cm]{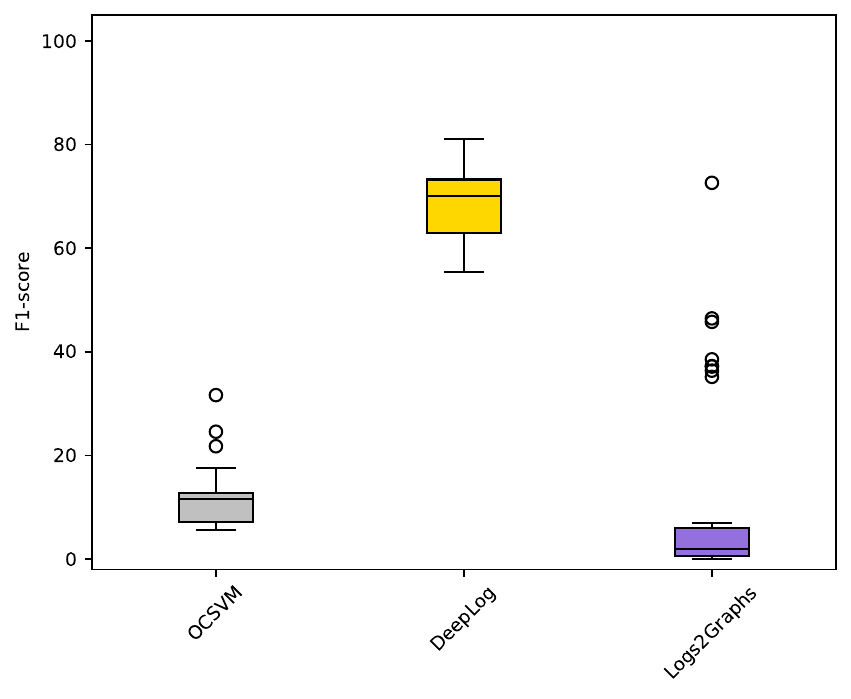}
  \caption{\new{\hdfs dataset}}
  \label{rq3Varocchdfs}
\end{subfigure}\begin{subfigure}{{.5\textwidth}}
  \includegraphics[width=6cm]{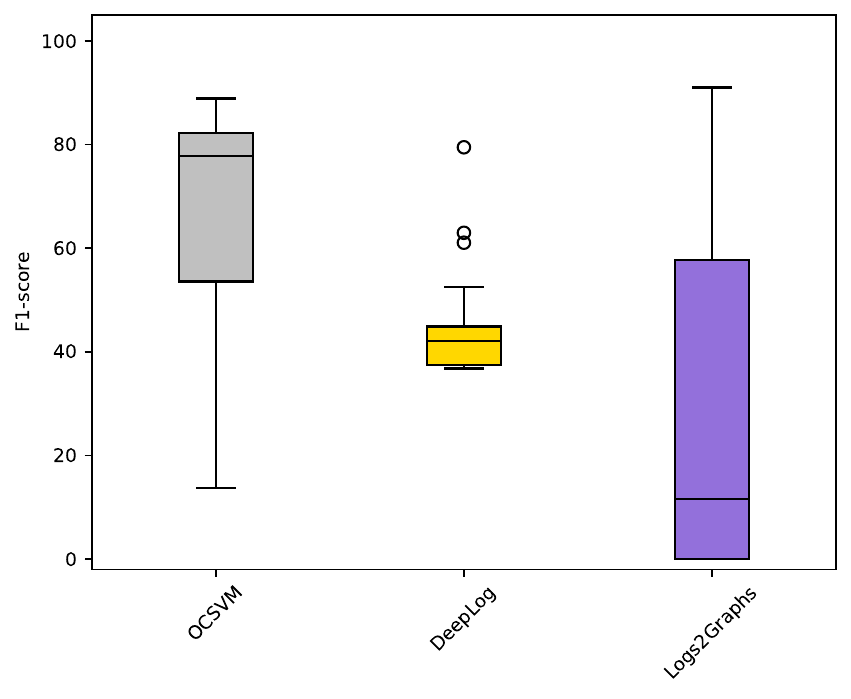}
  \caption{\new{\hadoop dataset}}
  \label{rq3Varocchadoop}
\end{subfigure}

\begin{subfigure}{{\textwidth}}
    \centering
  \includegraphics[width=6cm]{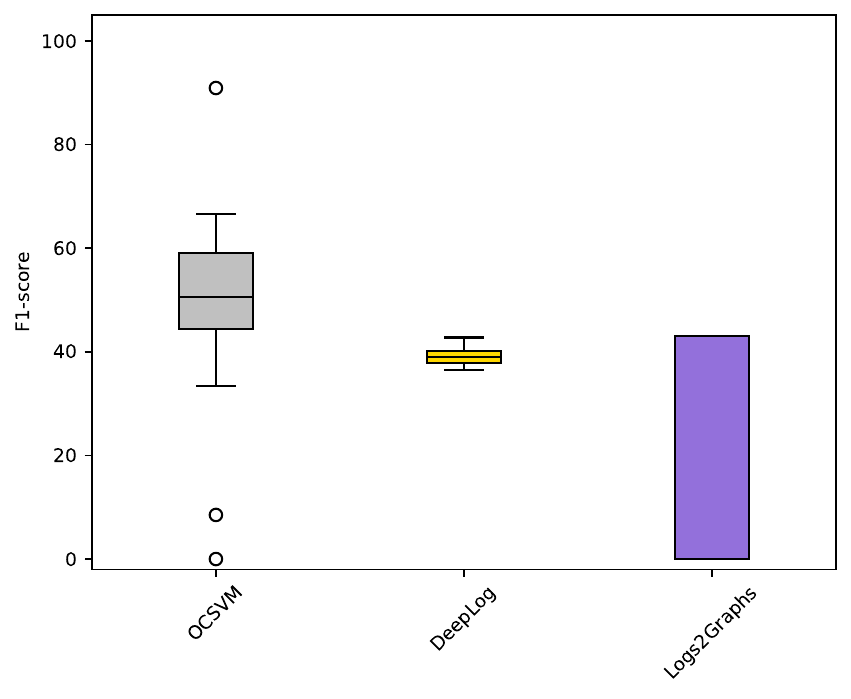}
  \caption{\new{\fdataset dataset}}
  \label{rq3VarAccfdataset}
\end{subfigure}

\caption{Sensitivity of the detection accuracy of semi-supervised traditional and deep ML techniques on session-based datasets}
\label{rq3VarAcc}
\end{figure}
 As depicted in \figurename~\ref{rq3VarAcc}, 
\new{the detection accuracy of Logs2Graphs is more sensitive to hyperparameter tuning than the remaining semi-supervised ML techniques across all the session-based datasets. 
For instance, on \hdfs (\figurename~\ref{rq3Varocchdfs}), the detection accuracy of Logs2Graphs ranges from 
$0.00$ to $72.61$ (avg $\approx 11.43$, stdDev $\approx 18.21$), whereas that of DeepLog ranges from 
$55.38$ to $81.12$ (avg $\approx 68.46$, stdDev $\approx 6.65$) and that of OC-SVM ranges from $5.67$ to $31.64$ (avg $\approx 11.92$, stdDev $\approx 5.66$).}
\new{Although the difference in sensitivity between OC-SVM and DeepLog on \hdfs is small (with outliers recorded for OC-SVM), OC-SVM is generally much more sensitive to hyperparameter tuning than DeepLog on both \hadoop and \fdataset.
For instance, on \hadoop, the detection accuracy of DeepLog ranges from $36.78$ to $79.47$ (avg $\approx 43.86$, stdDev $\approx 8.84$),
whereas the detection accuracy computed for OC-SVM ranges from $13.64$ to $88.89$ (avg $\approx 66.88$, stdDev $\approx 21.15$).}

\begin{figure}[p]
\centering
\begin{subfigure}{1\textwidth}
  \centering
  \includegraphics[height=3.9cm, width=14cm]
  {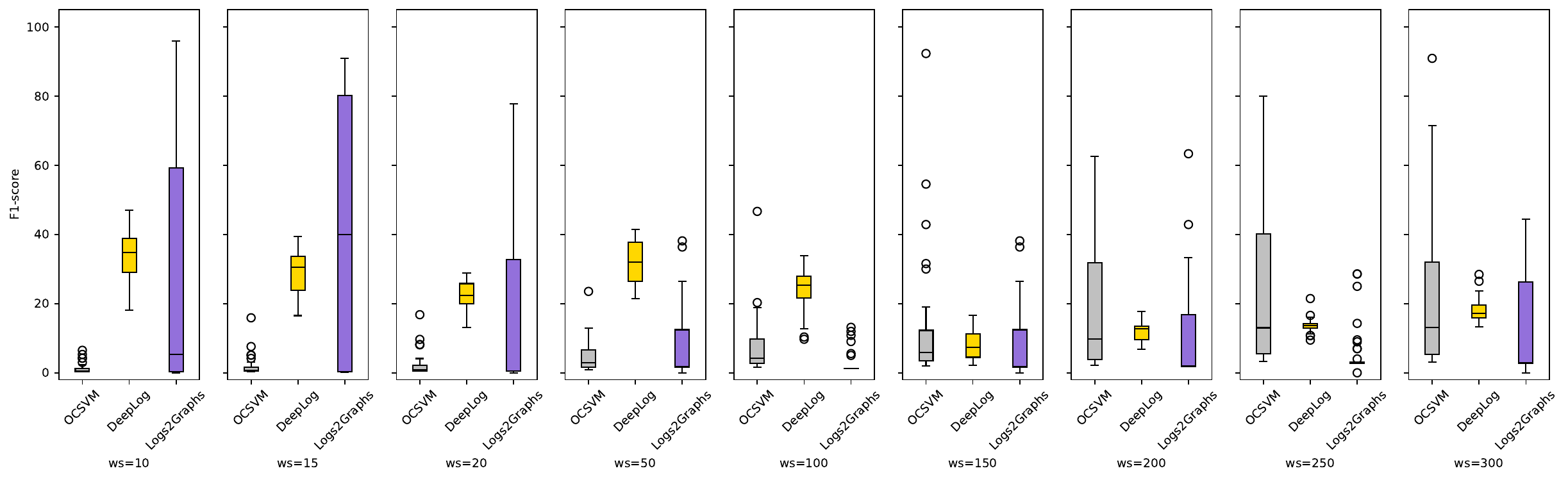}
  \caption{\new{Sensitivity on \hades dataset}}
  \label{figRQ3det:hades}
\end{subfigure}

\begin{subfigure}{1\textwidth}
  \centering
  \includegraphics[height=3.9cm, width=14cm]{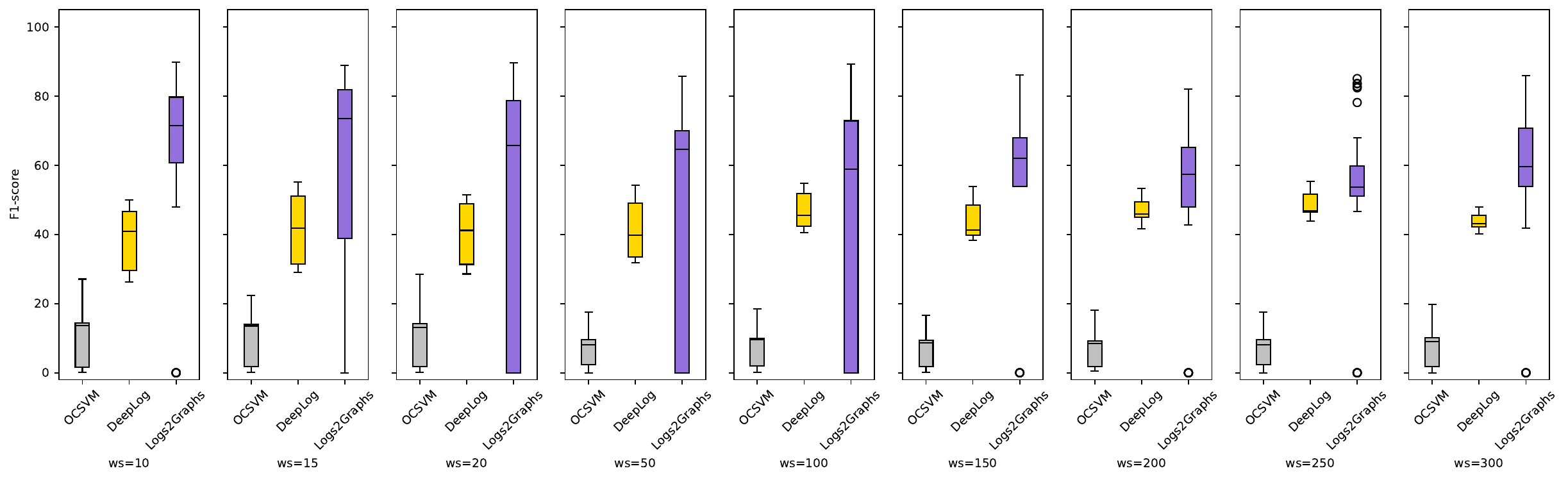}
  \caption{\new{Sensitivity on \bgl dataset}}
  \label{figRQ3det:bgl}
\end{subfigure}
\begin{subfigure}{1\textwidth}
  \centering
  \includegraphics[height=3.9cm, width=14cm]{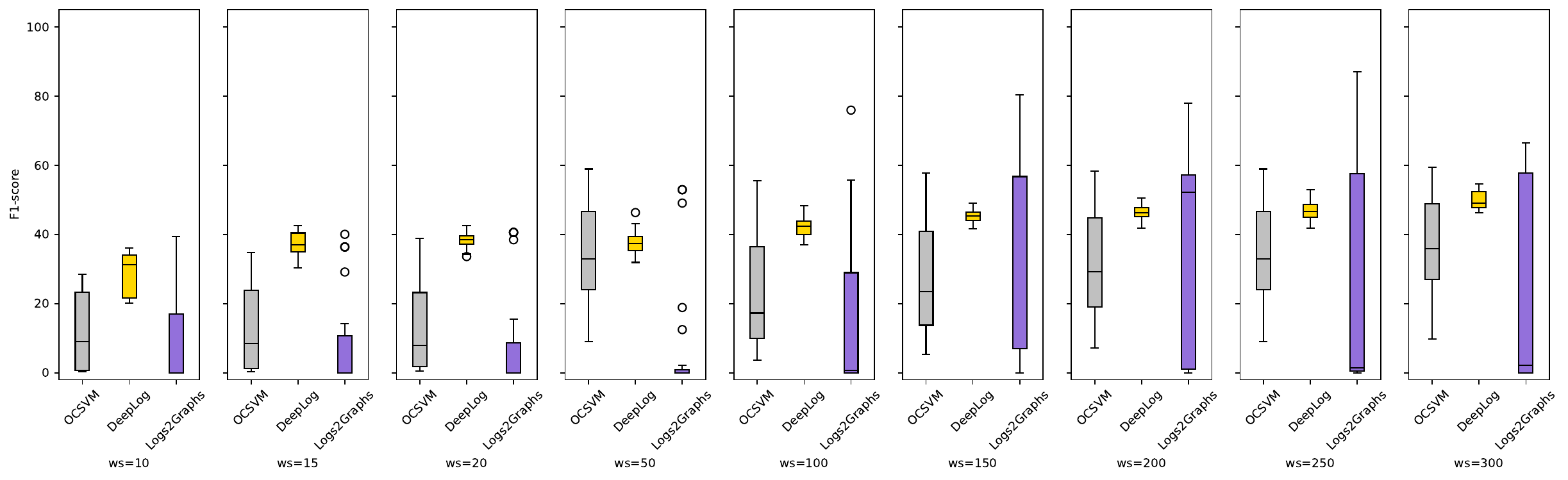}
  \caption{\new{Sensitivity on \tbird dataset}}
  \label{figRQ3det:tbird}
\end{subfigure}

\begin{subfigure}{1\textwidth}
  \centering
  \includegraphics[height=3.9cm, width=14cm]{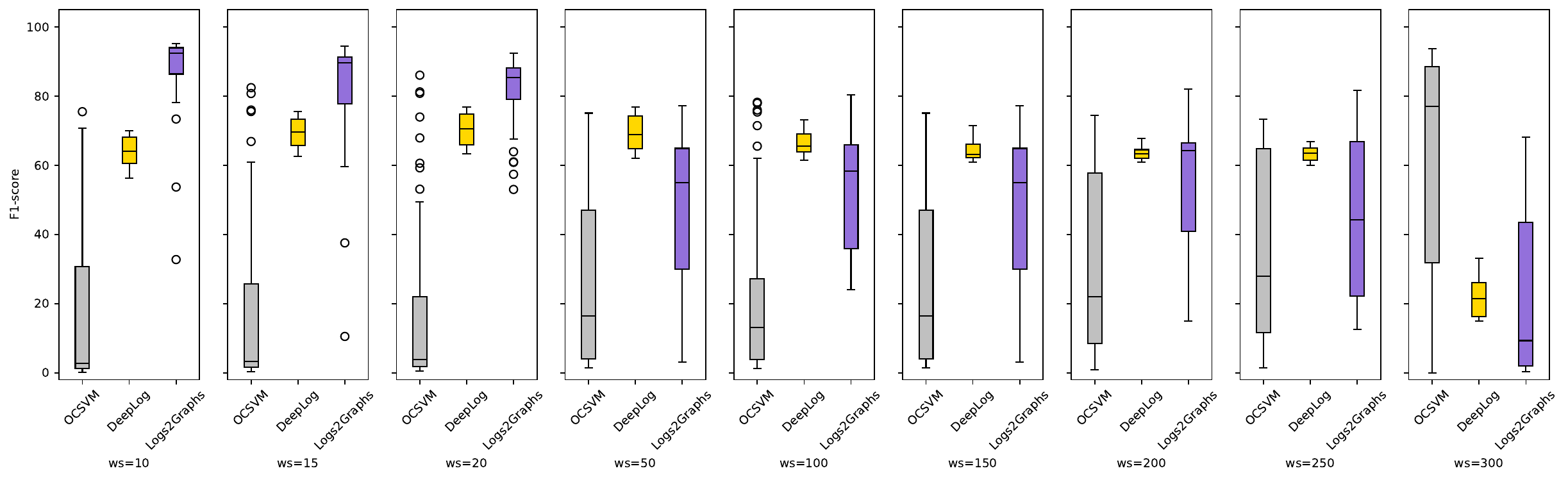}
  \caption{\new{Sensitivity on \spirit dataset}}
  \label{figRQ3det:spirit}
\end{subfigure}
\caption{Sensitivity of the detection accuracy of semi-supervised traditional and deep ML techniques on log message-based datasets}
\label{fig:sens-semi-detacc}
\end{figure}

\figurename~\ref{fig:sens-semi-detacc} shows the sensitivity of the detection accuracy of semi-supervised traditional and deep ML techniques across different window sizes on log message-based datasets.
\begin{itemize}
       \item \textit{Small window sizes $ \{10, 15, 20\}$.}
        \new{On small window sizes, the detection accuracy of semi-supervised deep ML techniques (notably Logs2Graphs) is more sensitive to hyperparameter tuning on more imbalanced datasets (\hades and \bgl) than the less imbalanced ones (\tbird and \spirit). 
        For instance, the detection accuracy (in terms of \emph{F1-score}) of Logs2Graphs on \hades (the most imbalanced dataset) ranges from $0.00$ to $96.00$ (avg $\approx 27.72$, stdDev $\approx 35.13$), whereas the detection accuracy of the latter technique ranges from $32.74$ to $95.12$ (avg $\approx 87.58$, stdDev $\approx 12.13$) on \spirit (the least imbalanced dataset).
        In contrast, OC-SVM is less sensitive to hyperparameter tuning on more imbalanced datasets than the less imbalanced ones.} For instance, on \hades, the detection accuracy of OC-SVM ranges from $0.27$ to $6.50$ (avg $\approx 1.23$, stdDev $\approx 1.53$) with $\mathit{ws}=10$, whereas the detection accuracy of DeepLog ranges from $18.05$ to $46.94$ (avg $\approx 34.21$, stdDev $\approx 7.29$) on the same window size. 
        Overall, the results show that data imbalance has an impact on sensitivity in terms of detection accuracy of semi-supervised, traditional and deep ML techniques. 
        \item \textit{Large window sizes $\{50, 100, 150, 200, 250, 300\}$.} 
        \new{
        The overall detection accuracy of OC-SVM (in terms of \emph{F1-score}) is more sensitive to hyperparameter tuning than that of the remaining semi-supervised deep ML techniques on large window sizes on \spirit and \hades datasets. 
        For instance, on \spirit, with $\mathit{ws}=300$, the detection accuracy of OC-SVM ranges from $0.00$ to $93.76$ (avg $\approx 61.39$, stdDev $\approx 32.31$), whereas the detection accuracy of Logs2Graphs ranges from $0.23$ to $68.13$ (avg $\approx 22.69$, stdDev $\approx 25.10$). 
        However, on \bgl and \tbird, Logs2Graphs shows more sensitivity of detection accuracy to hyperparameter tuning than the remaining semi-supervised techniques. For instance, on \bgl, with $\mathit{ws}=300$, the detection accuracy of Logs2Graphs ranges from $0.00$ to $85.95$ (avg $\approx 53.83$, stdDev $\approx 26.34$) whereas that of the OC-SVM ranges from $0.00$ to $19.84$ (avg $\approx 8.01$, stdDev $\approx 5.19$) and that of DeepLog ranges from $40.18$ to $47.94$ (avg $\approx 43.75$, stdDev $\approx 2.14$), indicating that the latter technique is the most suitable semi-supervised ML technique to detect log anomalies on log message-based datasets on larger contexts.} 
        
\end{itemize}

\new{The statistical test (see \S~\ref{sec:stat-analysis}) indicates that the sensitivity of the detection accuracy of the different semi-supervised ML techniques across hyperparameter settings is significantly different, showing a \emph{p-value} of $0.006$. The results of the post-hoc analysis based on Dunn's test  are shown in Table~\ref{tab:significant_pairs_sens_rq3}, which includes the pairs of ML techniques (Columns \emph{ML.1} and \emph{ML.2}) that show statistically significant differences in terms of the sensitivity of their detection accuracy to hyperparameter tuning.}

\begin{table}[!htbp]
\centering
\begin{tabular}{ccc}
\toprule
\emph{ML.1} & \emph{ML.2} & \emph{p-value} \\
\midrule
\text{DeepLog} & \text{Logs2Graphs} & 0.0077 \\
\text{DeepLog} & \text{OC-SVM} & 0.0476 \\
\bottomrule
\end{tabular}
\caption{Pairs of ML techniques with significant differences in the sensitivity of the \emph{F1-score} to hyperparameter tuning.}
\label{tab:significant_pairs_sens_rq3}
\end{table}

The answer to RQ3 is that \new{all} semi-supervised traditional \new{(OC-SVM)} and deep (DeepLog and Logs2Graphs) ML techniques do not fare well in terms of detection accuracy. 
Moreover, the overall detection accuracy of semi-supervised ML techniques and their sensitivity to hyperparameter tuning vary greatly across datasets.
We also observe that the detection accuracy of the semi-supervised techniques varies across log message-based datasets with different window sizes: 
OC-SVM performs better than DeepLog on small window sizes, when evaluated on \new{\tbird}. Its detection accuracy, however, reaches its maximum on large window sizes on the remaining log message-based datasets. 
\new{Further, Logs2Graphs outperforms OC-SVM and DeepLog on large window sizes, when evaluated on \bgl. 
Its detection accuracy, however, reaches its maximum on smaller window size on the remaining datasets.}

\subsection{RQ4 - Time performance of semi-supervised traditional and deep ML techniques}\label{RQ4results}
\subsubsection{Time Performance}\label{RQ4timeperformance}
Table~\ref{timePerformRQ4semi} shows that DeepLog performs much better, in terms of model re-training time and prediction time, than OC-SVM \new{and Logs2Graphs} on \hdfs: it takes $\approx \SI{40}{\minute}$ to re-train the corresponding model and \SI{100.23}{\second} to detect log anomalies, whereas OC-SVM \new{and Logs2Graphs} take \SI{53896.93}{\second} ($\approx \SI{15}{\hour}$) \new{and \SI{65495.75}{\second} ($\approx \SI{18}{\hour}$)} for the model re-training, and $\approx \SI{1.7}{\hour}$ \new{and $\approx \SI{2}{\minute}$} for log anomaly prediction, \new{respectively}. 
\new{The faster model re-training time of DeepLog is due to its index-based embedding of log event sequences, in contrast to the 300-dimensional vectors fed to other semi-supervised techniques (see Section~\ref{encoding}), resulting in lower-dimensional input.}
\new{However, DeepLog takes much longer than OC-SVM and Logs2Graphs in terms of model re-training time on \hadoop and \fdataset, respectively.}
The significantly higher model re-training time observed for DeepLog compared to OC-SVM \new{and Logs2Graphs} can be explained by the difference in the number of log event sequences processed during their respective model re-training processes: DeepLog is trained with $\num{130172}$ \new{and \num{704474}} log event sequences\footnote{The values \new{$\num{130172}$ and $\num{704474}$ correspond} to the number of sequences obtained by applying a sliding window of size 10 with a step size of 1 \new{on \hadoop and \fdataset, respectively} in DeepLog's implementation~\cite{deeploggithub}.}, whereas OC-SVM \new{ and Logs2Graphs} are trained with 648 \new{and 951} log event sequences only representing $80\%$ of the majority class (see Table~\ref{dataStrategies}) \new{on \hadoop and \fdataset, respectively}. 
However, the prediction time of \new{all semi-supervised} techniques on \new{\hadoop and \fdataset}  is very close ($< \SI{0.01}{\second}$ for OC-SVM, \SI{0.24}{\second} for DeepLog \new{and \SI{0.74}{\second} for Logs2Graphs on \hadoop, whereas $< \SI{0.08}{\second}$ for OC-SVM, \SI{13.13}{\second} for DeepLog and \SI{0.21}{\second} for Logs2Graphs on \fdataset}).
\begin{table}[!htbp]
\setlength\extrarowheight{3pt}
\centering
\caption{Time performance (in seconds) of semi-supervised traditional and deep ML techniques on all datasets} \label{timePerformRQ4semi} 
\begin{NiceTabular}
{m{1cm} m{1.8cm} m{1.4cm} m{1.5cm} m{1.3cm} m{1.8cm}}[hvlines-except-borders] 
\CodeBefore
\rowcolor{gray!15}{1-2}
\Body
\toprule
\Block{2-2}{\emph{Dataset}} & & \Block{2-1}{\emph{Metric}} & \Block{1-3}{\emph{Technique}} \\
& & & OC-SVM & DeepLog & \new{Logs2Graphs}\\

\Block{6-1}{\rotatebox[origin=c]{90}{\emph{Session}}} & \Block{2-1}{\hdfs} & \emph{Re-train.} & \num{53896.93} & \num{2398.04} & \new{\num{65495.75}} \\ 
& & \emph{Pred.} & \num{6180.44} & 100.23 & \new{122.19}\\
& \Block{2-1}{\hadoop} & 
\emph{Re-train.} & 0.01 & 944.49 & \new{132.99}\\ 
& & \emph{Pred.} & 0.00 & 0.24 & \new{0.74}\\
& \Block{2-1}{\new{\fdataset}} & 
\emph{\new{Re-train.}} & \new{0.05} & \new{\num{5110.91}} & \new{90.03}\\ 
& & \emph{\new{Pred.}} & \new{0.07} & \new{13.13} & \new{0.21}\\

\Block{8-1}{\rotatebox[origin=c]{90}{Log message}} & \Block{2-1}{\hades} & \emph{Re-train.} & 0.17  & 229.15 & \new{1170.75}\\ 
& & \emph{Pred.} & 0.04 & 1.85 & \new{4.66}\\ 
& \Block{2-1}{\bgl} & \emph{Re-train.} & 295.87  & \num{1014.52} & \new{54.91} \\ 
& & \emph{Pred.} & 45.20 & 4.56 & \new{1.05} \\ 
& \Block{2-1}{\tbird} & \emph{Re-train.} & 100.25  & 742.90 & \new{1310.45} \\ 
& & \emph{Pred.} & 108.81 & 60.43 & \new{65.37} \\ 
& \Block{2-1}{\spirit} & \emph{Re-train.} & \num{24110.32}  & 866.07 & \new{\num{12326.49}} \\ 
& & \emph{Pred.} & \num{11060.41} & 51.38 & \new{39.07} \\  
\bottomrule
\end{NiceTabular}
\end{table}

Recall that small window sizes lead to more sequences to train the supervised ML models (see Section ~\ref{RQ1detectionaccuracy}).
We observe that, on the \hades, \bgl and \tbird datasets, OC-SVM takes less model re-training time than DeepLog \new{and Logs2Graphs} due to the fewer sequences it uses for training, as compared to DeepLog \new{and Logs2Graphs}. 
More in detail, the highest detection accuracy of OC-SVM on \hades, \bgl and \tbird is associated with larger window sizes (300, 50 and 250, respectively) than DeepLog (10, 10 and 200, respectively) \new{and Logs2Graphs (10, 200 and 15, respectively)}, leading to fewer sequences fed to OC-SVM than those fed to \new{the remaining semi-supervised ML techniques} (see Table~\ref{newDatasetsStrategies}). 
\new{Logs2Graphs, however, takes less model re-training time than OC-SVM on \bgl due to the fewer sequences used for training with $ws=200$.}

We further study the impact of different window sizes on the time performance of semi-supervised traditional and deep ML techniques across log message-based datasets.

\begin{itemize}
       \item \textit{Small window sizes $\{10, 15, 20\}$.}
       As shown in Table~\ref{timePerformRQ4semi}, DeepLog \new{and Logs2Graphs} show a much higher model re-training time than  OC-SVM on \new{\hades.} 
       For instance, on \hades, the model re-training time of DeepLog \new{and Logs2Graphs} is \SI{229.15}{\second} \new{and \SI{1170.75}{\second}} with $\mathit{ws}=10$, whereas OC-SVM takes only \SI{0.17}{\second} with $\mathit{ws}=300$. 
       The longer model re-training time of \new{both semi-supervised deep ML techniques} is expected as \new{they} achieve \new{their} highest detection accuracy with smaller window sizes, resulting in more log event sequences fed to the corresponding ML model\new{s} (see Table~\ref{bestWSFourdatasets} for the window sizes associated with the highest detection accuracy and Table~\ref{newDatasetsStrategies} for the number of log event sequences generated across different window sizes).
       More in detail, on \hades, DeepLog \new{and Logs2Graphs} take \num{83774} sequences in input, generated with $\mathit{ws}=10$, whereas OC-SVM is fed with only \num{2751} sequences generated with $\mathit{ws}=300$
       However, on \spirit, DeepLog outperforms \new{both} OC-SVM \new{and Logs2Graphs} in terms of model re-training time, since it is fed with much fewer log event sequences (\num{184060}, generated with $\mathit{ws}=15$) than OC-SVM \new{and Logs2Graphs}, which \new{are} fed with \num{282913} log event sequences, generated with $\mathit{ws}=10$.
       
       In terms of prediction time, \new{Logs2Graphs} outperforms \new{both} OC-SVM \new{and DeepLog} on \new{\spirit}. For instance, \new{Logs2Graphs} takes only \new{\SI{39.07}{\second}} with $\mathit{ws}=10$, whereas \new{OC-SVM takes \SI{11060.41}{\second} ($\approx \SI{3}{\hour}$) with the same window size and DeepLog takes \SI{51.38}{\second} with $\mathit{ws}=15$.} 
       Further, although OC-SVM is quicker (for prediction) than  DeepLog \new{and Logs2Graphs} on \hades, the difference in prediction time is not significant. This suggests that \new{deep ML techniques (notably Logs2Graphs)} are more suitable at predicting log anomalies, especially for small window sizes.
    \item \textit{Large window sizes $\{50, 100, 150, 200, 250, 300\}$.}
        On large window sizes, \new{Logs2Graphs} is \new{much slower} than \new{OC-SVM and} DeepLog on \tbird. More in detail, \new{Logs2Graphs takes \SI{1310.45}{\second} with $\mathit{ws}=15$, whereas OC-SVM and} 
        DeepLog take \SI{742.9}{\second} with $\mathit{ws}=200$ \new{and \SI{100.25}{\second} with $\mathit{ws}=250$, respectively.} 
        The shorter model re-training time of OC-SVM \new{and DeepLog} is due to the fewer log event sequences fed to the \new{corresponding} model\new{s} on large window sizes.
\end{itemize}

Statistical analysis (see \S~\ref{sec:stat-analysis}) indicates that the time performance of both semi-supervised traditional and deep ML techniques in terms of model re-training and prediction time is not significantly different, showing a \emph{p-value} of $\new{0.4437}$ and $\new{0.9744}$, respectively.

\subsubsection{Sensitivity of Training Time}\label{RQ4SensTimePerformance}
\begin{figure}[tb]
\begin{subfigure}{{.5\textwidth}}
  \includegraphics[width=6cm]{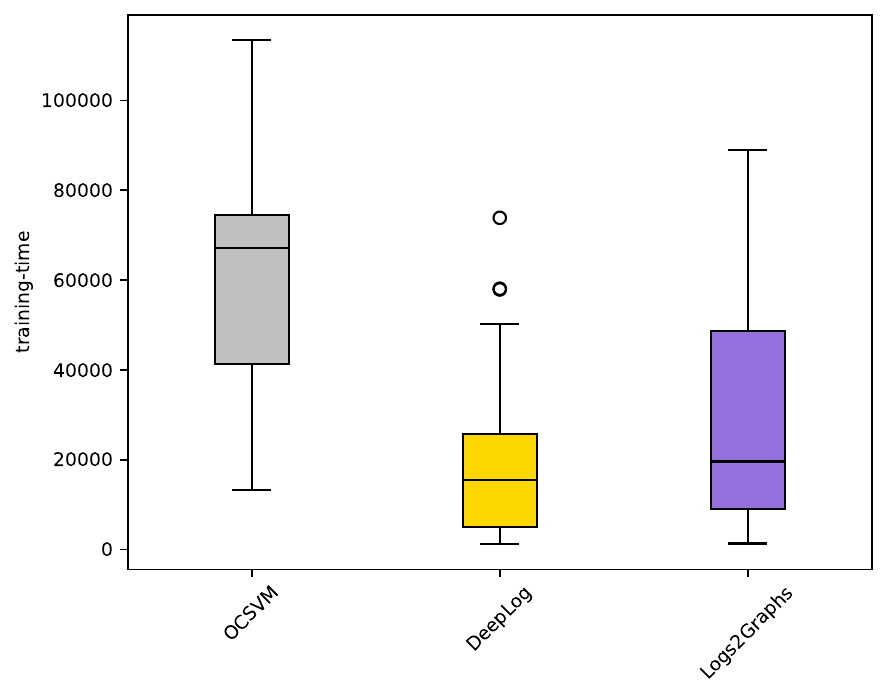}
  \caption{\new{\hdfs dataset}}
  \label{rq4Varocchdfs}
\end{subfigure}\begin{subfigure}{{.5\textwidth}}
  \includegraphics[width=6cm]{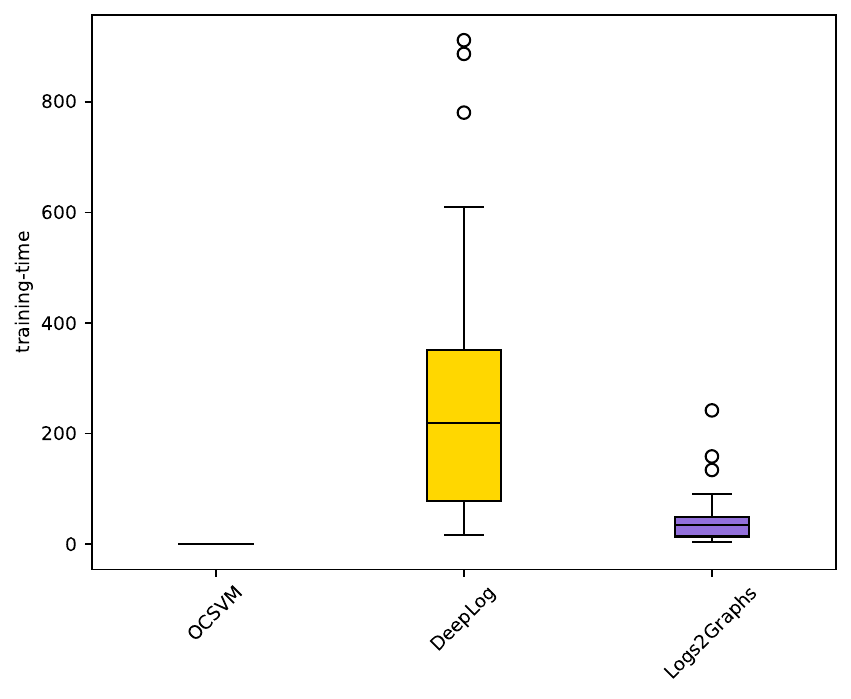}
  \caption{\new{\hadoop dataset}}
  \label{rq4Varocchadoop}
\end{subfigure}

\begin{subfigure}{{\textwidth}}
    \centering
  \includegraphics[width=6cm]{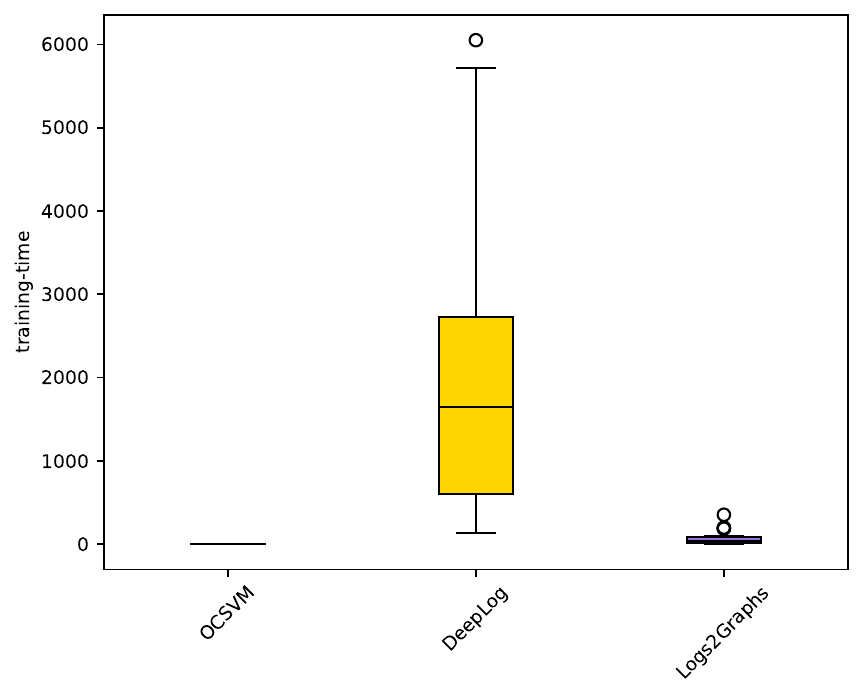}
  \caption{\new{\fdataset dataset}}
  \label{rq4VarTimefdataset}
\end{subfigure}

\caption{Sensitivity of the time performance (in seconds) of semi-supervised traditional and deep ML techniques on session-based datasets (The y-scale of the \new{three} plots is different due to the different training size of the three datasets)}
\label{rq4semiVarTime}
\end{figure}

As depicted in \figurename~\ref{rq4Varocchdfs}, DeepLog is less sensitive to hyperparameter tuning \new{(with outliers)} than OC-SVM \new{and Logs2Graphs} on \hdfs: its model training time ranges from 
$\SI{1183.43}{\second}$ to $\SI{73879.67}{\second}$ (avg $\approx \SI{20133.51}{\second}$, stdDev $\approx \SI{18286.96}{\second}$), whereas OC-SVM takes from $\SI{13157.73}{\second}$ to $\SI{113438.99}{\second}$ (avg $\approx \SI{60369.42}{\second}$, stdDev $\approx \SI{25798.33}{\second}$) \new{and Logs2Graphs takes  from $\SI{1346.73}{\second}$ to $\SI{89010.09}{\second}$ (avg $\approx \SI{30675.87}{\second}$, stdDev $\approx \SI{25202.10}{\second}$)} on the same dataset.
However, on \hadoop \new{and \fdataset} (\figurename~\ref{rq4Varocchadoop} \new{and \figurename~\ref{rq4VarTimefdataset}, respectively}), the time performance of OC-SVM is far less sensitive to hyperparameter tuning than Deeplog \new{and Logs2Graphs, with outliers of the latter techniques on both datasets}.
\new{For instance, on \fdataset, the} model training \new{of OC-SVM} ranges from \new{$\SI{0.02}{\second}$ to $\SI{0.16}{\second}$ (avg $\approx \SI{0.10}{\second}$, stdDev $\approx \SI{0.04}{\second}$)}, whereas DeepLog takes from \new{$\SI{139.69}{\second}$ to $\SI{6050.34}{\second}$ (avg $\approx \SI{1972.07}{\second}$, stdDev $\approx \SI{1665.83}{\second}$) and Logs2Graphs takes from $\SI{4.06}{\second}$ to $\SI{353.71}{\second}$ (avg $\approx \SI{65.05}{\second}$, stdDev $\approx \SI{74.23}{\second}$)} for its model training.

\begin{figure}[p]
\centering
\begin{subfigure}{1\textwidth}
  \centering
  \includegraphics[height=3.9cm, width=14cm]{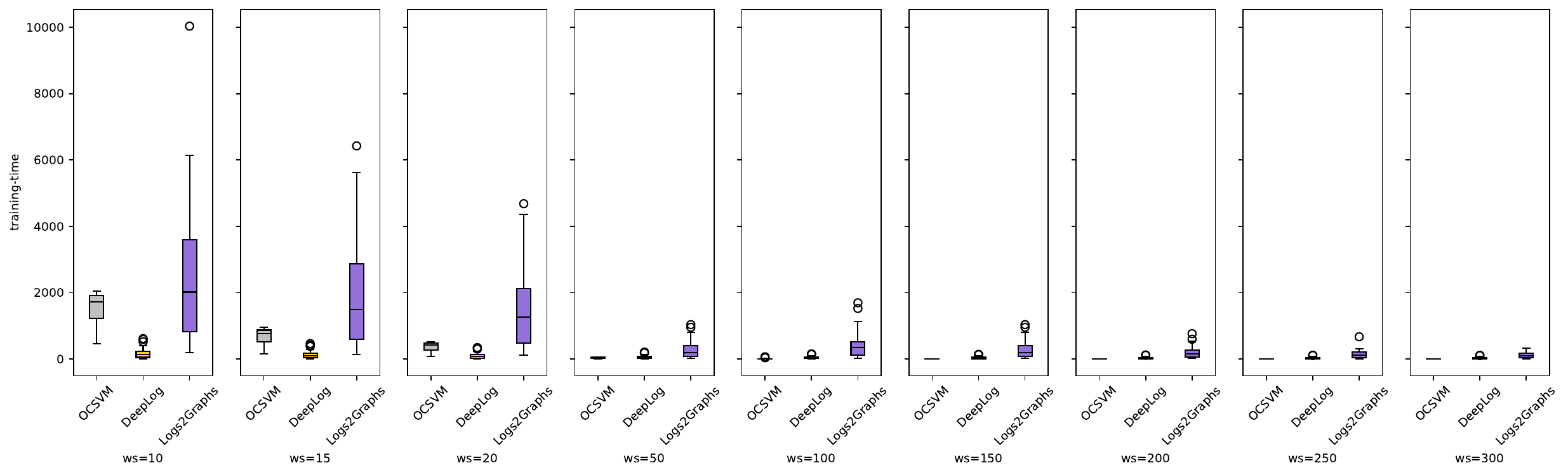}
  \caption{\new{Sensitivity on \hades dataset}}
  \label{figRQ4time:hades}
\end{subfigure}

\begin{subfigure}{1\textwidth}
  \centering
  \includegraphics[height=3.9cm, width=14cm]{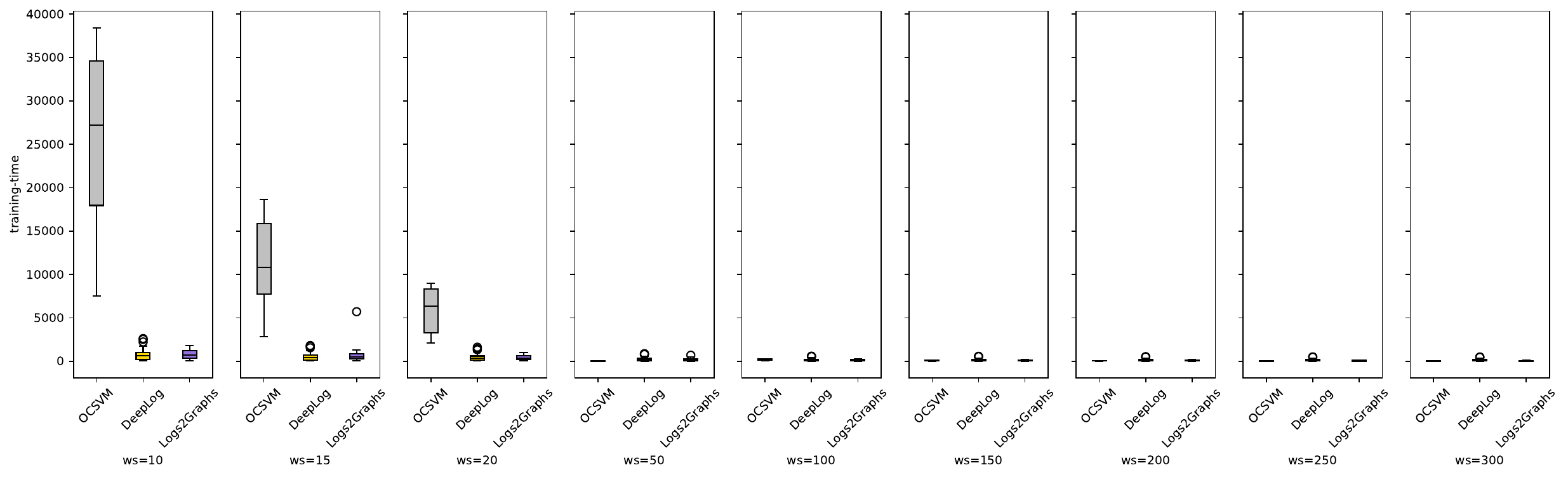}
  \caption{\new{Sensitivity on \bgl dataset}}
  \label{figRQ4time:bgl}
\end{subfigure}
\begin{subfigure}{1\textwidth}
  \centering
  \includegraphics[height=3.9cm, width=14cm]{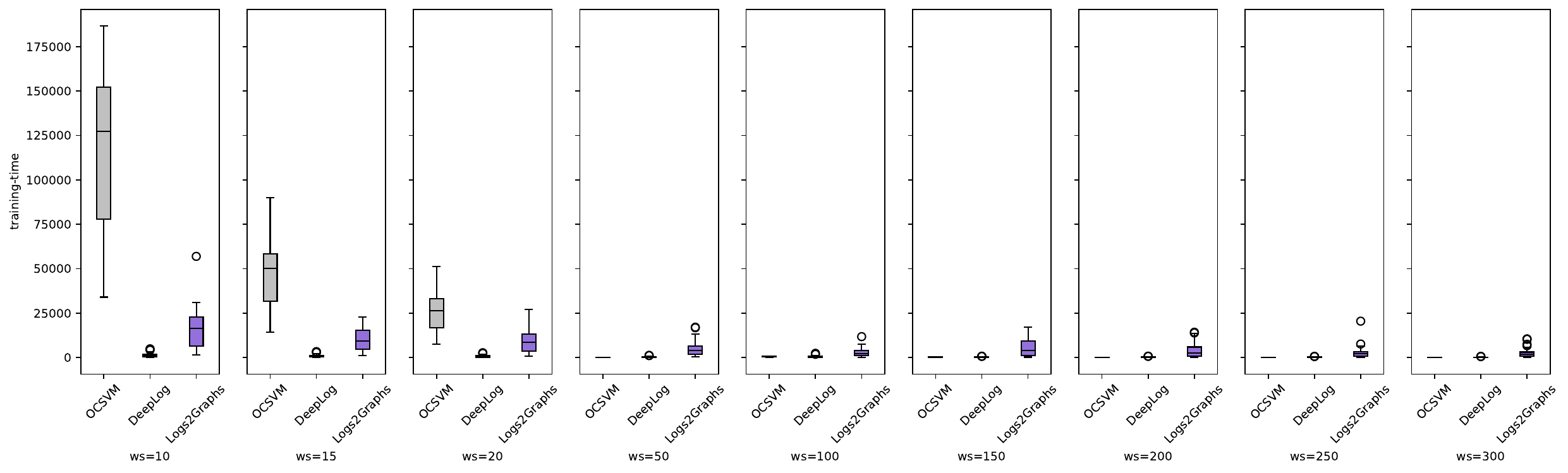}
  \caption{\new{Sensitivity on \tbird dataset}}
  \label{figRQ4time:tbird}
\end{subfigure}

\begin{subfigure}{1\textwidth}
  \centering
  \includegraphics[height=3.9cm, width=14cm]{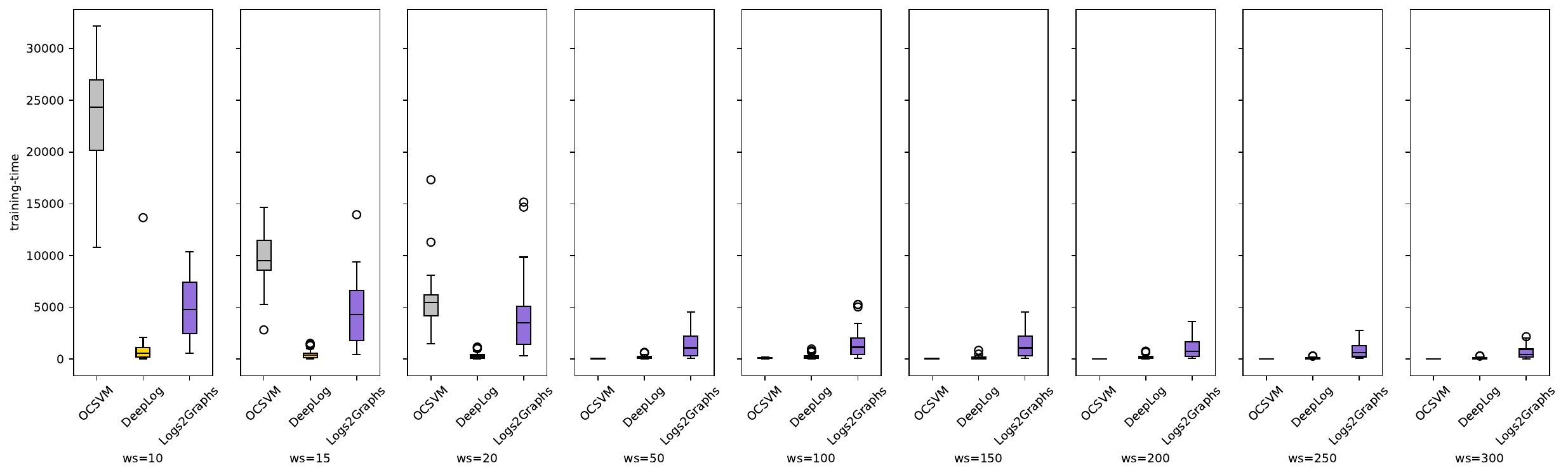}
  \caption{\new{Sensitivity on \spirit dataset}}
  \label{figRQ4time:spirit}
\end{subfigure}
\caption{Sensitivity of the time performance (in seconds) of semi-supervised traditional and deep ML techniques on log message-based datasets}
\label{fig:sens-semi-time}
\end{figure}

\figurename~\ref{fig:sens-semi-time} shows the sensitivity of training time of semi-supervised traditional and deep ML
techniques across different window sizes on log message-based datasets.
\begin{itemize}
    \item \textit{Small window sizes $ \{10, 15, 20\}$.}
 \new{Overall,} DeepLog shows  much less sensitivity in terms of model training time to hyperparameter tuning than OC-SVM \new{and Logs2Graphs}, on small window sizes across all the log message-based datasets. For instance, on \hades (see \figurename~\ref{figRQ4time:hades}) with $\mathit{ws}=10$,  \new{DeepLog takes from $\SI{12.18}{\second}$ to $\SI{618.90}{\second}$ (avg $\approx \SI{184.27}{\second}$, stdDev $\approx \SI{162.77}{\second}$) for its model training whereas}
    OC-SVM takes from $\SI{460.82}{\second}$ to 
    $\SI{2054.20}{\second}$ (avg $\approx \SI{1498.93}{\second}$, stdDev $\approx \SI{511.85}{\second}$),  while \new{Logs2Graphs takes from $\SI{193.34}{\second}$ to $\SI{10034.02}{\second}$ (avg $\approx \SI{2520.01}{\second}$, stdDev $\approx \SI{2194.36}{\second}$)} on the same dataset and window size. 
    This indicates that training size has more impact on the sensitivity of OC-SVM \new{and Logs2Graphs}  to hyperparameter tuning than that of DeepLog in terms of model training time.
    
    \item \textit{Large window sizes $\{50, 100, 150, 200, 250, 300\}$.}
    Overall, \new{all} the semi-supervised ML techniques show less sensitive model training time to hyperparameter tuning on large window sizes, ranging from $\mathit{ws}=50$ to $\mathit{ws}=300$ across all log message-based datasets. 
    The model training time of OC-SVM, however, is less sensitive to hyperparameter tuning than that of \new{the remaining semi-supervised techniques}. For instance, on \new{\spirit, with $\mathit{ws}=150$, OC-SVM takes from $\SI{10.44}{\second}$ to $\SI{61.20}{\second}$ (avg $\approx \SI{38.83}{\second}$, stdDev $\approx \SI{15.07}{\second}$), whereas the training time of DeepLog ranges from $\SI{13.52}{\second}$ to $\SI{847.51}{\second}$ (avg $\approx \SI{154.71}{\second}$, stdDev $\approx \SI{166.61}{\second}$) and Logs2Graphs ranges from $\SI{83.91}{\second}$ to $\SI{4523.87}{\second}$ (avg $\approx \SI{1381.98}{\second}$, stdDev $\approx \SI{1281.64}{\second}$).
    Thus, training size affects more significantly the sensitivity of the model training time of DeepLog and Logs2Graphs than OC-SVM.
    }
    
\end{itemize}

Statistical analysis (see \S~\ref{sec:stat-analysis}) shows that sensitivity of the training time of the semi-supervised traditional and deep ML techniques across hyperparameter settings is not significantly different, with a \emph{p-value} of $\new{0.2}$.

The answer to RQ4 is that the time performance of semi-supervised traditional and deep ML techniques and the sensitivity of their model training time to hyperparameter tuning greatly vary across datasets. We therefore cannot draw general conclusions with that respect. 

More in detail, OC-SVM shows a better performance in terms of i) model re-training time than 
\new{DeepLog on \hadoop, \fdataset, \hades, \bgl and \tbird and Logs2Graphs on \hdfs, \hadoop, \fdataset, \hades and \tbird and ii) prediction time than DeepLog and Logs2Graphs on \hadoop, \fdataset, and \hades.}
DeepLog, however, is faster than OC-SVM and Logs2Graphs in terms of i) model re-training on \hdfs and \spirit and ii) prediction time on \hdfs and \tbird. 

Further, the time performance of OC-SVM is less sensitive to hyperparameter tuning than that of DeepLog \new{and Logs2Graphs} on the session-based datasets (\hadoop and \fdataset) and all log message-based datasets (\hades, \bgl, \tbird and \spirit) with large window sizes, ranging from 50 to 300.
\new{In contrast, the model training time of DeepLog is less sensitive to hyperparameter tuning than that of OC-SVM and Logs2Graphs on \hdfs and all log message-based datasets on small window sizes, ranging from 10 to 20.}
Besides, OC-SVM is faster in terms of model re-training than DeepLog \new{and Logs2Graphs} on large window sizes, as the former technique is fed with fewer log event sequences.
\new{Logs2Graphs}, however, is faster at predicting log anomalies than OC-SVM \new{and DeepLog} on small window sizes. 


\subsection{Discussion}\label{discussion}

\subsubsection{Findings and Implications}\label{findingsAndImplications}
In this empirical study, we have systematically evaluated and compared 
\new{nine} supervised and semi-supervised, traditional and deep ML techniques in terms of detection accuracy, time performance, and sensitivity of their i) detection accuracy and ii) time performance to hyperparameter tuning on \new{seven} benchmark datasets. Overall, the answer to the four research questions addressed in this study suggests that more attention should be given to traditional supervised ML techniques when it comes to the detection of \ep log anomalies before considering more complex, deep ML techniques.

Our findings indicate that supervised traditional and deep ML techniques i) perform very closely in terms of detection accuracy and prediction time \new{on most of the benchmark datasets} and ii) far outperform semi-supervised ML techniques. 
Further, supervised traditional ML techniques show less sensitivity to hyperparameter tuning compared to deep ML techniques.
Despite the considerable emphasis, across the scientific literature, on deep learning-based ML techniques for addressing the LAD problem, our study shows that traditional ML techniques (notably RF) are competitive with \new{deep learning} ones on \new{seven} benchmark datasets w.r.t. four evaluation criteria that are relevant in practice. \emph{RF is therefore our recommendation to detect \ep log anomalies from a practical standpoint.} 

Moreover, our findings show that most of the supervised ML techniques yield their highest detection accuracy on small window sizes (10, 15 and 20). This is a rather useful insight as small window sizes enable the earlier detection of anomalies in practice. 
\new{We also remark that deep semi-supervised ML techniques perform better on small window sizes across most of the log message-based datasets. However, this trend is not applicable to the traditional semi-supervised OC-SVM, which tends to perform better on larger window sizes (from 50 to 300) across most of the datasets.
Overall, our findings suggest that for supervised ML techniques, smaller window sizes are recommended as they generate more training sequences, leading to higher detection accuracy due to the data-hungry nature of these models. For semi-supervised techniques, practitioners should adjust the window size based on the model: smaller window sizes work better for deep semi-supervised models, while larger window sizes are more effective for traditional methods like OC-SVM to achieve high detection accuracy.
}

\new{Further, in terms of the best hyperparameter settings observed for the different ML techniques in our experiments, RF tends to achieve its highest detection accuracy with higher numbers of decision trees ($\mathit{dTr} \geq 50$) across all datasets, indicating that a large number of decision trees is generally required to guarantee a high detection accuracy. 
For SVM, the best value of hyperparameter $\mathit{C}$ was 1000 across most of the datasets, indicating that a strong regularization parameter is often required to handle the complexity of decision boundaries in log-based anomaly detection datasets. Further, the highest detection accuracy of LSTM is observed on the $\mathit{adam}$ optimizer on six out of the seven datasets, except for BGL, on which the best optimizer was $\mathit{rmsprop}$. LogRobust showed its highest detection accuracy with the $\mathit{rmsprop}$ optimizer across most datasets, except for BGL and Spirit, where $\mathit{adam}$ was preferred, indicating that the choice of optimizer is dataset-dependent. For the number of epochs ($\mathit{epN}$), the optimal values ranged from 10 to 150, with most datasets requiring higher values ($\mathit{epN}\geq 100$), suggesting that a sufficient number of epochs is necessary for LogRobust to achieve an optimal detection accuracy across datasets. Similarly, the two versions of NeuralLog (NeuralLog1 and NeuralLog2) showed their highest detection accuracy with higher number of epochs ($\mathit{epN}$) within the range {100, 150} across most of the datasets, indicating that transformer-based techniques require a large number of epochs to effectively capture the intricate log patterns in log-based data to detect log anomalies.}
\new{
For semi-supervised ML techniques (OC-SVM, DeepLog, and Logs2Graphs), no consistent trends in best hyperparameter settings were observed across datasets. For instance, in OC-SVM, the hyperparameter $\nu$ varied greatly across datasets, with values such as 0.2, 0.1, and 0.9, reflecting a wide range of optimal regularization values depending on the dataset. 
}

These findings may guide AIOps engineers in selecting the right ML technique, to find a trade-off between detection accuracy and time performance when addressing the LAD problem.
The hyperparameter tuning conducted in this study allows AIOps engineers to assess the suitability of a specific ML technique to detect log anomalies for a specific context and dataset w.r.t. their overall detection accuracy, their time performance (model training time and prediction time) and their sensitivity to hyperparameter tuning. 
Moreover, AIOps engineers can prioritize the tuning of hyperparameters that have the most significant impact on the detection accuracy and time performance of the model of their choice, thus reducing the required time and computational resources.

\subsubsection{Threats to validity}\label{threats}
Two types of 
threats to validity can affect the findings of our study. 

\textit{Internal threats.} 
We relied on publicly available implementations of DeepLog~\cite{deeploggithub}, LogRobust~\cite{logadempirical}, \new{ NeuralLog~\cite{neuralloggithub} and Logs2Graphs~\cite{logs2graphsgithub}}.
These third-party implementations might be faulty and could introduce bias in the results. To mitigate this, we carefully performed code reviews and used the replication package of the existing empirical study~\cite{le2022logHowFar}.
We remark that most of the results reported in LAD studies~\cite{le2022logHowFar, liu2021lognads, qi2022adanomaly, du2021log, xie2022loggd}, for most of the ML techniques we used in our work (e.g., DeepLog, LogRobust, SVM, RF), 
are not reproducible, mostly because hyperparameter settings are not fully shared by these studies. We also note that most of the LAD studies do not share their code, making it more difficult for us to reproduce the same results. This has been also confirmed by~\cite{landauer2024critical}, who studied the characteristics of common benchmark datasets and their impact on the effectiveness of existing ML techniques at detecting \ep log anomalies.    
The internal validity of our empirical study could also be threatened by the choice of specific window sizes; other window size values could lead to different results in terms of detection accuracy and time performance. To mitigate this, we considered various fixed window sizes, including the ones that have been adopted in existing studies (see Table~\ref{bestF1_windowSizes}).

Another threat is the choice of different hyperparameter settings for each ML technique, which we had to limit due to the high computational cost (notably the model training time) of our experiments.
To mitigate this, we motivated the choice of different hyperparameter settings for each ML technique based on the literature (\S~\ref{hyperParameterSettings}).
Different results in terms of detection accuracy and time performance could be obtained with different hyperparameter settings.

\new{Further, in this paper, we have not considered ways to enhance the detection accuracy, such as improving data preprocessing. This omission could also impact the detection accuracy and generalizability of the results.
}
\new{
While we acknowledge the impact of data preprocessing on the detection accuracy of ML techniques, recent empirical studies~\cite{khan2024impact, wu2023effectiveness} suggest that certain preprocessing improvements, such as refining log parsing and log representation techniques, may not significantly enhance the detection accuracy for log-based anomaly detection techniques. For instance, ~\citet{khan2024impact} found no strong correlation between log parsing accuracy and the anomaly detection accuracy, and ~\citet{ wu2023effectiveness} showed that semantic-based log representations yielded similar detection accuracy across different techniques. 
}

\textit{External threats.} 
The selection of only \new{three} semi-supervised and \new{five} supervised traditional and deep ML techniques may limit the generalization of our findings. To mitigate this threat, we relied on commonly adopted \new{and diverse} supervised and semi-supervised, traditional (\S~\ref{traditional}) and deep (\S~\ref{deep}) ML techniques, \new{including RNN, Transformers and GNN-based learning models} from \new{recent studies}.

 \section{Conclusion and Future Work} \label{conclusion}

In this large empirical study, we assessed the anomaly detection accuracy and the time performance, in terms of model training and log anomaly prediction, of different semi-supervised and supervised, traditional and deep ML techniques. We further studied the sensitivity of detection accuracy and model training time, for each of these techniques, to hyperparameter tuning across datasets. This is of significant importance for practitioners as using techniques that are less sensitive reduces the effort entailed by applying them. 

Our study shows that supervised traditional and deep ML techniques fare similarly in terms of detection accuracy and prediction time \new{on most of the benchmark datasets}. 
Further, as expected, supervised traditional ML techniques far outperform supervised deep learning ones in terms of re-training time. Among the former, Random Forest shows the least sensitivity to hyperparameter tuning regarding its detection accuracy and time performance.

Though they offer advantages when dealing with datasets containing few anomalies, semi-supervised techniques yield significantly worse detection accuracy than supervised ML techniques. The time performance and sensitivity to hyperparameter tuning of semi-supervised traditional and deep ML techniques widely vary across datasets.

The results of this study enable system and AIOps engineers to select the most accurate ML technique for detecting log anomalies, taking into account time performance which has significant practical implications. 
Though they need to be confirmed with further studies, our results are of practical importance because they suggest---when accounting for accuracy, training and prediction time, and sensitivity to hyperparameter tuning---that supervised, traditional techniques are a better option for log anomaly detection, with a preference for Random Forest. Given the emphasis on the use of deep learning in the research literature, this may come as a surprise. 

\new{As part of future directions, we plan to study the impact of different data distributions and model complexity on the detection accuracy and time performance of the different LAD techniques. Further, }
\new{d}etecting log anomalies is not sufficient for system engineers as it does not provide them with enough details about the cause(s) of anomalies. This warrants the design of solutions to facilitate the diagnosis of anomalies, which we will address as part of future work. 
 
\backmatter

\bmhead{Acknowledgements}
The experiments conducted in this work were enabled by support provided by the Digital Research Alliance of Canada. 
We thank Nathan Aschbacher for his feedback on earlier drafts of this article.

\section*{Declarations}

\bmhead{Funding}
This work was supported by the Natural Sciences and Research Council of Canada (NSERC) Discovery Grant program, the Canada Research Chairs (CRC) program, the Mitacs Accelerate program, the Ontario Graduate Scholarship (OGS) program, and the Luxembourg National Research Fund (FNR), grant reference C22/IS/17373407/LOGODOR; Lionel Briand was partly funded by the Research Ireland grant 13/RC/2094-2.
For the purpose of open access, and in fulfillment of the obligations arising from the grant agreement, the authors have applied a Creative Commons Attribution 4.0 International (CC BY 4.0) license to any Author Accepted Manuscript version arising from this submission.
\bmhead{Ethical approval} not applicable.
\bmhead{Informed consent} not applicable.
\bmhead{Author Contributions}
Shan Ali: Conceptualization, Investigation, Data Curation, Formal analysis, Software, Writing - original draft. Chaima Boufaied: Conceptualization, Investigation, Data Curation, Formal analysis, Writing - original draft. Domenico Bianculli: Conceptualization, Supervision, Writing - review \& editing. Paula Branco: Conceptualization, Supervision, Writing - review \& editing. Lionel Briand: Supervision, Writing - review \& editing, Funding acquisition.

\bmhead{Data availability Statement}
The replication package accompanying this work is available 
at \url{https://figshare.com/articles/software/LADEmpStudy/22756871?file=50577753}. 
We make available i) the pre-processed datasets, as well as the corresponding pre-processing scripts; ii) the implementations of the different alternative traditional and deep ML techniques considered in this study; and iii) the detailed results.

\bmhead{Conflicts of Interest} The authors declare that they have no known competing financial interests or personal relationships that could have appeared
to influence the work reported in this paper.

\bmhead{Clinical Trial Number} not applicable.

\end{document}